\documentclass[12pt]{article}

\usepackage[utf8]{inputenc}
\usepackage{textcomp}
\usepackage{mathtools}
\usepackage{amsmath}
\usepackage[shortlabels]{enumitem}
\usepackage{chetnew}
\usepackage{tikz}
\usepackage{amssymb,amsfonts,mathrsfs,dsfont,yfonts,bbm}\usetikzlibrary{calc}
\usepackage{xcolor}
\usepackage{braket}
\usepackage[normalem]{ulem}
\usepackage{import}
\usepackage{booktabs}
\usepackage{varioref}
\usepackage{bm}

\newcommand{\be}{\begin{equation}}
\newcommand{\ee}{\end{equation}}
\newcommand{\bea}{\begin{eqnarray}}
\newcommand{\eea}{\end{eqnarray}}

\newcommand{\DZ}{\mathbb{Z}}
\newcommand{\CL}{\mathcal{L}}
\newcommand{\CA}{\mathcal{A}}

\newcommand{\CQ}{\mathcal{Q}}

\newcommand{\CB}{\mathcal{B}}
\newcommand{\CC}{\mathcal{C}}

\newcommand{\CZ}{\mathcal{Z}}
\newcommand{\CO}{\mathcal{O}}
\newcommand{\CT}{\mathcal{T}}
\newcommand{\CV}{\mathcal{V}}
\newcommand{\CI}{\mathcal{I}}
\newcommand{\CN}{\mathcal{N}}
\newcommand{\CS}{\mathcal{S}}
\newcommand{\CM}{\mathcal{M}}

\newcommand{\mathdash}{\relbar\mkern-20mu\relbar}

\newcommand*{\boxcoloro}{orange}
\makeatletter
\newcommand{\boxedo}[1]{\textcolor{\boxcoloro}{%
\tikz[baseline={([yshift=-1ex]current bounding box.center)}] \node [rectangle, minimum width=1ex,rounded corners,draw] {\normalcolor\m@th$\displaystyle#1$};}}
 \makeatother
\newcommand*{\boxcolorr}{red}
\makeatletter
\newcommand{\boxedr}[1]{\textcolor{\boxcolorr}{%
\tikz[baseline={([yshift=-1ex]current bounding box.center)}] \node [rectangle, minimum width=1ex,rounded corners,draw] {\normalcolor\m@th$\displaystyle#1$};}}
 \makeatother
\newcommand*{\boxcolorb}{blue}
\makeatletter
\newcommand{\boxedb}[1]{\textcolor{\boxcolorb}{%
\tikz[baseline={([yshift=-1ex]current bounding box.center)}] \node [rectangle, minimum width=1ex,rounded corners,draw] {\normalcolor\m@th$\displaystyle#1$};}}
 \makeatother
\newcommand*{\boxcolorg}{green}
\makeatletter
\newcommand{\boxedg}[1]{\textcolor{\boxcolorg}{%
\tikz[baseline={([yshift=-1ex]current bounding box.center)}] \node [rectangle, minimum width=1ex,rounded corners,draw] {\normalcolor\m@th$\displaystyle#1$};}}
 \makeatother
 \newcommand*{\boxcolorp}{purple}
\makeatletter
\newcommand{\boxedp}[1]{\textcolor{\boxcolorp}{%
\tikz[baseline={([yshift=-1ex]current bounding box.center)}] \node [rectangle, minimum width=1ex,rounded corners,draw] {\normalcolor\m@th$\displaystyle#1$};}}
 \makeatother
  \newcommand*{\boxcolorc}{cyan}
\makeatletter
\newcommand{\boxedc}[1]{\textcolor{\boxcolorc}{%
\tikz[baseline={([yshift=-1ex]current bounding box.center)}] \node [rectangle, minimum width=1ex,rounded corners,draw] {\normalcolor\m@th$\displaystyle#1$};}}
 \makeatother
  \newcommand*{\boxcolory}{yellow}
\makeatletter
\newcommand{\boxedy}[1]{\textcolor{\boxcolory}{%
\tikz[baseline={([yshift=-1ex]current bounding box.center)}] \node [rectangle, minimum width=1ex,rounded corners,draw] {\normalcolor\m@th$\displaystyle#1$};}}
 \makeatother

\usepackage{amsmath}

\title{On the Classification of Bosonic and \\ Fermionic One-Form Symmetries in $2+1$d \\[5mm] and 't Hooft Anomaly Matching}

\author{Mahesh Balasubramanian,$^{\between}$ Matthew Buican,$^{\between}$ and Rajath Radhakrishnan$^{\boxtimes}$}

\affiliation{$^{\between}$\smallskip CTP and Department of Physics and Astronomy\\
Queen Mary University of London, London E1 4NS, UK \\[2mm]$^{\boxtimes}$ International Centre for Theoretical Physics, Strada Costiera 11, Trieste 34151, Italy}

\abstract{Motivated by the fundamental role that bosonic and fermionic symmetries play in physics, we study (non-invertible) one-form symmetries in $2+1$d consisting of topological lines with bosonic and fermionic self-statistics. We refer to these lines as Bose-Fermi-Braided (BFB) symmetries and argue that they can be classified. Unlike the case of generic anyonic lines, BFB symmetries are closely related to groups. In particular, when BFB lines are non-invertible, they are non-intrinsically non-invertible. Moreover, BFB symmetries are, in a categorical sense, weakly group theoretical. Using this understanding, we study invariants of renormalization group flows involving non-topological QFTs with BFB symmetry.}

\begin{document}
\setcounter{tocdepth}{2}
\maketitle
\toc

\newsec{Introduction}
Bosonic and fermionic groups and algebras play foundational roles in constraining QFTs in various dimensions (e.g., see the classic results of \cite{Coleman:1967ad,Haag:1974qh,Nahm:1977tg}). Recently, considerable attention has been paid to a vast generalization of the notion of symmetry in which one replaces groups with categories of topological defects implementing symmetries that are generally non-invertible (e.g., see \cite{Freed:2022qnc,Carqueville:2023jhb,Schafer-Nameki:2023jdn} for recent reviews).

In the context of $2+1$d QFTs, quasi-particles and their braided worldlines often obey an interesting and relatively \lq\lq wild" anyonic, or fractional, generalization of bose / fermi statistics characterized by complex phases and, more generally, unitary matrices \cite{Leinaas:1977fm,Wilczek:1982wy}. When they are topological, these $2+1$d anyonic lines generate one-form symmetries that are typically non-invertible (e.g., as in the case of Wilson lines in generic non-Abelian Chern-Simons theories).\footnote{Although the original definition of one-form symmetry \cite{Gaiotto:2014kfa} involved invertible symmetries, we define one-form symmetry to also cover the case of a non-invertible categorical symmetry generated (in $2+1$d) by topological lines.}

Given this picture, a natural first question is to try to classify the one-form symmetries arising from topological lines in $2+1$d that are bosonic and fermionic (here we define bosonic and fermionic lines to have self-statistics, or topological spin, $\theta=1$ and $\theta=-1$ respectively).\footnote{Since these lines have real self-statistics, this question amounts, in some sense, to classifying \lq\lq non-Anyonic" symmetries. However, this notion is somewhat misleading because of the possibility of non-trivial (but, as we will see, still real) mutual statistics (e.g., consider Kitaev's toric code).} One might imagine that, unlike more generic one-form symmetries, these one-form symmetries are closely related to groups. Indeed, we will show this intuition is correct by demonstrating that:

\bigskip
\noindent
{\it Any symmetry category, $\CB$, consisting solely of bosonic and fermionic lines is  related to groups in (at least) two ways: {\bf(1)} $\CB$ is weakly group theoretical (in the categorical sense \cite{etingof2016tensor}) and {\bf(2)} If $\CB$ is non-invertible, it is non-intrinsically non-invertible.\footnote{We define a non-invertible one-form symmetry, $\CB$, to be non-intrinsically non-invertible if there is a topological manipulation, $\rho$, such that $\rho(\CB)$ only contains invertible genuine line operators (note that, throughout this work, when we refer to line operators, we have in mind genuine lines). This definition is closely related to the notion of non-intrinsically non-invertible symmetry in 1+1d QFTs studied in \cite{Kaidi:2023maf,Kaidi:2022cpf,Sun:2023xxv}. However, note that in 1+1d, this definition is equivalent to a fusion category being group-theoretical in the sense of \cite{etingof2016tensor} (see also \cite{Sun:2023xxv}). This statement is no longer true in 2+1d. For example, $\text{Ising} \boxtimes \overline{\text{Ising}}$ is non-intrinsically non-invertible in 2+1d because it can be obtained form gauging the $\DZ_2$ electromagnetic duality symmetry of the 2+1d $\DZ_2$ Dijkgraaf-Witten theory. However, $\text{Ising} \boxtimes \overline{\text{Ising}}$ is not group-theoretical. \label{definitionNINI}}} 

\bigskip
\noindent
Depending on the context, we will refer to such $\CB$ symmetries as Bose-Fermi-Braided (BFB) symmetries or BFB categories. Roughly speaking, the properties described in the italics above mean that any BFB category can, through topological manipulations, be related to invertible objects forming a group.\footnote{Interestingly, two of the present authors recently showed that non-invertible zero-form symmetries in $2+1$d generated by condensing lines are also non-intrinsically non-invertible \cite{Buican:2023bzl}. In that context, non-invertibility arises from the presence of non-trivial bosons that can be condensed in all of spacetime. Below, we will see that such bosons are also necessary (though, unlike in the zero-form case, not sufficient) to realize non-invertible BFB categories. \label{NINI}} Note that general anyonic one-form symmetries are neither weakly group theoretical nor intrinsically non-invertible (a particularly simple example is the Fibonacci category arising from the two lines in $(G_2)_1$ Chern-Simons theory).

The relative \lq\lq tameness" of BFB categories allows us to get a handle on these symmetries and argue that:
\begin{itemize}
\item All BFB symmetries can be classified (with the classification of BFB categories lacking a transparent fermion being particularly explicit) and realized. It is rare that infinite families of (non-Abelian) symmetry categories can be classified (exceptions include categories with trivial braiding, which are simple examples of BFB categories, and \lq\lq metaplectic" modular categories \cite{ardonne2016classification}\footnote{Metaplectic modular categories are closely related to the categories we consider here in the sense that they are also related to Chern-Simons theories with ${\rm Spin}(N)$ gauge groups (although with $N$ odd).}). We put this classification to work by deriving invariants of renormalization group (RG) flows involving QFTs that have BFB symmetries. We interpret these invariants as relatives of the spectator sectors 't Hooft used in his original anomaly matching arguments \cite{tHooft:1980xss}.
\item An important subclass of BFB symmetries are full-fledged (spin) TQFTs. We can connect any such BFB (spin) TQFT with a non-topological UV completion. In other words, we can construct explicit RG flows that result in any BFB (spin) TQFT as a gapped IR phase. For general topological phases, such an explicit connection seems out of reach, but we hope that our results can serve as simple stepping stones for connections between classes of more general topological phases and the RG flow.
\end{itemize}

The plan of this paper is as follows. In the next section we introduce further details of BFB symmetries and focus on the case that $\CB$ corresponds to a (spin) TQFT. Then, in Section \ref{UVcompletion}, we provide some simple UV completions of these TQFTs via circular quivers and decoupled product QCD theories. We move on to more general $\CB$ in Section \ref{general} and give a proof of the italicized statement above.  Then, in Section \ref{coupling}, we give a concrete characterization of general BFB categories lacking transparent fermions. In Section \ref{RG} we consider the RG consequences of our analysis, and we conclude with a discussion of open problems.

\newsec{Bose-Fermi-Braided (BFB) Categories}\label{BFB}
In this section, we characterize one-form symmetries consisting of bosons and fermions. Note that we {\it do not} assume these symmetries are invertible. As described in the introduction, we interchangeably refer to such symmetries as BFB symmetries or BFB categories depending on the context. They consist of line operators with bosonic or fermionic self-statistics that are closed under fusion and have a notion of braiding. In a more mathematical language, BFB symmetries are \lq\lq premodular" categories with real twists (e.g., see \cite{etingof2016tensor} for a definition of a premodular category).

Roughly speaking, we would like to classify collections of line operators that can have non-trivial mutual statistics but are not \lq\lq genuinely" anyonic (in the perhaps misleading sense of not having fractional self-statistics; note that these lines are genuine line operators and are not attached to surfaces). Examples of such collections of line operators include Kitaev's toric code modular tensor category (MTC) \cite{kitaev2006anyons}, which is one of the simplest examples of BFB topological order and will play an important role below.

To describe BFB symmetries, we begin with the modular data\footnote{This $S$-matrix is unitary up to a normalization factor, $D:=\sqrt{\sum_{\ell_i} d_{\ell_i}^2}$.}
\begin{equation}\label{STgen}
\theta(\ell_i)\in\left\{\pm1\right\}~,\ \ \ S_{\ell_i\ell_j}=\sum_{\ell_k}N_{\ell_i\ell_j}^{\ell_k}{\theta(\ell_k)\over\theta(\ell_i)\theta(\ell_j)}d_{\ell_k}~,
\end{equation}
which characterize the self-statistics and mutual-statistics of the lines, $\ell_{i,j}\in\CB$, of the BFB category, $\CB$, respectively. In writing the modular $S$-matrix, we sum over simple lines, $\ell_k\in\CB$, and weight the sum by the non-negative integer fusion coefficients
\begin{equation}\label{fus}
\ell_i\times\ell_j=\sum_{\ell_k}N_{\ell_i\ell_j}^{\ell_k}\ell_k~,
\end{equation} 
and real quantum dimensions, $d_{\ell_k}\in\mathbb{R}_{\ge1}$.\footnote{Since we will mostly focus on unitary theories, the quantum dimensions are just the categorical Frobenius-Perron dimensions,
\begin{equation}
d_{\ell_k}:={\rm FPDim}(\ell_k)~.
\end{equation}
More physically, we can think of the quantum dimensions as $S^3$ expectation values of loops.
} Note that the quantum dimensions satisfy the fusion rules
\begin{equation}\label{qdimFus}
d_{\ell_i}\cdot d_{\ell_j}=\sum_{\ell_k}N_{\ell_i\ell_j}^{\ell_k}d_{\ell_k}~.
\end{equation}
Therefore, non-invertible lines (i.e., those satisfying $\ell\times\bar\ell=1+\cdots$, with non-trivial lines in the ellipses) have $d_{\ell}>1$, while invertible lines have $d_{\ell}=1$.

One obvious fact that follows from \eqref{STgen} is that $S$ is real. As a result, in BFB categories, both the self-statistics and mutual-statistics of lines are governed by real numbers (see Fig. \ref{fig:BFB modular data}). 
\begin{figure}
    \centering

\tikzset{every picture/.style={line width=0.75pt}} 

\begin{tikzpicture}[x=0.75pt,y=0.75pt,yscale=-1,xscale=1]

\draw    (499.64,181.64) .. controls (481.86,168.28) and (479,140.43) .. (510.75,131.75) ;
\draw    (476.31,192.33) .. controls (528.5,190.13) and (524.24,146.04) .. (503.9,136.8) ;
\draw    (476.31,192.33) .. controls (427.11,188.93) and (437.32,121.33) .. (496.17,132.82) ;
\draw    (510.75,131.75) .. controls (580.23,112.51) and (579.64,204.81) .. (508.53,187.88) ;
\draw    (162.37,142.1) .. controls (200.37,109.1) and (205.37,187.1) .. (236.37,185.1) ;
\draw    (206.37,152.1) .. controls (249.37,97.1) and (284.37,182.1) .. (236.37,185.1) ;
\draw    (203.37,161.1) .. controls (193.37,207.1) and (133.37,175.1) .. (162.37,142.1) ;

\draw (435,154) node    {$\ell_i$};
\draw (577,156) node    {$\ell_j$};
\draw (360,150.4) node [anchor=north west][inner sep=0.75pt]    {$S_{\ell_i\ell_j} =$};
\draw (54.9,145.4) node [anchor=north west][inner sep=0.75pt]    {$\theta(\ell_i) \ =$};
\draw (116,145.4) node [anchor=north west][inner sep=0.75pt]    {$\frac{1}{d_{\ell_i}}$};
\draw (270,149) node    {$\ell_i$};

\end{tikzpicture}
    \caption{BFB categories have $\mathbb{R}$-valued modular data.}
    \label{fig:BFB modular data}
\end{figure}
Another trivial fact following from this discussion is that all lines in a BFB category {\it with invertible $S$} are self-dual
\begin{equation}\label{SD}
\ell_i\times\ell_i\ni1~,
\end{equation}
where \lq\lq$1$" denotes the trivial line. In more general cases of BFB categories, \eqref{SD} does not necessarily hold (e.g., consider $\CB\cong{\rm Rep}(G)$ for a non-ambivalent group, $G$).

We can contemplate two extremes for the $S$ matrix of $\CB$, namely that it is completely degenerate or that it is invertible. In the case that it is degenerate, a theorem of Deligne \cite{deligne2002categories} guarantees that the lines form the representation category of some finite (super) group, $\CB\cong{\rm Rep}(G_z)$ (see also the work of Doplicher and Roberts \cite{doplicher1990there,doplicher1989new}). We will describe such cases in more detail below. We should think of such a $\CB$ as corresponding to a sub-sector of a non-topological QFT, $\CQ$, rather than as characterizing a topological phase of matter. For example, in one of its guises, $\CB\cong{\rm Rep}(\mathbb{Z}_2)$ appears as the one-form symmetry of pure $2+1$d $SU(2)$ Yang-Mills (YM) theory \cite{Gaiotto:2014kfa}.
 
When $S$ is non-degenerate, $\CB$ describes a non-spin TQFT, i.e. a TQFT that does not depend on a spin structure (for convenience, we will drop the \lq\lq non-spin" modifier). In the language of category theory, $\CB$ corresponds (as a 1-category) to an MTC, $\CB\cong\CM$.

To get an idea of what is possible in the non-degenerate case, let us first consider the case in which all non-trivial lines, $\ell_i\in\CB$ (i.e., $\ell_i\ne1$), are fermions. An example of such an MTC is the \lq\lq 3-fermion" MTC, $F_2$ (using the notation of \cite{wang2020and}), described by the following topological spins and $S$ matrix\footnote{Note that for the $S$ matrix we are using the normalization in \eqref{STgen}.}
\begin{equation}\label{3fST}
\theta(1)=1~,\ \theta(\psi_1)=\theta(\psi_2)=\theta(\psi_3)=-1~,\ \ \ S=\begin{pmatrix}
1 & 1 & 1& 1\\
1 & 1 & -1& -1\\
1 & -1 & 1& -1\\
1 & -1 & -1& 1\\
\end{pmatrix}~.
\end{equation}
This theory consists of invertible / Abelian lines with $\mathbb{Z}_2\times\mathbb{Z}_2$ fusion rules. In fact, from the list of prime Abelian MTCs in \cite{wang2020and},\footnote{A prime MTC is an MTC that cannot be written as the (Deligne) product of two or more other MTCs. Any Abelian MTC, $\CM$, can be written in terms of a (not always unique) prime factorization
\begin{equation}
\CM\cong\CM_1\boxtimes\CM_2\boxtimes\cdots\boxtimes\CM_n~,
\end{equation}
where the $\CM_i$ are prime MTCs. Each factor in the above decomposition is closed under fusion, and each $\CM_i$ braids trivially with any $\CM_j$ labeled by $j\ne i$.
} it is easy to see that this is the only Abelian theory whose non-trivial lines are all fermions. In the language of Chern-Simons (CS) theory, we can obtain such an MTC from ${\rm Spin}(N)_1$ with $N=8\ {\rm mod}\ 16$
\begin{equation}\label{3fTO}
{\rm Spin}(N)_1\ \leftrightarrow\ {\rm 3-fermion\ MTC}\cong F_2\ {\rm MTC}~, \ \ \ N=8\ {\rm mod}\ 16~.
\end{equation}

Next let us consider theories in which all non-trivial lines are fermions and we also allow for non-Abelian fusion. Using the general expression for the $S$ matrix in \eqref{STgen}, and requiring the $S$ matrix to be invertible, it is easy to see that the only possibility has three simple lines with
\begin{equation}\label{2fNA}
S=\begin{pmatrix}
1 & 1 & \sqrt{2}\\
1 & 1 & -\sqrt{2}\\
\sqrt{2} & -\sqrt{2} & 0\\
\end{pmatrix}~.
\end{equation}
This result follows from the fact that the combination of topological spins, $\theta(\ell_k)/\theta(\ell_i)\theta(\ell_j)$, entering the expression for the $S$ matrix in \eqref{STgen} is equal to minus one for $\ell_{i,j,k}\ne1$ and one otherwise. However, the above theory is inconsistent. Indeed, it has Ising fusion rules: $\sigma\times\sigma=1+\epsilon$, with $1$ and $\epsilon$ invertible (where we write $S$ in \eqref{2fNA} in the basis $\left\{1,\epsilon,\sigma\right\}$). The issue is that the non-invertible (Kramers-Wannier duality) line, $\sigma$, cannot be fermionic in such a theory but rather must have anyonic self-statistics (given by a sixteenth root of unity).\footnote{One can look at the full set of rank-three MTCs (i.e., MTCs with three simple objects) in \cite{rowell2009classification} to see this theory is inconsistent. Another way to argue that this theory is inconsistent is to note that the quantum dimensions of such a theory would live in an extension of the rationals, $\mathbb{Q}(\sqrt{2})$ (since $d_{\sigma}=\sqrt{2}$). In an MTC, this extension is determined by the conductor, which is the smallest $N>0$ such that $\theta(\ell_i)^N=1$ for all $\ell_i$. In particular, we should have $d_{\ell_i}\in\mathbb{Q}(\xi_N)$ for some primitive $N$th root of unity, $\xi_N$ \cite{DeBoer:1990em}. However, $\sqrt{2}\not\in\mathbb{Q}(\xi_2)\cong\mathbb{Q}$.\label{numtheory}} 

 As a result, we arrive at the following simple theorem:

\bigskip
\noindent
{\bf Theorem 1:} The only MTC whose non-trivial simple lines are fermions is the 3-fermion (a.k.a. $F_2$) MTC. One realization of this MTC is via the Wilson lines of any ${\rm Spin}(N)_1$ CS TQFT with $N=8$ mod $16$.

\bigskip

What can we say about the most general BFB TQFT? In this case, the simple argument involving the $S$ matrix below \eqref{2fNA} no longer works because the combination of topological spins that we use is less constrained. To avoid complications from the topological spins, we would like to study an observable built from modular data that is quadratic in spins, so that the spin dependence drops out for BFB theories. Another hint regarding which observable to use arises from the fact that all lines in our theory are in fact self-dual and therefore have a non-vanishing Frobenius-Schur (FS) indicator, $\nu_2(\ell_i)=\pm1$. The fact that the FS indicator has value $\pm1$ arises from the fact that it can be understood as a $\mathbb{Z}_2$ action on the $a\times a\ni 1$ fusion space.\footnote{This same fact is behind the appearance of the FS indicator in the quantum mechanical addition of spins \cite{Simon:2022ohj,Simon:2023hdq}.}

\begin{figure}[h!]
    \centering

\tikzset{every picture/.style={line width=0.75pt}} 

\begin{tikzpicture}[x=0.75pt,y=0.75pt,yscale=-1,xscale=1]

\draw    (118,142) -- (298,142) ;
\draw  [fill={rgb, 255:red, 0; green, 0; blue, 0 }  ,fill opacity=1 ] (206,142) .. controls (206,140.9) and (206.9,140) .. (208,140) .. controls (209.1,140) and (210,140.9) .. (210,142) .. controls (210,143.1) and (209.1,144) .. (208,144) .. controls (206.9,144) and (206,143.1) .. (206,142) -- cycle ;
\draw  [fill={rgb, 255:red, 0; green, 0; blue, 0 }  ,fill opacity=1 ] (527,149) .. controls (527,147.9) and (527.9,147) .. (529,147) .. controls (530.1,147) and (531,147.9) .. (531,149) .. controls (531,150.1) and (530.1,151) .. (529,151) .. controls (527.9,151) and (527,150.1) .. (527,149) -- cycle ;
\draw    (415.5,142) .. controls (562.5,84) and (631,152) .. (531,149) ;
\draw    (641.5,158) .. controls (522.5,222) and (414,154) .. (527,149) ;
\draw    (341,143) -- (381,143) ;
\draw [shift={(383,143)}, rotate = 180] [color={rgb, 255:red, 0; green, 0; blue, 0 }  ][line width=0.75]    (10.93,-3.29) .. controls (6.95,-1.4) and (3.31,-0.3) .. (0,0) .. controls (3.31,0.3) and (6.95,1.4) .. (10.93,3.29)   ;
\draw  [fill={rgb, 255:red, 0; green, 0; blue, 0 }  ,fill opacity=1 ] (147.6,140) -- (151.6,142) -- (147.6,144) -- (149.6,142) -- cycle ;
\draw  [fill={rgb, 255:red, 0; green, 0; blue, 0 }  ,fill opacity=1 ] (273.89,144.1) -- (270,141.89) -- (274.1,140.11) -- (272,142) -- cycle ;
\draw  [fill={rgb, 255:red, 0; green, 0; blue, 0 }  ,fill opacity=1 ] (505.02,117.67) -- (509.35,118.77) -- (505.87,121.58) -- (507.4,119.2) -- cycle ;
\draw  [fill={rgb, 255:red, 0; green, 0; blue, 0 }  ,fill opacity=1 ] (540.17,187.03) -- (536.2,184.97) -- (540.23,183.03) -- (538.2,185) -- cycle ;

\draw (289,146.4) node [anchor=north west][inner sep=0.75pt]    {$\ell $};
\draw (116,147.4) node [anchor=north west][inner sep=0.75pt]    {$\ell $};
\draw (630,167.4) node [anchor=north west][inner sep=0.75pt]    {$\ell $};
\draw (412,148.4) node [anchor=north west][inner sep=0.75pt]    {$\ell $};
\draw (56,136.4) node [anchor=north west][inner sep=0.75pt]    {$C_{\ell} :$};

\end{tikzpicture}
    \caption{$C_\ell: V_{\ell\ell}^{1} \to V_{\ell\ell}^{1}$ acts on the fusion vertex by rotating it clockwise by an angle $\pi$.}
    \label{fig:FS indicator}
\end{figure}
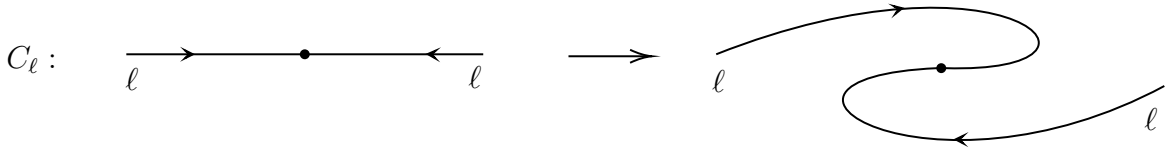

More precisely, the FS indicator is defined as
\be\label{FSdef}
\nu_2(\ell):= \text{Tr}(C_\ell)~,
\ee
where $C_\ell$ is the action on the fusion space, $V_{\ell\ell}^1$, given in Fig. \ref{fig:FS indicator}. From the pivotal property of the category (which follows from unitarity, see \cite[Fig. 16]{kitaev2006anyons}) rotating a fusion vertex by an angle $2\pi$ must be the identity operation. Therefore, $C_\ell$ is order two. It follows that the eigenvalues of $C_\ell$ are valued in $\pm 1$. In fact, since $V_{\ell \ell}^{1}$ is at most 1-dimensional, $C_\ell = \pm 1$, and $\nu_2(\ell)=\pm 1$.\footnote{The FS indicator can be defined for line operators in any unitary fusion category. Indeed, if a line operator, $\ell$, is non-self-dual, then $\nu_2(\ell):=0$.\label{NSDFS}} 

Since the FS indicator gives a gauge-invariant measure of the total angular momentum in an anyonic system \cite{kitaev2006anyons}, we expect to find an expression in terms of the modular data. Indeed, according to \cite{bantay1997frobenius,ng2007frobenius,bruillard2012rank}, we have\footnote{Clearly $\nu_2(\ell_i)=+1$ in unitary MTCs with $\theta=\pm1$. We thank Brandon Rayhaun for discussing this point with us.} 
\be\label{FS2}
\nu_2(\ell_i)= \frac{1}{D^2} \sum_{\ell_{j},\ell_k\in\CB} N_{\ell_j\ell_k}^{\ell_i} d_{\ell_j} d_{\ell_k} \bigg (\frac{\theta(\ell_j)}{\theta(\ell_k)} \bigg )^2~.
\ee

From the expression in \eqref{FS2}, it is then easy to see that since a BFB TQFT only has bosonic and fermionic spins
\be\label{Simplev2}
\nu_2(\ell_i)= \frac{1}{D^2} \sum_{\ell_j,\ell_k\in\CB} N_{\ell_j\ell_k}^{\ell_i} d_{\ell_j} d_{\ell_k} = \frac{1}{D^2} \sum_{\ell_j,\ell_k\in\CB} \left(N_{\ell_i \ell_j}^{\ell_k} d_{\ell_k}\right) d_{\ell_j} = \frac{1}{D^2} \left(\sum_{\ell_j\in\CB} d_{\ell_j}^2\right) d_{\ell_i}=d_{\ell_i}~.
\ee
In the third equality, we have used the fact that quantum dimensions satisfy fusion rules as in \eqref{qdimFus}. Since the FS indicator is plus or minus one (recall that all lines are self-dual), we learn that $d_{\ell_i}=\pm1$, but, in a unitary theory (which we assume throughout), $d_{\ell_i}=1$, and we have
\begin{equation}
d_{\ell_i}=1~,\ \ \ \forall\ell_i\in\CB~.
\end{equation}
In other words, we see that all BFB TQFTs consist of Abelian / invertible lines! In fact, this statement was already derived in \cite{bruillard2012modular,wan2021classification} using essentially the same arguments.

From the classification of Abelian MTCs in \cite{wang2020and}, it is easy to see that the most general BFB MTC we can write down involves a stacking of MTCs corresponding to the 3-fermion MTC we encountered around \eqref{3fTO} and the $E_2$ MTC in the nomenclature of \cite{wang2020and}. This latter MTC can be equivalently realized by (among other quantum systems) Kitaev's toric code at low energies, the untwisted $\mathbb{Z}_2$ Dijkgraaf-Witten (DW) theory, or ${\rm Spin}(N)_1$ CS theory with $N=0$ mod $16$. In other words, we have
\begin{equation}\label{tcTO}
{\rm Spin}(N)_1~,\ \mathbb{Z}_2\ {\rm untwisted\ DW\ theory}\ \leftrightarrow\ {\rm toric\ code\ MTC}\cong E_2\ {\rm MTC}~, \ \ \ N=0\ {\rm mod}\ 16~,
\end{equation}
and, we arrive at the following theorem:

\bigskip
\noindent
{\bf Theorem 2:} The most general BFB TQFT corresponds to the following MTC (see also \cite{wan2021classification})
\begin{equation}\label{GenMTC}
\CM\cong (E_2)^{\boxtimes n}\boxtimes (F_2)^{\boxtimes m}~,\ \ \ (X)^{\boxtimes p}:=\overbrace{X\boxtimes X\boxtimes\cdots\boxtimes X}^{\rm p\ times}~.
\end{equation}
In fact, using the equivalence $E_2\boxtimes E_2\cong F_2\boxtimes F_2$, we can simplify $\CM$ as follows
\begin{equation}
\CM\cong (E_2)^{\boxtimes n}\boxtimes (F_2)^{\boxtimes m}~,\ \ \ n=0,1~,\ \ \ m\ge0~.
\end{equation}
At the level of CS theories, such an MTC can be realized by, for example, stacking $n$ ${\rm Spin}(N)_1$ CS theories ($N=0$ mod $16$) with $m$ ${\rm Spin}(N')_1$ CS theories ($N'=8$ mod 16).

\bigskip
Let us now consider BFB symmetries corresponding to a degenerate $S$ matrix. Here it is useful to use a more general expression for the FS indicator in premodular categories \cite{ng2007frobenius,bruillard2012rank}
\bea\label{FS2PreMod}
\nu_2(\ell_i)&=&\frac{1}{D^2} \sum_{\ell_{j},\ell_k\in\CB} N_{\ell_j\ell_k}^{\ell_i} d_{\ell_j} d_{\ell_k} \bigg (\frac{\theta(\ell_j)}{\theta(\ell_k)} \bigg )^2-\theta(\ell_i)\sum_{\ell\in\CZ_M(\CB), \ell\ne1}{\rm Tr}(R_{\ell_i\ell_i}^{\ell})\cdot d_{\ell}\cr&=&d_{\ell_i}-\theta(\ell_i)\sum_{\ell\in\CZ_M(\CB), \ell\ne1}{\rm Tr}(R_{\ell_i\ell_i}^{\ell})\cdot d_{\ell}~,
\eea
where $\CZ_M(\CB)$ is the so-called \lq\lq M\"uger center" of $\CB$ \cite{muger2003structure}, and $R_{\ell_i\ell_j}^{\ell_k}$ is the braiding matrix. Physically $\CZ_M(\CB)$ is the set of line operators in $\CB$ that braid trivially with all lines in $\CB$ (i.e., the set of \lq\lq transparent" lines). It forms a fusion subcategory of $\CB$ \cite[Lemma 2.8]{muger2003structure}. In the second equality we have used logic similar to that around \eqref{Simplev2}. The formula \eqref{FS2PreMod} for the FS indicator can be applied to both self-dual and non-self-dual lines. This statement holds because the argument leading to this formula in \cite{bruillard2012rank} can be repeated for non-self dual lines (where, as in footnote \ref{NSDFS}, we take $\nu_2(\ell):=0$ when $\ell\ne\bar\ell$).

We can simplify \eqref{FS2PreMod} further. Indeed, we know from Deligne's theorem that, for $\ell$ to be transparent, it should be a boson or a fermion (i.e., it cannot have anyonic self statistics). However, a transparent fermion, $\ell=\psi$, cannot appear in $\ell_i\times\ell_i$. Indeed, otherwise $\psi\times\bar\ell_i\ni\ell_i$, and $\psi$ would braid non-trivially with $\ell_i$. Therefore, we arrive at
\begin{equation}\label{FS2final}
\nu_2(\ell_i)=d_{\ell_i}-\theta(\ell_i)\sum_{\ell\in\CZ^{\rm bos}_M(\CB), \ell\ne1}{\rm Tr}(R_{\ell_i\ell_i}^{\ell})\cdot d_{\ell}~,
\end{equation}
where the \lq\lq bos" superscript in the summation refers to the fact that only transparent bosons contribute.

We can motivate the formula in \eqref{FS2final} as follows. At a basic level, the correction term arising from the M\"uger center is required in order to reproduce what we already know: in the case of symmetric $\CB$ governed by Deligne's theorem, $\CZ_M(\CB)\cong\CB$, we can have non-Abelian  lines, $\ell_i\in\CB\cong{\rm Rep}(G)$, when $G$ is non-Abelian (i.e., $d_{\ell_i}={\rm dim}(\pi)>1$ for a non-Abelian irrep $\pi\in{\rm Rep}(G)$). Moreover, we can also have non-self-dual lines (when $\CB$ is symmetric, such lines occur whenever $G$ is not ambivalent) and lines with negative FS indicator (this situation occurs for lines labeled by pseudo-real representations of $G$).

To give further motivation for the correction term in \eqref{FS2final}, note that contributions from $\ell\in\CZ_M^{\rm bos}(\CB)$ satisfying $\ell\in\ell_i\times\ell_i$ turn out to be crucial because these are precisely the bosonic $\ell$ that satisfy $\ell\times\bar\ell_i\ni \ell_i$. When $\ell_i$ is self-dual, condensing $\ell$ can produce Abelian lines. Moreover, when $\ell_i\ne\bar\ell_i$, condensing $\ell$ can produce self-dual lines.  Since condensing $\CZ_M^{\rm bos}(\CB)$ gives a (possibly trivial) TQFT, this discussion is consistent with Theorem 2 which requires that all BFB TQFT lines are self-dual and Abelian.\footnote{Note that, from the perspective of higher gauging \cite{Roumpedakis:2022aik}, the difference between the FS indicator and the correction term in \eqref{FS2final} can be understood in terms of 2-gauging. Indeed, up to normalization, this difference arises from 2-gauging $J:=\sum_{\ell} d_{\ell} ~\ell$ on a cycle surrounding $\ell_i \times \ell_i$. This maneuver projects onto $\CZ_{M}^{\rm bos}(\CB)$. \label{higher}}

We should distinguish between two separate cases of degenerate $S$:
\begin{enumerate}
\item $\CB$ with a \lq\lq slightly" degenerate $S$ matrix (see the general discussion in \cite{Johnson-Freyd:2021chu}). In this case, we have a \lq\lq super-MTC" with a single transparent line: a fermion, $\psi$, that generates $\CZ_M(\CB)$
\begin{equation}
\CZ_M(\CB)\cong\langle\psi\rangle\cong{\rm SVec}~,
\end{equation}
where ${\rm SVec}$ is the category of finite-dimensional super vector spaces.

A super-MTC is the algebraic realization of the line operators in a spin TQFT (i.e., a TQFT that depends on the spin structure of spacetime). In this setting, $\psi$ is a transparent fermion. We can for example realize ${\rm SVec}$ via $SO(N)_1\cong{\rm Spin}(N)_1/\mathbb{Z}_2$ CS theory, where we condense a fermionic $\mathbb{Z}_2$ line in ${\rm Spin}(N)_1$ (e.g., see \cite{Seiberg:2016rsg}).
\item $\CB$ with any other degenerate $S$ matrix. In this case, we should understand $\CB$ as part of some non-topological QFT.
\end{enumerate}

We will finish this section by describing case 1 above, and we will leave a discussion of case 2 to Section \ref{general}. To that end, in the first case, we have a super-MTC with the only non-trivial transparent line being the fermion, $\psi$. As follows from \eqref{FS2final}, this line does not contribute to $\nu_2(\ell_i)$, and we get
\be
\nu_2(\ell_i)=d_{\ell_i}~.
\ee
In a unitary super-MTC, $d_{\ell_i}\geq1$ for all $\ell_i \in \CB$. However, for non-self-dual lines $\nu_2(\ell_i)=0$. Therefore, the above equality implies that in a super-MTC with real spins all lines must be self-dual.\footnote{Another way to see this is to use the fact that all (pseudo-unitary) super-MTCs admit a minimal modular extension \cite{Johnson-Freyd:2021chu}. The modular extension is a $\DZ_2$-graded category, $\CM=\CM_0 + \CM_1$, where $\CM_0=\CB$. The modular $S$ matrix of $\CM$ is given by \eqref{STgen}. Therefore, it is clear that $S|_{\CM_0}$ is real. Now, from \cite[Theorem 3.5]{bruillard2017fermionic}, we have $S|_{\CM_0}= \hat S \otimes S_{\text{SVec}}$, where $S_{\text{SVec}}$ is the $S$ matrix of SVec. Furthermore, \cite[Proposition 2.7]{bruillard2017classification} shows that $\hat S \overline{\hat S}=\frac{D^2}{2} I$, and $\hat S^2 = \frac{D^2}{2} C$, where $C$ is the charge conjugtion matrix. Since $\hat S$ is real in our case, $C=I$. Therefore, all lines in a super-MTC with real spins are self-dual.} As a result, we learn that in a super-MTC, we again have (see also \cite{wan2021classification})
\begin{equation}
d_{\ell_i}=1~,\ \ \ \forall\ell_i\in\CB~.
\end{equation}
In other words, all BFB spin TQFTs are also Abelian. Moreover, all Abelian super-MTCs are split, which means any such super-MTC, $\CM$, can be written as $\CM\cong\widehat\CM\boxtimes{\rm SVec}$, where $\widehat\CM$ is an MTC (see, for example, \cite{bruillard2017fermionic,PhysRevB.99.075143}).

Therefore, using the classification in \cite{wang2020and}, we arrive at the following theorem:
 
 \bigskip
 \noindent
 {\bf Theorem 3:} The most general BFB spin TQFT corresponds to the following super-MTC (see also \cite{wan2021classification})
\begin{equation}\label{GenSMTC}
\CM\cong (E_2)^{\boxtimes n}\boxtimes (F_2)^{\boxtimes m}\boxtimes{\rm SVec}~,\ \ \ (X)^{\boxtimes p}:=\overbrace{X\boxtimes X\boxtimes\cdots\boxtimes X}^{\rm p\ times}~.
\end{equation}
In fact, using the equivalences $E_2\boxtimes {\rm SVec}\cong F_2\boxtimes {\rm SVec}$ and $E_2\boxtimes E_2\cong F_2\boxtimes F_2$, we can write any $\CM$ as follows
\begin{equation}
\CM\cong (E_2)^{\boxtimes n}\boxtimes {\rm SVec}\cong(F_2)^{\boxtimes n}\boxtimes {\rm SVec}~.
\end{equation}
At the level of CS theories, such an MTC can be realized by, for example, stacking $n$ ${\rm Spin}(N)_1$ CS theories ($N=0$ mod $16$) with an $SO(M)_1$ CS theory or $n$ ${\rm Spin}(N')_1$ CS theories ($N'=8$ mod 16) with $SO(M')_1$ CS theory.

\bigskip
A few remarks are in order:
\begin{enumerate}
\item Theorems 2 and 3 hold even in a non-unitary (super-) MTC (see Appendix \ref{ap:non-unitary BFB} for an argument).
\item The converses of Theorems 2 and 3 guarantee that any (spin) TQFT with non-invertible lines contains line operators with complex spins. Assuming the low-energy description of a general topological phase is a (spin) MTC, we have shown that if a topological phase contains anyons with non-invertible fusion rules, then it must contain anyons with complex spins. 
\item In \cite{Buican:2023bzl}, two of the present authors asked whether time-reversal symmetry of a non-Abelian TQFT can act trivially on line operators. Theorems 2 and 3 answer this question in the negative.\footnote{This statement is to be contrasted with non-trivial unitary symmetries of a non-Abelian TQFT which can act trivially on all line operators \cite{davydov2014bogomolov}.} 
\end{enumerate}

\bigskip
\noindent
In the next section, we discuss how to realize the above BFB (spin) TQFTs via RG flows. Then, in Section \ref{general}, we discuss the case of more general BFB symmetries and give a classification.

\subsec{BFB (spin) TQFTs and UV completions}\label{UVcompletion}
In general, given a (spin) TQFT, it is interesting to ask what kind of UV completion one can find. In our context, we have in mind a UV Poincar\'e-invariant and non-topological QFT, $\CQ_{UV}$, that flows to the (spin) TQFT in the IR.\footnote{It is of course also interesting to think instead in terms of UV completions via lattice models, but we will not do so here.}  Indeed, by better understanding such flows, one hopes to elucidate the structure of the space of QFTs. However, given a general class of abstract (spin) TQFTs, we do not expect it to be straightforward to find such a UV completion. On the other hand, we have seen in \eqref{GenMTC} and \eqref{GenSMTC} that, by imposing bose / fermi statistics on (spin) TQFT lines, we get a remarkably simple class of theories.

Even in this context, a UV completion is wildly non-unique due to dualities of non-topological theories (e.g., see \cite{Cordova:2017vab} for some relevant dualities in our class of theories). Examples of these dualities include IR dualities (i.e., where distinct UV theories flow to the same IR theory) and more trivial examples where distinct UV theories differ by some matter fields that can be made massive and integrated out.

Let us turn to some examples. Note from the discussion around \eqref{GenMTC} that we can realize the $E_2$ BFB MTC via ${\rm Spin}(N)_1$ CS theory for $N=0$ mod $16$. However, when we couple this theory to matter, it becomes non-topological and generally distinct for different values of $N$. For example, we can take ${\rm Spin}(N)$ YM theory with CS level $k=1$ and couple $N_f$ real scalars, $\phi_a$ (with $a=1,\cdots,N_f$), in the vector representation of the gauge group. This UV theory is clearly distinct for all $N=0$ mod $16$. Then, giving a large mass to each of the matter fields, $\delta\CL=-{m^2\over2}\phi_a^2$, with $m\gg g^2$ (where $g$ is the gauge coupling) results in dual ${\rm Spin}(N)_1$ CS theories in the IR.

As a result, it is trivial to find UV completions for all BFB (spin) TQFTs. For example, we can engineer all MTCs in \eqref{GenMTC} by taking
\begin{equation}\label{NT}
\CQ_{UV}:=({\rm Spin}(N)_1\ {\rm with}\ N_f\ \phi\in {\bf N})^{\boxtimes n}\boxtimes ({\rm Spin}(N')_1\ {\rm with}\ N_f\ \phi\in {\bf N'})^{\boxtimes m}~,
\end{equation}
where $N=0$ mod $16$, $N'=8$ mod $16$, and the scalars transform in the vector representation. Note that in these theories, a $\CB_{UV}\cong{\rm Rep}(\mathbb{Z}_2^{n+m})$ subcategory of lines is topological in the UV. This statement holds because the fundamental Wilson line can end on the matter fields and so the other would-be topological lines of each ${\rm Spin}(N)_1$ and each ${\rm Spin}(N')_1$ factor become non-topological (they braid non-trivially with at least one fundamental Wilson line while the fundamental Wilson lines braid trivially with themselves; e.g., see  \cite{Rudelius:2020orz}).

\begin{figure}[h!]
    \centering

\tikzset{every picture/.style={line width=0.75pt}} 

\begin{tikzpicture}[x=0.75pt,y=0.75pt,yscale=-1,xscale=1]

\draw   (264,157) .. controls (264,121.65) and (292.65,93) .. (328,93) .. controls (363.35,93) and (392,121.65) .. (392,157) .. controls (392,192.35) and (363.35,221) .. (328,221) .. controls (292.65,221) and (264,192.35) .. (264,157) -- cycle ;
\draw  [fill={rgb, 255:red, 255; green, 255; blue, 255 }  ,fill opacity=1 ] (322,93) .. controls (322,89.69) and (324.69,87) .. (328,87) .. controls (331.31,87) and (334,89.69) .. (334,93) .. controls (334,96.31) and (331.31,99) .. (328,99) .. controls (324.69,99) and (322,96.31) .. (322,93) -- cycle ;
\draw  [fill={rgb, 255:red, 0; green, 0; blue, 0 }  ,fill opacity=1 ] (274,116) .. controls (274,112.69) and (276.69,110) .. (280,110) .. controls (283.31,110) and (286,112.69) .. (286,116) .. controls (286,119.31) and (283.31,122) .. (280,122) .. controls (276.69,122) and (274,119.31) .. (274,116) -- cycle ;
\draw  [fill={rgb, 255:red, 0; green, 0; blue, 0 }  ,fill opacity=1 ] (322,221) .. controls (322,217.69) and (324.69,215) .. (328,215) .. controls (331.31,215) and (334,217.69) .. (334,221) .. controls (334,224.31) and (331.31,227) .. (328,227) .. controls (324.69,227) and (322,224.31) .. (322,221) -- cycle ;
\draw  [fill={rgb, 255:red, 255; green, 255; blue, 255 }  ,fill opacity=1 ] (386,163) .. controls (386,159.69) and (388.69,157) .. (392,157) .. controls (395.31,157) and (398,159.69) .. (398,163) .. controls (398,166.31) and (395.31,169) .. (392,169) .. controls (388.69,169) and (386,166.31) .. (386,163) -- cycle ;
\draw  [fill={rgb, 255:red, 0; green, 0; blue, 0 }  ,fill opacity=1 ] (280,205) .. controls (280,201.69) and (282.69,199) .. (286,199) .. controls (289.31,199) and (292,201.69) .. (292,205) .. controls (292,208.31) and (289.31,211) .. (286,211) .. controls (282.69,211) and (280,208.31) .. (280,205) -- cycle ;
\draw  [fill={rgb, 255:red, 255; green, 255; blue, 255 }  ,fill opacity=1 ] (365,204) .. controls (365,200.69) and (367.69,198) .. (371,198) .. controls (374.31,198) and (377,200.69) .. (377,204) .. controls (377,207.31) and (374.31,210) .. (371,210) .. controls (367.69,210) and (365,207.31) .. (365,204) -- cycle ;
\draw  [color={rgb, 255:red, 255; green, 255; blue, 255 }  ,draw opacity=1 ][fill={rgb, 255:red, 255; green, 255; blue, 255 }  ,fill opacity=1 ] (286,162.5) .. controls (286,150.63) and (276.37,141) .. (264.5,141) .. controls (252.63,141) and (243,150.63) .. (243,162.5) .. controls (243,174.37) and (252.63,184) .. (264.5,184) .. controls (276.37,184) and (286,174.37) .. (286,162.5) -- cycle ;
\draw  [fill={rgb, 255:red, 0; green, 0; blue, 0 }  ,fill opacity=1 ] (263.5,148.5) .. controls (263.5,148.22) and (263.72,148) .. (264,148) .. controls (264.28,148) and (264.5,148.22) .. (264.5,148.5) .. controls (264.5,148.78) and (264.28,149) .. (264,149) .. controls (263.72,149) and (263.5,148.78) .. (263.5,148.5) -- cycle ;
\draw  [fill={rgb, 255:red, 0; green, 0; blue, 0 }  ,fill opacity=1 ] (263.5,162.5) .. controls (263.5,162.22) and (263.72,162) .. (264,162) .. controls (264.28,162) and (264.5,162.22) .. (264.5,162.5) .. controls (264.5,162.78) and (264.28,163) .. (264,163) .. controls (263.72,163) and (263.5,162.78) .. (263.5,162.5) -- cycle ;
\draw  [fill={rgb, 255:red, 0; green, 0; blue, 0 }  ,fill opacity=1 ] (265.5,175.5) .. controls (265.5,175.22) and (265.72,175) .. (266,175) .. controls (266.28,175) and (266.5,175.22) .. (266.5,175.5) .. controls (266.5,175.78) and (266.28,176) .. (266,176) .. controls (265.72,176) and (265.5,175.78) .. (265.5,175.5) -- cycle ;
\draw  [color={rgb, 255:red, 255; green, 255; blue, 255 }  ,draw opacity=1 ][fill={rgb, 255:red, 255; green, 255; blue, 255 }  ,fill opacity=1 ] (396.5,114) .. controls (396.5,102.13) and (386.87,92.5) .. (375,92.5) .. controls (363.13,92.5) and (353.5,102.13) .. (353.5,114) .. controls (353.5,125.87) and (363.13,135.5) .. (375,135.5) .. controls (386.87,135.5) and (396.5,125.87) .. (396.5,114) -- cycle ;
\draw  [fill={rgb, 255:red, 0; green, 0; blue, 0 }  ,fill opacity=1 ] (365.5,105.5) .. controls (365.5,105.22) and (365.72,105) .. (366,105) .. controls (366.28,105) and (366.5,105.22) .. (366.5,105.5) .. controls (366.5,105.78) and (366.28,106) .. (366,106) .. controls (365.72,106) and (365.5,105.78) .. (365.5,105.5) -- cycle ;
\draw  [fill={rgb, 255:red, 0; green, 0; blue, 0 }  ,fill opacity=1 ] (374.5,113.5) .. controls (374.5,113.22) and (374.72,113) .. (375,113) .. controls (375.28,113) and (375.5,113.22) .. (375.5,113.5) .. controls (375.5,113.78) and (375.28,114) .. (375,114) .. controls (374.72,114) and (374.5,113.78) .. (374.5,113.5) -- cycle ;
\draw  [fill={rgb, 255:red, 0; green, 0; blue, 0 }  ,fill opacity=1 ] (382.5,124.5) .. controls (382.5,124.22) and (382.72,124) .. (383,124) .. controls (383.28,124) and (383.5,124.22) .. (383.5,124.5) .. controls (383.5,124.78) and (383.28,125) .. (383,125) .. controls (382.72,125) and (382.5,124.78) .. (382.5,124.5) -- cycle ;

\end{tikzpicture}
    \caption{The circular quiver above describes a UV completion for each of the topological phases in \eqref{GenMTC} and \eqref{GenSMTC} (note that we can couple in an $SO(M)$ node with corresponding matter and appropriate level as well if desired). Black nodes denote ${\rm Spin}(N)$ gauge groups (with $N=0\ {\rm mod}\ 16$), and white nodes denote ${\rm Spin}(N')$ gauge groups (with $N'=8\ {\rm mod}\ 16$). For each gauge group, we turn on appropriate CS levels as described in the main text. Lines connecting nodes in the quiver correspond to appropriate matter fields (described in the main text) transforming as vectors under each of the corresponding two gauge groups. In this particular UV completion, we have made a choice to place the $n$ black nodes in one grouping and the $m$ white nodes in another (other arrangements are also acceptable; ours is universal in the sense that it exists for any $m$ and $n$). Slightly away from zero gauge coupling, this theory has no decoupled sectors. Turning on large masses for the matter fields takes us to the topological phases described by \eqref{GenMTC} and \eqref{GenSMTC}.}
    \label{fig:circular quiver}
\end{figure}
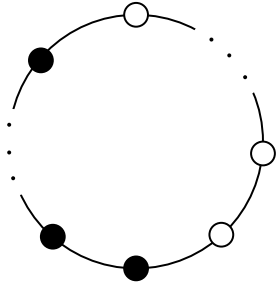

Now, giving large mass to the scalars (compared to the squares of the individual gauge couplings) results in the theory described in \eqref{GenMTC} (the additional topological lines that emerge are accidental symmetries). If we want, we can also consider a $\CQ_{UV}$ that does not have decoupled sectors in the UV. A simple way to do this is to consider gauge group ${\rm Spin}(N)^n\times {\rm Spin}(N')^m$ and construct a circular quiver (i.e., a circular graph with nodes corresponding to gauge groups and edges corresponding to matter) with $\phi$'s charged under successive gauge groups as bi-vector representations (i.e., for successive gauge nodes, $\phi\in{\bf(A,B)}$ for gauge group ${\rm Spin}(A)\times{\rm Spin}(B)$; see Fig. \ref{fig:circular quiver}).\footnote{Related quiver theories appear in the context of dimensional deconstruction \cite{Arkani-Hamed:2001kyx} (see also \cite{Aitken:2018joz} for a more closely related study in $2+1$d where circular quivers are related to a fourth space-time dimension). It would be interesting to understand if there are consequences for $3+1$d physics via these types of constructions.}

To engineer UV completions for all super-MTCs in \eqref{GenMTC}, we can instead consider
\begin{equation}\label{NTSpin}
\CQ_{UV}:=({\rm Spin}(N)_{1+N_f/2}\ {\rm with}\ N_f\ \psi\in {\bf N})^{\boxtimes n}\boxtimes ({\rm Spin}(N')_{1+N_f/2}\ {\rm with}\ N_f\ \psi\in {\bf N'})^{\boxtimes m}~,
\end{equation}
where the $\psi_a$ are Majorana fermions, and the $N_f/2$ shifts in the UV levels arise from the massless fermion determinants. If we give large negative masses to the Majorana fermions, $\delta\CL=-M\psi_a\bar\psi_a$, with $|M|\gg g^2$ and $M<0$, we obtain the theory in \eqref{GenSMTC}. As in the previous case, if we want a theory with a unique stress tensor, we can consider a circular quiver but replace the bosons with fermions and adjust the bare CS level (taking into account the number of fermions charged under each node; see Fig. \ref{fig:circular quiver}).

\subsec{Non-(super-) modular BFB symmetries}\label{general}
In this section, we discuss the more general case of BFB symmetries in which the $S$ matrix is non-degenerate and the M\"uger center satisfies
\begin{equation}
\CZ_M(\CB)\not\cong{\rm Vec}~,\ {\rm SVec}~.
\end{equation}
In other words, we are interested in BFB symmetries in which we have transparent lines other than the trivial line and the transparent fermion. As we will discuss in more detail below, we should physically think of such $\CB$ symmetries as corresponding to sectors of non-topological QFTs.\footnote{Indeed, we have already encountered such examples in the non-topological QFTs of \eqref{NT} and \eqref{NTSpin}. Note that the case $\CZ_M(\CB)\cong{\rm Vec}$ corresponds to $\CB$ being an MTC, while $\CZ_M(\CB)\cong{\rm SVec}$ corresponds to $\CB$ being a super-MTC.}

Let us describe these $\CB$ categories more carefully. As alluded to above, we know from Deligne's theorem that $\CZ_{M}\cong{\rm Rep}(G_{z})$ for some non-trivial (super-) group, $G_{z}$. Here we use $G_{z}$ to refer to either a discrete group or a discrete super-group. In the case that there are no transparent fermions, $G_{z}$ is a discrete group, and we will write the group as $G_1$. Otherwise, we write $G_{\psi}$. Since the braiding of $\CZ_M(B)$ is trivial, we always have a closed subcategory that includes all the bosonic lines \cite{Sawin:1999xt}
\begin{equation}
\CZ_{M}^{\rm bos}(\CB):={\rm Rep}(G)\le{\rm Rep}(G_z)~.
\end{equation}
When there are no transparent fermions, $G\cong G_1$, and $\CZ_M^{\rm bos}(\CB)\cong\CZ_M(\CB)$.

Therefore, given a general BFB category, $\CB$, we are always free to condense a non-anomalous $\CZ_{M}^{\rm bos}(\CB)<\CB$ (in the math literature this is referred to as de-equivariantization; e.g., see \cite{etingof2016tensor}). Indeed, this condensation is precisely the process that removes the correction terms in \eqref{FS2final} (this condensation corresponds to a 0-gauging that is closely related to the $2$-gauging described in footnote \ref{higher}; in the $0$-gauging, the relevant object is a restriction of $J$ to elements in $\CZ_M^{\rm bos}(\CB)$). Clearly, this procedure produces an associated (super-) MTC
\begin{equation}
\CM:=\CB/\CZ_{M}^{\rm bos}(\CB)~.
\end{equation}
By the results in Section \ref{BFB}, $\CM$ has the form given in \eqref{GenMTC} or \eqref{GenSMTC} depending on whether there is a transparent fermion or not.

To better understand the landscape of allowed $\CB$ categories, we should start with an $\CM$ of the form in \eqref{GenMTC} or \eqref{GenSMTC}, construct an action of a discrete group, $G$, on this category and equivariantize (i.e., perform the inverse operation of condensation / de-equivariantization).\footnote{The fact that $G$ is unique follows from Tannaka-Krein reconstruction applied to $\CZ_M^{\rm bos}(\CB)$.} This mathematical maneuver amounts to dropping twisted sectors and considering the splitting and fusing of the lines in $\CM$ under the action of $G$. The reason we throw out twisted sectors is that we perform the inverse procedure of condensing a subcategory, $\CZ_{M}^{\rm bos}(\CB)$, that braids trivially with all the lines in $\CB$ (and hence no lines are projected out in the de-equivariantization). However, there are sometimes obstructions to equivariantization (e.g., see \cite{galindo2017categorical}), which we will soon discuss from a physical perspective.

In any case, if we have an unobstructed $G$-equivariantization of an $\CM$ of the form in \eqref{GenMTC} or \eqref{GenSMTC}, then the quantum dimensions in $\CB$ are clearly integers (this statement is also consistent with the discussion in footnote \ref{numtheory}). As a result, $\CB$ is said to be \lq\lq integral" \cite{etingof2016tensor}.

It is also interesting to understand to what extent the theory of groups underlies our $\CB$ categories. To that end, we note that the $\CM$ (super-) MTCs in \eqref{GenMTC} and \eqref{GenSMTC} are \lq\lq group theoretical" since they are Morita equivalent to the representation category ${\rm Rep}(\mathbb{Z}_2^p)$ for some $p>0$ (for super-MTCs we have ${\rm Rep}(\mathbb{Z}_2^p)\to{\rm Rep}(\mathbb{Z}_2^p)\boxtimes{\rm SVec}$).\footnote{This statement can be easily seen by invoking \cite[Theorem 3.1]{etingof2011weakly} and noting that Morita equivalence is an equivalence of Drinfeld centers. Then, we can invoke the relation $F_2\boxtimes F_2\cong E_2\boxtimes E_2$.} These super-MTCs are therefore also \lq\lq weakly" group theoretical in the sense of \cite{etingof2016tensor}. Since this latter property is preserved under equivariantization \cite[Prop. 9.8.4]{etingof2016tensor}, we have arrived at the following theorem:

\bigskip
\noindent
{\bf Theorem 4:} Given any BFB category, $\CB$, there exists a unique discrete group, $G$, such that $\CB$ can be constructed via a consistent $G$-equivariantization of a (super-) MTC, $\CM$, of the form in \eqref{GenMTC} or \eqref{GenSMTC}. As a result, all BFB symmetries are integral and weakly group theoretical (in the sense of \cite{etingof2016tensor}). Moreover, since we can condense $\CZ_M^{\rm bos}(\CB)$ to get an invertible one-form symmetry, it also follows that BFB categories are non-intrinsically non-invertible (in the sense of footnote \ref{definitionNINI}).

\bigskip
\noindent
In the next section, we would like to understand more precisely when a \lq\lq consistent" $G$-equivariantization exists. To do so, we will mostly focus on the case $\CZ_{M}^{\rm bos}(\CB)\cong\CZ_{M}(\CB)$ and invoke some physical reasoning. At the end of the next section, we will also discuss the case with transparent fermions.

\subsec{Coupling to QFTs, anomaly cancelation, and general BFB categories}\label{coupling}
In this section, we would like to understand the possible BFB symmetry categories more concretely. Except for some comments and an explicit example at the end of this section, we will mostly focus on the case
\begin{equation}
\CZ_{M}^{\rm bos}(\CB)\cong\CZ_{M}(\CB)\not\cong{\rm Vec}~.
\end{equation}
In other words, we will primarily study $\CB$ symmetries that have no transparent fermions. We call such $\CB$ symmetries \lq\lq non-spin" BFB categories (generalizing  the case of non-spin TQFTs). However, in what follows, we will often drop this modifier (instead, when there are cases with a transparent fermion, we will call such $\CB$, \lq\lq spin" BFB categories).

The main issue we wish to address is that the description of BFB categories via Theorem 4 is implicit and depends on the existence of a consistent $G$-equivariantization. Moreover, since $G$-equivariantization is closely related to gauging $G$, one may wonder about the role of $G$ 't Hooft anomalies. Therefore, we would like to first understand such categories more physically before deriving further theorems.

To that end, note that a $2+1$d QFT, $\CQ$, should make sense on manifolds with $T^2$ spatial slices. On these manifolds, line operators in $\CQ$ should be able to \lq\lq detect" each other through their mutual statistics (e.g., see \cite{Kong:2014qka,Lan:2018bui,Lan:2018vjb,Johnson-Freyd:2020usu} and also \cite{Shi:2023kwr} in the (spin) TQFT case). In particular, lines should not be completely \lq\lq invisible." When $\CQ$ is a TQFT, this is the statement that the modular $S$ matrix is invertible. When $\CQ$ is a spin TQFT, $S$ no longer needs to be invertible, but the mild degeneracy in this case reflects the existence of a transparent fermion that can be associated with the spin structure of the spacetime on which $\CQ$ lives. This discussion explains our analysis of the $\CB$ categories of Section \ref{BFB}.

In the case of more general $\CB$ categories, the lines in $\CZ_M^{\rm bos}(\CB)$ are, by definition, \lq\lq undetectable" to the degrees of freedom in $\CB$. Therefore, in order for the lines in $\CZ_M^{\rm bos}(\CB)$ to not be invisible, they must act non-trivially on some non-topological lines in $\CQ$. In other words, $\CB<\CQ$ should be a symmetry (sub) category for the non-topological QFT, $\CQ$.

We can arrange for precisely such an embedding of $\CB\hookrightarrow\CQ$ as a symmetry by coupling $\CB$ to a non-topological QFT, $\CT$. In particular, recall that, by Theorem 4, any $\CB$ is a $G$-equivariantization of an MTC, $\CM$. This statement implied throwing out the twisted sectors of the $G$ gauging of $\CM$. We would like to not throw out such twisted sectors by hand and instead find a use for them.

\begin{figure}
    \centering

\tikzset{every picture/.style={line width=0.75pt}} 

\begin{tikzpicture}[x=0.75pt,y=0.75pt,yscale=-1,xscale=1]

\draw    (156,8) -- (155.93,55.08) ;
\draw    (118.37,43.74) .. controls (118.37,64.16) and (190.45,62.46) .. (192.48,44.87) ;
\draw    (118.37,43.74) .. controls (118.37,35.23) and (139.69,31.26) .. (150.86,32.39) ;
\draw    (192.48,44.87) .. controls (192.48,36.36) and (175.22,31.82) .. (163.04,32.39) ;
\draw    (155.93,61.89) -- (156,95) ;
\draw    (154.92,182.86) -- (154.92,264.55) ;
\draw    (117.35,309.18) .. controls (117.35,329.6) and (189.44,327.9) .. (191.47,310.31) ;
\draw    (117.35,309.18) .. controls (117.35,300.67) and (132.58,294.43) .. (155.93,295) ;
\draw    (191.47,310.31) .. controls (191.47,301.81) and (173.19,294.43) .. (155.93,295) ;
\draw    (534.63,11) -- (534.63,92.69) ;
\draw    (321,62) -- (350.94,62.1) ;
\draw [shift={(352.94,62.11)}, rotate = 180.2] [color={rgb, 255:red, 0; green, 0; blue, 0 }  ][line width=0.75]    (10.93,-3.29) .. controls (6.95,-1.4) and (3.31,-0.3) .. (0,0) .. controls (3.31,0.3) and (6.95,1.4) .. (10.93,3.29)   ;
\draw    (536.66,190.29) -- (536.66,271.98) ;
\draw    (155,132) -- (154.91,159.98) ;
\draw [shift={(154.9,161.98)}, rotate = 270.19] [color={rgb, 255:red, 0; green, 0; blue, 0 }  ][line width=0.75]    (10.93,-3.29) .. controls (6.95,-1.4) and (3.31,-0.3) .. (0,0) .. controls (3.31,0.3) and (6.95,1.4) .. (10.93,3.29)   ;
\draw    (320,237) -- (349.88,236.82) ;
\draw [shift={(351.88,236.81)}, rotate = 179.66] [color={rgb, 255:red, 0; green, 0; blue, 0 }  ][line width=0.75]    (10.93,-3.29) .. controls (6.95,-1.4) and (3.31,-0.3) .. (0,0) .. controls (3.31,0.3) and (6.95,1.4) .. (10.93,3.29)   ;
\draw  [color={rgb, 255:red, 0; green, 0; blue, 0 }  ,draw opacity=1 ][fill={rgb, 255:red, 0; green, 0; blue, 0 }  ,fill opacity=1 ] (154,95) .. controls (154,94.09) and (154.76,93.35) .. (155.69,93.35) .. controls (156.62,93.35) and (157.38,94.09) .. (157.38,95) .. controls (157.38,95.91) and (156.62,96.65) .. (155.69,96.65) .. controls (154.76,96.65) and (154,95.91) .. (154,95) -- cycle ;
\draw  [color={rgb, 255:red, 0; green, 0; blue, 0 }  ,draw opacity=1 ][fill={rgb, 255:red, 0; green, 0; blue, 0 }  ,fill opacity=1 ] (532.63,92.69) .. controls (532.63,91.77) and (533.39,91.03) .. (534.32,91.03) .. controls (535.25,91.03) and (536.01,91.77) .. (536.01,92.69) .. controls (536.01,93.6) and (535.25,94.34) .. (534.32,94.34) .. controls (533.39,94.34) and (532.63,93.6) .. (532.63,92.69) -- cycle ;
\draw  [color={rgb, 255:red, 0; green, 0; blue, 0 }  ,draw opacity=1 ][fill={rgb, 255:red, 0; green, 0; blue, 0 }  ,fill opacity=1 ] (534.98,273.63) .. controls (534.98,272.72) and (535.73,271.98) .. (536.66,271.98) .. controls (537.6,271.98) and (538.35,272.72) .. (538.35,273.63) .. controls (538.35,274.54) and (537.6,275.28) .. (536.66,275.28) .. controls (535.73,275.28) and (534.98,274.54) .. (534.98,273.63) -- cycle ;
\draw  [color={rgb, 255:red, 0; green, 0; blue, 0 }  ,draw opacity=1 ][fill={rgb, 255:red, 0; green, 0; blue, 0 }  ,fill opacity=1 ] (153.54,264.55) .. controls (153.54,263.64) and (154.3,262.9) .. (155.23,262.9) .. controls (156.16,262.9) and (156.92,263.64) .. (156.92,264.55) .. controls (156.92,265.46) and (156.16,266.2) .. (155.23,266.2) .. controls (154.3,266.2) and (153.54,265.46) .. (153.54,264.55) -- cycle ;

\draw (465.86,29.59) node [anchor=north west][inner sep=0.75pt]    {$\frac{S_{\ell _{i} \ell _{j}}}{S_{1\ell _{j}}}$};
\draw (542.87,45.51) node [anchor=north west][inner sep=0.75pt]    {$\ell _{j}$};
\draw (158.17,67.77) node [anchor=north west][inner sep=0.75pt]    {$\ell _{j}$};
\draw (161.12,216.24) node [anchor=north west][inner sep=0.75pt]    {$\ell _{j}$};
\draw (95.13,300.58) node [anchor=north west][inner sep=0.75pt]    {$\ell _{i}$};
\draw (98.16,33.24) node [anchor=north west][inner sep=0.75pt]    {$\ell _{i}$};
\draw (538.01,96.09) node [anchor=north west][inner sep=0.75pt]    {$\mathcal{O}$};
\draw (159.91,95) node [anchor=north west][inner sep=0.75pt]    {$\mathcal{O}$};
\draw (160.91,261.32) node [anchor=north west][inner sep=0.75pt]    {$\mathcal{O}$};
\draw (542.87,223.67) node [anchor=north west][inner sep=0.75pt]    {$\ell _{j}$};
\draw (540.35,277.03) node [anchor=north west][inner sep=0.75pt]    {$\mathcal{O}$};
\draw (496.21,223.46) node [anchor=north west][inner sep=0.75pt]    {$d_{\ell _{i}}$};

\end{tikzpicture}
    \caption{Shrinking a loop of topological line operator $\ell_{i}$ on $\ell_j$, we get $\frac{S_{\ell _{i} \ell _{j}}}{S_{1\ell _{j}}}$. Since $\ell_j$ is an endable line operator, we can move the $\ell_i$ loop down and then shrink it to get $d_{\ell_i}$. This shows that $\frac{S_{\ell _{i} \ell _{j}}}{S_{1\ell _{j}}}=d_{\ell_i}$. Therefore, if $\ell_i$ acts non-trivially on $\ell_j$, $\ell_i$ must be non-topological. These facts allow us to recover the lines thrown out in $G$ equivariantization as lines with non-trivial displacement operators.}
    \label{fig:action on endable line}
\end{figure}
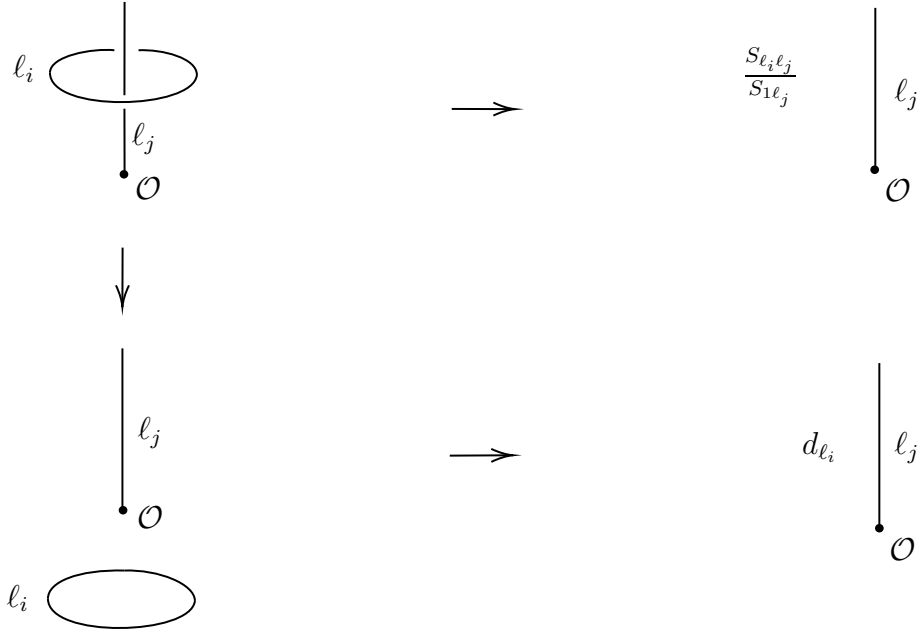

To that end, suppose both $\CM$ and $\CT$ have an action of $G$ (i.e., we suppose there are some local operators in $\CT$ that transform as faithful irreps of $G$; this situation can always be arranged by considering a collection of free bosons and invoking Cayley's theorem). Then, we can consider gauging the diagonal group, $G_{\rm diag}\cong G\cong{\rm diag}(G\times G)$, that couples $\CM$ to $\CT$ to produce the theory $\CQ$. If the gauging is well-defined, then the twisted sector lines we produce from gauging $\CM$ become non-topological lines in $\CQ$, and only $\CB$ lines remain as symmetries of $\CQ$. The reason the twisted sector lines become non-topological is that these lines braid non-trivially with the \lq\lq quantum" $\CZ_M(\CB)\cong{\rm Rep}(G)$ one-form symmetry that arises, and the ${\rm Rep}(G)$ Wilson lines can end on local operators, $\CO_i$, transforming under the corresponding faithful irrep, $\pi_i\in {\rm Rep}(G)$ \cite{Rudelius:2020orz} (note that the $\CB$ lines braid trivially with the quantum symmetry; see Fig. \ref{fig:action on endable line}). For example, if $\CQ$ is a CFT, then we expect the twisted sector lines to become one-dimensional defect CFTs (DCFTs). Therefore, we have precisely accomplished, at the level of the topological lines, a $G$-equivariantization by endowing the twisted sector lines with non-trivial displacement operators. 

But we should be careful to insist that the $G_{\rm diag}$ gauging is well-defined. Indeed, this amounts to studying a theory in which the 't Hooft anomaly, $\omega_4\in H^4(G_{\rm diag},U(1))$ is vanishing in cohomology, and the so-called Postnikov class, $\beta\in H^3_{\rho}(G_{\rm diag},\CA)$ is vanishing (here $\CA$ consist of the invertible one-form symmetry lines).\footnote{See the physical discussion in \cite{Barkeshli:2014cna,Benini:2018reh} for more details. Technically speaking, $\beta$ need not vanish if we are willing to gauge the one-form symmetry as well (i.e., gauge the corresponding 2-group).}

Let us consider $\omega_4\in H^4(G_{\rm diag},U(1))$ first. We have that
\begin{equation}
\omega_4=\omega_4|_{\CM}+\omega_4|_{\CT}~,
\end{equation}
where $\omega_4|_{\CM}$ and $\omega_4|_{\CT}$ are the contributions to the anomaly from $\CM$ and $\CT$ respectively. In general, it may happen that $\omega_4|_{\CM}\ne[0]$ in cohomology. In this case, we will need the contribution from $\CT$ to cancel it. We claim we can always arrange for such a cancelation to occur through an appropriate modification of $\CT$. Indeed, since $H^4(G_{\rm diag},U(1))$ is a finite Abelian group, $\omega_4|_{\CM}$ and $\omega_4|_{\CT}$ have order $n\ge1$ and $m\ge1$ respectively. We can now redefine
\begin{equation}\label{stackT}
\CT\to(\CT)^{\boxtimes m}~.
\end{equation}
In other words, we stack $m$ copies of $\CT$ together and call this our new theory, $\CT$. We can extend the definition of $G_{\rm diag}$ to act diagonally on each of the $m$ copies of $\CT$ as well so that $\omega_4|_{\CT}=[0]$. To get rid of the anomaly from $\CM$, we can now deform
\begin{equation}\label{stack1}
\CT\to\CT\boxtimes(\CM)^{\boxtimes(n-1)}~,
\end{equation}
and extend the action of $G_{\rm diag}$ to the $(n-1)$ copies of $\CM$ we have stacked onto $\CT$. If we wish to render each copy of $\CM$ non-topological, we can always couple it to sufficiently many free real scalars as we did in Section \ref{UVcompletion}. In any case, we now have
\begin{equation}\label{stack2}
\omega_4=\omega_4|_{\CM}+\omega_4|_{\CT}=[0]~.
\end{equation}
In particular, we see that the existence of an $H^4(G,U(1))$ anomaly in $\CM$ does not matter. This statement is in fact consistent with the mathematical literature on equivariantization, which ignores such anomalies.

Next, let us discuss the Postnikov class, $\beta$. In this case, even after performing the modifications in \eqref{stack1} and \eqref{stack2}, we cannot cancel $\beta\ne0$. The reason is that this quantity is operator valued. In particular, $\beta$ is valued in $\CA$, the invertible (part of the) one-form symmetry of $\CM$. Adding additional copies of $\CM$ and $\CT$ therefore cannot cancel $\beta$. As a result, we should still insist on $\beta=0$. This statement is again consistent with the mathematical literature. According to the results in \cite{Benini:2018reh}, the Postnikov class vanishes in gaugings of Abelian MTCs. As a result, we arrive at the following theorem:

\bigskip
\noindent
{\bf Theorem 5:} Assuming the results in \cite{Benini:2018reh} (see also \cite{Delmastro:2019vnj} and references therein) on the vanishing of the Postnikov class in Abelian TQFTs (at least in theories of the type \eqref{GenMTC}), we find a one-to-one correspondence between non-spin BFBs and $G$-equivariantizations for any finite group, $G$, of an MTC of the form in \eqref{GenMTC}.

\bigskip
\noindent
One consequence of this theorem is that, in the non-(super-) modular case, $\CB$ can have non-Abelian lines {\it and} non-trivial braiding. In other words, generic $\CB$ BFB categories can, like the symmetric categories described by Deligne, have non-Abelian lines.

Moreover, Theorem 5 leads to an explicit description of the modular data of any non-spin BFB category in terms of the data of an MTC of the form in \eqref{GenMTC}  and a $G$-action on it. Let $\CM$ be an MTC of the form in \eqref{GenMTC}. Consider an action of a 0-form symmetry, $G$, on $\CM$ specified by a group homomorphism
\be
\rho : G \to \text{Aut}(\CM)~,
\ee
where $\text{Aut}(\CM)$ is the group of 0-form symmetries of $\CM$. Let $\eta_\ell(g,h)$ be the symmetry fractionalization class. From Theorem 5, we know that, upon gauging $G$, we get a non-spin BFB category, $\CB$. We also have the freedom to stack a $G$-SPT while gauging the symmetry $G$. This maneuver changes the correlation functions of the twisted-sector line operators and leaves the data of the genuine topological line operators unchanged. Since the twisted-sector lines operators are non-topological in our case, stacking a $G$-SPT do not change the details of the topological line operators obtained after gauging $G$. 

The line operators in $\CB$ can be written as
\be
([\ell],\pi_\ell)~,
\ee
where $[\ell]$ is an orbit of line operators in $\CB$ under $G$-action with representative $\ell$, and $\pi_\ell$ are irreducible projective representations of the centralizer $C_\ell$ of $\ell$ satisfying
\be
\pi_\ell(g) \pi_\ell(h) = ~\eta_\ell(g,h)~ \pi_\ell(gh) ~ \forall ~ g,h \in C_\ell~.
\ee
Note that the representations $\pi_{\ell}$ depend crucially on the choice of fractionalization class. In particular, if $\eta_{\ell}(g,h)$ is a non-trivial 2-cocycle on $C_{\ell}$, then all irreducible projective representations are higher-dimensional. The topological spin of these line operators is given by
\be
\theta((\ell,\pi_\ell))=\theta(\ell)~,
\ee
where $\theta(\ell)$ is the spin of $\ell\in\CM$.  Also, the normalized $S$ matrix of $\CB$ is given by \cite{Barkeshli:2014cna}
\be
S_{([\ell_1],\pi_{\ell_1}),([\ell_2],\pi_{\ell_2})}= \frac{1}{|G|} \sum_{t\in G/G_{\ell_1}, s\in G/G_{\ell_2}} S_{{}^t{\ell_1},{}^s{\ell_2}} ~\text{dim}(\pi_{\ell_1}) \text{dim}(\pi_{\ell_2}) ~,
\ee
where $G_{\ell}$ is the normal subgroup of $G$ that stabilizes $\ell$. Finally, using the expression for fusion rules of equivariantizations of fusion categories in \cite{burciu2013fusion} (see also \cite{Barkeshli:2014cna}), the fusion rules of $\CB$ can be explicitly determined. 

To conclude this section, let us comment on the case of spin BFB categories. From Theorem 4, we know that these can be obtained from equivariantization of a consistent $G$-action on a super-MTC of the form in \eqref{GenSMTC}. Even though the super-MTCs in \eqref{GenSMTC} are all Abelian, unlike the case of Abelian MTCs, an arbitrary group $G$ does not always act consistently on an Abelian super-MTC.

In particular, the generalization of the Postnikov class to $G$-actions on Abelian super-MTCs can be non-trivial. Several examples of this phenomenon are given in \cite{galindo2017categorical}. Let us construct an explicit example of $G$-action on a BFB spin TQFT with non-trivial Postnikov class.

To that end, consider the super-MTC, $\CM := E_2\boxtimes \{1,\psi\} = \{1,e,m,f\}\boxtimes \{1,\psi\}$, where $\psi$ is the transparent fermion (recall that $E_2$ is the toric code MTC). Let us focus on the $G_f \cong\DZ_{2}\times \DZ^{F}_{2} $ 0-form symmetry generated by $g_1$ and $(-1)^F$, where $g_1$ implements the transformation 
\be1\rightarrow 1~,\ \ \  m\rightarrow m~,\ \ \ e \rightarrow f\psi~,\ \ \  f\rightarrow e\psi~,
\ee
and $\DZ^{F}_{2}$ is the fermion parity generated by $(-1)^F$. We prescribe the following action of the symmetry on the fusion spaces 
\be
\rho_{g_1}[V^{a,b}] := U_{g_1}(a,b)V^{g_1(a),g_1(b)}~,\ \ \ U_{g_1}(a,b) := (-1)^{a_eb_e+a_eb_{\psi}}~,
\ee
where we denote the anyons of $\CM$ as $a := (a_e,a_m,a_\psi)$ with $a_i \in \{0,1\}$ (so that we have $a=e^{a_e}m^{a_m}\psi^{a_{\psi}}$).\footnote{This symmetry action can be extended consistently to at least one minimal modular extension of $\CM$ (in particular, to $\widetilde\CM := E_2\boxtimes E_2$).} The fractionalization class, $\nu(g,h) = (gh,0,0)$, has a fermionic Postnikov class given by $\textswab{O}(g,h,k):= (d_{\rho}\nu)(g,h,k) := (0,ghk,ghk)$ (here $d_{\rho}$ is the twisted differential of $C^{\star}_{\rho}(\DZ_{2},\CM)$ complexes). In particular, $\textswab{O}$ is non-trivial in $H^{3}(\DZ_{2},\CM/\{1,\psi\})$ (this fermionic Postnikov class corresponds to the obstruction discussed in \cite[Section V.B.2]{BulmashBarkeshli1}; see also \cite{Aasen:2021vva}). This discussion shows that the one-form symmetry of $\CM$ forms a fermionic version of a non-trivial 2-group with the $\DZ_{2}\times \DZ^{F}_{2} $ 0-form symmetry. Therefore, this 0-form symmetry cannot be gauged on its own. 

In the next section we will use the above results and study continuous deformations of non-topological $\CQ$ with $\CB$ symmetry.

\newsec{Continuous Deformations and RG Flows}\label{RG}

In Section \ref{BFB}, we saw that all BFB categories can be obtained by gauging a discrete $G$ symmetry of a highly restricted set of Abelian (super-) MTCs in \eqref{GenMTC} and \eqref{GenSMTC} and coupling to a non-topological QFT.\footnote{Or, from a simpler though more mathematical perspective, we can obtain any BFB by equivariantizing the Abelian (super-) MTCs in question with respect to $G$.} Here we make use of this fact and discuss its implications for continuous deformations of non-topological QFTs. We will focus primarily on RG flows.

Let us suppose that we start with some UV QFT, $\CQ_{UV}$, that has a one-form symmetry category, $\CB_{UV}<\CQ_{UV}$ (recall that, in our terminology, one-form symmetry can also be non-invertible). For now, we will assume that this category is a general premodular fusion category.\footnote{In more general cases, we can think of $\CB_{UV}$ as being a premodular fusion subcategory of a larger (possibly non-semisimple) UV one-form symmetry category, $\CC_{UV}$.} We will return to the case where $\CB_{UV}$ is a BFB category (or, slightly more generally, a BFB subcategory) soon.

As we have seen in the previous section, via condensation of the transparent bosonic lines in the M\"uger center, $\CZ_{M}^{\rm bos}(\CB_{UV})\le\CB_{UV}$, we obtain a (super-) MTC, $\CM_{UV}$
\begin{equation}\label{MUV}
\CM_{UV}:=\CB_{UV}/\CZ_M^{\rm bos}(\CB_{UV})~.
\end{equation}
Almost by definition, $\CM_{UV}$ and $\CB_{UV}$ share equivalent one-form symmetry 't Hooft anomalies
\begin{equation}\label{AnomEquiv}
\CA^{(1)}(\CM_{UV})\cong\CA^{(1)}(\CB_{UV})~.
\end{equation}
This statement holds because the one-form anomaly is invariant under condensing transparent lines (and therefore also under the inverse operation of equivariantization). Said more concretely, by \eqref{AnomEquiv}, we mean that the modular data of these two categories agree up to condensation of transparent lines (or, depending on which side one starts with, equivariantization\footnote{More generally, one can equate anomalies of categories related by a combination of condensation, equivariantization, and the application of invertible zero-form symmetries.}). We will describe more general notions of anomaly matching later.

Now, let us consider an RG flow which, for concreteness, starts from some UV-complete QFT (e.g., a CFT) and involves turning on vacuum expectation values for local operators (e.g., as is common in free bosonic theories or when going onto the moduli space of vacua in supersymmetric theories; this expectation value could also be the result of dynamics of the QFT) and / or turning on various local deformations. We expect such a flow to respect $\CB_{UV}$, in the sense that the $\CB_{UV}$ symmetry defects remain topological and remain as genuine lines (although some defects may become trivial).

More generally, our results apply to any flow that preserves $\CB_{UV}$. It might seem surprising that we consider flows that allow some of the defects in $\CB_{UV}$ to condense / trivialize (e.g., this can happen explicitly through turning on vacuum expectation values for defect endpoint operators and condensing corresponding lines). The point is that by \lq\lq preserving $\CB_{UV}$," we mean that only condensation / trivialization of operators in $\CZ_{M}(\CB_{UV})$ can occur. In particular, none of the $\CB_{UV}$ lines are (in the condensed matter language) confined or (in the high-energy theory language) rendered non-genuine as a result of the RG flow.\footnote{Rendering a genuine line non-genuine changes the topology of the line, since the image under condensation must be attached to a surface.}

The result of this flow can be summarized as follows for the degrees of freedom of interest to us
\begin{equation}\label{RGB}
\CB_{UV}<\CQ_{UV}\ \xrightarrow[]{\text{  RG  }}\ F_{RG}(\CB_{UV})<\CQ_{IR}~,
\end{equation}
where $F_{RG}$ is a functor, some of whose properties we will describe below, that implements the RG.\footnote{In $1+1$d, when the gapped IR phase is trivial, we have $\CB_{IR}\cong \text{Vec}$ making the RG functor, $F_{RG}$, a fiber functor. Constraints on $1+1$d RG flows from the existence of fiber functors were studied in \cite{Chang:2018iay,Thorngren:2019iar,Thorngren:2021yso}. RG functors for RG flows from $1+1$d CFTs to possibly non-trivial IR gapped or gapless phase were studied in \cite{Kikuchi:2022rco,Kikuchi:2021qxz,Kikuchi:2022gfi,Kikuchi:2022ipr}. In $3+1$d, RG functors for BPS line defects were studied in \cite{Gaiotto:2024fso}.} In other words, $F_{RG}$ maps the UV QFT to the IR QFT
\begin{equation}
F_{RG}:\CQ_{UV}\to\CQ_{IR}~.
\end{equation}
The image, $F_{RG}(\CB_{UV})$, of the UV one-form symmetry is in general only a proper subcategory of the IR one-form symmetry
\begin{equation}
F_{RG}(\CB_{UV})\le\CB_{IR}~.
\end{equation}
This statement holds because, although $F_{RG}(\CB_{UV})$ is closed, we typically expect additional accidental / emergent symmetries in the IR (we have already seen examples of this common phenomenon in Section \ref{UVcompletion}), and so $F_{RG}(\CB_{UV})$ constitutes the IR symmetries that are \lq\lq visible" in the UV.\footnote{$F_{RG}(\CB_{UV})$ is closed because otherwise, as we go back up the RG flow from the IR to the UV, a product of two topological lines in $F_{RG}(\CB_{UV})$ would produce non-topological lines.} Of course, depending on the 't Hooft anomalies, parts or all of $F_{RG}(\CB_{UV})$ may be trivial at long distances.

\begin{figure}[h!]
    \centering

\tikzset{every picture/.style={line width=0.75pt}} 

\begin{tikzpicture}[x=0.75pt,y=0.75pt,yscale=-1,xscale=1]

\draw  [color={rgb, 255:red, 37; green, 118; blue, 212 }  ,draw opacity=1 ][fill={rgb, 255:red, 30; green, 94; blue, 151 }  ,fill opacity=0.51 ] (280.99,178.47) -- (281.01,63.85) -- (358.99,41.35) -- (358.97,155.97) -- cycle ;

\draw (206,100.4) node [anchor=north west][inner sep=0.75pt]    {$\CQ_{UV}$};
\draw (407,100.4) node [anchor=north west][inner sep=0.75pt]    {$\CQ_{IR}$};
\draw (330,51.4) node [anchor=north west][inner sep=0.75pt]    {$\CI_{RG}$};

\end{tikzpicture}
    \caption{An RG interface, $\CI_{RG}$, between the UV and IR QFTs. In principle, this interface encodes the mapping of all UV operators to IR operators.}
    \label{fig:RG interface}
\end{figure}
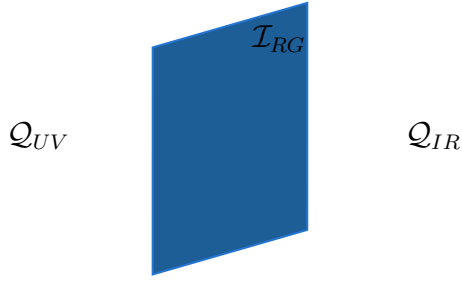

Another useful perspective on $F_{RG}$ is to think of it as a non-topological interface, $\CI_{RG}$, between $\CQ_{UV}$ and $\CQ_{IR}$ (see Fig. \ref{fig:RG interface}). When $\CQ_{UV}$ is a CFT, we have a particularly simple picture:\footnote{Although conceptually simple, implementing the RG interface in practice is typically non-trivial (e.g., see \cite{Brunner:2007ur,Gaiotto:2012np,Andrei:2018die,Konechny:2023xvo}).} we can generate $\CI_{RG}$ by integrating a relevant deformation of $\CQ_{UV}$ (and / or turning on a vacuum expectation value) on half the spacetime and flowing to the IR.  't Hooft anomaly matching for the one-form symmetry is the statement that\footnote{Formally, the RG interface, $\CI_{RG}$, describes how all local and extended operators of the UV QFT, $\CQ_{UV}$, map to those in the IR QFT, $\CQ_{IR}$. For example, $\CI_{RG}$ specifies how topological surface and line operators in the UV are mapped to their counterparts in the IR. The topological surfaces and lines of a 2+1d QFT form a 2-category. Therefore, studying RG flows in this general setting requires understanding monoidal 2-functors between two 2-categories. We leave this study for future work and  instead focus on the topological lines and corresponding 1-categories.}
\begin{equation}\label{FuncAnom}
\CA^{(1)}(\CB_{UV})\cong\CA^{(1)}(F_{RG}(\CB_{UV}))~.
\end{equation}

Let us describe $F_{RG}$ in more detail and summarize some of its properties. In the following, we will denote simple line operators in $\CB_{UV}$ by $x,y,z$ and simple line operators in $\CB_{IR}$ by $a,b,c$. Since we are considering RG flows which preserve the $\CB_{UV}$ symmetry, $F_{RG}(x)$ must be a topological operator in $\CB_{IR}$ for all $x\in \CB_{UV}$. In general, the RG functor, $F_{RG}$, maps a UV simple line operator to a direct sum of simple line operators in the IR.   Indeed, we have 
\be
F_{RG}(x)= \sum_{a\in \CB_{IR}} N_{F_{RG}(x)}^a ~ a~,
\ee
where $N_{F_{RG}(x)}^a$ are non-negative integers. The image of $\CB_{UV}$ under $F_{RG}$, denoted $F_{RG}(\CB_{UV})$, is a subcategory of $\CB_{IR}$. It is defined as the subcategory generated by all simple line operators, $a \in \CB_{IR}$, such that $N_{F_{RG}(x)}^a\neq 0$ for some $x \in \CB_{UV}$. What properties should $F_{RG}$ (and $\CI_{RG}$) have in order to reproduce \eqref{FuncAnom}? One natural class of functors that satisfy this condition are braided monoidal functors \cite{baez2004some}. In Appendix \ref{Ap:FRG} we describe these functors in more detail and prove that they lead to the anomaly matching conditions \eqref{FuncAnom}. We also describe the implications for $\CI_{RG}$. In particular, we prove the following theorem:
\vspace{0.2cm}

\noindent {\bf Theorem 6:} Assuming that the RG functor, $F_{RG}$, is a braided monoidal functor, it satisfies 
\begin{enumerate}
\item[(i)] $N_{F_{RG}(x)F_{RG}(y)}^{F_{RG}(z)}=N_{xy}^z$.
\item[(ii)] $d_{F_{RG}(x)}=d_x$.
\item[(iii)] If $N_{F_{RG}(x)}^a\neq 0$, then $\theta(a)=\theta(x)$.
\end{enumerate}

\bigskip
\noindent
Equations (i) and (ii) follow from the monoidal property of the functor, $F_{RG}$. In particular, condition (ii) above implies that if there are topological line operators in $\CB_{UV}$ with irrational quantum dimensions, then $\CB_{IR}$ cannot be trivial. Therefore, this condition is an obstruction to having a trivially gapped IR phase.\footnote{In 1+1d, obstructions to gapped IR phases from quantum dimensions of UV line operators were studied in \cite{Chang:2018iay,Thorngren:2019iar,Thorngren:2021yso}.} When all line operators in $\CB_{UV}$ have integer quantum dimensions, then (i) and (ii) alone do not give an obstruction to a trivially gapped IR phase. However, combining (i), (ii), and (iii), we get the following important corollary:
\bigskip

\noindent {\bf Corollary (one-form symmetry 't Hooft anomaly matching):} The braiding of line operators in $\CB_{UV}$ and $\CB_{IR}$ are related as follows
\be
\label{eq:anomaly matching}
S_{xy} = \sum_{a,b} N_{F_{RG}(x)}^a N_{F_{RG}(y)}^b~ S_{ab}~.
\ee

\bigskip
\noindent
In fact, the RHS of the above equation is equal to $S_{F_{RG}(x),F_{RG}(y)}$ \cite[Lemma 2.4]{muger2003structure}. This statement agrees with the expectation that the braiding between the line operators $x$ and $y$ in the UV must be equal to the braiding between (generically non-simple) line operators $F_{RG}(x)$ and $F_{RG}(y)$ in the IR. This relation explicitly demonstrates the anomaly matching condition \eqref{FuncAnom}. Clearly, if some line operators in $\CB_{UV}$ braid non-trivially with each other, then the IR phase cannot be trivially gapped. 

What more can we say in general? Let us study the image of $\CM_{UV}$ under $F_{RG}$
\begin{eqnarray}\label{MIR}
F_{RG}(\CM_{UV})&\cong& F_{RG}(\CB_{UV}/\CZ_M^{\rm bos}(\CB_{UV}))\cong F_{RG}(\CB_{UV})/ F_{RG}(\CZ^{\rm bos}_M(\CB_{UV}))\cr&\cong& F_{RG}(\CB_{UV})/ \CZ_M^{\rm bos}(F_{RG}(\CB_{UV}))~,
\end{eqnarray}
where, in the second equivalence, we used the fact that RG flows commute with topological operations. In the third equality, we used \eqref{FuncAnom}.\footnote{Recall that $\CZ_M^{\rm bos}(F_{RG}(\CB_{UV}))$ is the subcategory of bosonic line operators that are transparent in $F_{RG}(\CB_{UV})$. This subcategory must be the same as $F_{RG}(\CZ^{\rm bos}_M(\CB_{UV}))$. Indeed, any $\ell\in\CZ_M^{\rm bos}(F_{RG}(\CB_{UV}))$ must come from lines in $ F_{RG}(\CZ^{\rm bos}_M(\CB_{UV}))$. Otherwise, when we go back up the RG flow to the UV, $\ell$ is mapped to lines that have non-trivial 't Hooft anomalies, which is inconsistent. Moreover, by 't Hooft anomaly matching, any $\ell\in F_{RG}(\CZ^{\rm bos}_M(\CB_{UV}))$ must braid trivially with $F_{RG}(\CB_{UV})$ and hence $\ell$ must be a member of $\CZ_M^{\rm bos}(F_{RG}(\CB_{UV}))$.} 
Using \eqref{FuncAnom}, the equivalence in \eqref{AnomEquiv}, and the fact that \eqref{MIR} implies
\begin{equation}
\CA^{(1)}(F_{RG}(\CM_{UV}))=\CA^{(1)}(F_{RG}(\CB_{UV}))~,
\end{equation}
we find 
\begin{equation}
\CA^{(1)}(\CM_{UV})\cong\CA^{(1)}(F_{RG}(\CM_{UV}))~.
\end{equation}
Since the categories in question are (super-) MTCs, the above equivalence implies that the modular $S$ and $T$ matrices match between the two categories. However, modular data does not determine a (super-) MTC \cite{mignard2021modular}, and so this matching is insufficient to conclude that $\CM_{UV}\cong F_{RG}(\CM_{UV})$.\footnote{A reasonable view to take is that, in the general case, more (gauge-invariant) data should be associated with the one-form symmetry 't Hooft anomaly than just the modular data (perhaps even the entire $\CM_{UV}$ MTC itself). We will offer some further comments on this point after we describe the next theorem, but we leave the details for future work. \label{reasonable}}

To proceed further, let us instead apply our discussion to the case we have focused on in Section \ref{BFB}, namely the situation in which $\CB_{UV}$ is a BFB category.\footnote{More generally, $\CB_{UV}$ may contain a closed BFB subcategory, $\tilde\CB_{UV}$. It is straightforward to modify our argument for this case by focusing on $\tilde\CB_{UV}$ instead.} By the above discussion, $F_{RG}(\CB_{UV})$ is also a BFB category. Therefore, both $\CM_{UV}$ and $F_{RG}(\CM_{UV})$ are of the form \eqref{GenMTC} or \eqref{GenSMTC}. Since these are Abelian (super-) MTCs, they are determined by their modular data and so
\begin{equation}\label{MUVMIR}
\CM_{UV}\cong F_{RG}(\CM_{UV})\cong\CM~,
\end{equation}
where the equivalence should be understood up to the action of an invertible topological surface. It is then natural to think of $\CI_{RG}$ and the associated functor, $F_{RG}$, as acting on the above MTC data as an invertible (possibly trivial) zero-form symmetry. In particular, $\CM$ is an invariant of the RG flow. We therefore arrive at the following statement:

\bigskip
\noindent
{\bf Theorem 7:} Consider an RG flow of the type discussed around \eqref{RGB} from a UV theory, $\CQ_{UV}$, with a UV BFB one-form symmetry category, $\CB_{UV}<\CQ_{UV}$. Then, $\CM:=\CB_{UV}/\CZ_M(\CB_{UV})$ is an invariant of the RG flow. Moreover, if there are no emergent one-form symmetries, then $\CM_{IR}:=\CB_{IR}/\CZ_M(\CB_{IR})$ is isomorphic to $\CM$.\footnote{We believe it is likely that Theorem 7 holds more generally (see footnote \ref{reasonable}). One interesting question to better understand is how to fruitfully characterize 't Hooft anomaly matching for general pre-modular categories (here we see that for the BFB case, these anomalies are particularly simple to characterize). Indeed, when $\CB_{UV}$ is a general pre-modular category, $\CB_{UV}/\CZ_M(\CB_{UV})$ is typically a non-Abelian (super-) MTC. A general (super-) MTC is not determined by its modular data. However, we can be somewhat more concrete whenever $\CB_{UV}/\CZ_M(\CB_{UV})$ is multiplicity free. Indeed, using geometric invariant theory, one can show that such modular categories are determined by a finite number of basis-independent polynomial combinations of the $F$ and $R$ matrices \cite{hagge2015geometric,titsworth2016arithmetic} (see also \cite{bonderson2019invariants}). A study of such invariants will be crucial in deriving a concrete statement for anomaly matching for general $\CB_{UV}$.}\footnote{For invertible one-form symmetries, the 't Hooft anomaly can be cancelled using  a 3+1d invertible TQFT (an invertible one-form SPT phase) \cite{Gaiotto:2014kfa,Thorngren:2015gtw,Hsin:2018vcg,Jian:2020qab}. In this case, the invariance of the anomaly under the RG flow is a consequence of the fact that the 3+1d theory is a TQFT. Theorem 7 applies also to more general (non-invertible) one-form symmetry, and the RG invariant TQFT, $\CM$, is a 2+1d TQFT. It will be interesting to understand the anomaly inflow picture for (non-invertible) one-form symmetries.}
\bigskip

Let us study some consequences of this result. By definition we have (up to the action of an invertible topological surface)
\begin{equation}\label{DefRel}
\CB_{UV}/\CZ_M^{\rm bos}(\CB_{UV})\cong F_{RG}(\CB_{UV})/\CZ_M^{\rm bos}(F(\CB_{UV}))~.
\end{equation}
Now, by Deligne's theorem \cite{deligne2002categories} 
\begin{equation}
\CZ_M^{\rm bos}(\CB_{UV})\cong{\rm Rep}(G^{UV})~,\ \ \ \CZ_M^{\rm bos}(F_{RG}(\CB_{UV}))\cong F_{RG}({\rm Rep}(G^{UV}))\cong{\rm Rep}(F_{RG}(G^{UV}))~.
\end{equation}
Using the invertible map in \eqref{MUVMIR}, we can take the groups $G^{UV}$ and $F_{RG}(G^{UV})$ to act on $\CM$. Moreover, along the RG flow, some lines forming a subcategory, $\CC_{\rm triv}\le\CZ_M(\CB_{UV})$, may become trivial in the IR (other lines in $\CB_{UV}$ cannot trivialize due to their non-trivial braiding). More precisely, a line operator, $x \in \CB_{UV}$, is said to trivialize in the IR if $F_{RG}(x)= d_{x}\cdot 1$. Note that the line operators in $\CC_{\rm triv}$ must be closed under fusion. To understand this statement, consider the line operator $x \times y$, where $x,y \in \CC_{\rm triv}$. Let us then study the braiding of $x \times y$ with some general (possibly non-topological) line operator, $z$. 
From Fig. \ref{fig:IR trivial lines}, we find that the action of all simple lines in $x \times y$ on $z$ must be trivial in the IR for any $z$. Therefore, $x \times y$ is trivial in the IR, and
\begin{figure}[h!]
    \centering

\tikzset{every picture/.style={line width=0.75pt}} 

\begin{tikzpicture}[x=0.75pt,y=0.75pt,yscale=-1,xscale=1]

\draw    (156,7) -- (155.93,54.08) ;
\draw    (118.37,43.74) .. controls (118.37,64.16) and (190.45,62.46) .. (192.48,44.87) ;
\draw    (118.37,43.74) .. controls (118.37,35.23) and (139.69,31.26) .. (150.86,32.39) ;
\draw    (192.48,44.87) .. controls (192.48,36.36) and (175.22,31.82) .. (163.04,32.39) ;
\draw    (155.93,60.89) -- (156,88) ;
\draw    (563.63,11) -- (564,123) ;
\draw    (237,73) -- (266.94,73.1) ;
\draw [shift={(268.94,73.11)}, rotate = 180.2] [color={rgb, 255:red, 0; green, 0; blue, 0 }  ][line width=0.75]    (10.93,-3.29) .. controls (6.95,-1.4) and (3.31,-0.3) .. (0,0) .. controls (3.31,0.3) and (6.95,1.4) .. (10.93,3.29)   ;
\draw    (117.37,76.74) .. controls (117.37,97.16) and (189.45,95.46) .. (191.48,77.87) ;
\draw    (191.48,77.87) .. controls (191.48,69.36) and (174.22,64.82) .. (162.04,65.39) ;
\draw    (117.37,76.74) .. controls (117.37,68.23) and (138.69,64.26) .. (149.86,65.39) ;
\draw    (154,197) -- (153.93,265.08) ;
\draw    (116.37,254.74) .. controls (116.37,275.16) and (188.45,273.46) .. (190.48,255.87) ;
\draw    (116.37,254.74) .. controls (116.37,246.23) and (137.69,242.26) .. (148.86,243.39) ;
\draw    (190.48,255.87) .. controls (190.48,247.36) and (173.22,242.82) .. (161.04,243.39) ;
\draw    (153.93,273.89) -- (154,319) ;
\draw    (155.88,97.88) -- (156,123) ;
\draw    (416,11) -- (415.93,58.08) ;
\draw    (378.37,47.74) .. controls (378.37,68.16) and (450.45,66.46) .. (452.48,48.87) ;
\draw    (378.37,47.74) .. controls (378.37,39.23) and (399.69,35.26) .. (410.86,36.39) ;
\draw    (452.48,48.87) .. controls (452.48,40.36) and (435.22,35.82) .. (423.04,36.39) ;
\draw    (415.93,64.89) -- (416,92) ;
\draw    (377.37,80.74) .. controls (377.37,101.16) and (449.45,99.46) .. (451.48,81.87) ;
\draw    (451.48,81.87) .. controls (451.48,73.36) and (434.22,68.82) .. (422.04,69.39) ;
\draw    (377.37,80.74) .. controls (377.37,72.23) and (398.69,68.26) .. (409.86,69.39) ;
\draw    (415.88,101.88) -- (416,127) ;
\draw    (234,259) -- (263.94,259.1) ;
\draw [shift={(265.94,259.11)}, rotate = 180.2] [color={rgb, 255:red, 0; green, 0; blue, 0 }  ][line width=0.75]    (10.93,-3.29) .. controls (6.95,-1.4) and (3.31,-0.3) .. (0,0) .. controls (3.31,0.3) and (6.95,1.4) .. (10.93,3.29)   ;
\draw    (493,195) -- (492.93,263.08) ;
\draw    (455.37,252.74) .. controls (455.37,273.16) and (527.45,271.46) .. (529.48,253.87) ;
\draw    (455.37,252.74) .. controls (455.37,244.23) and (476.69,240.26) .. (487.86,241.39) ;
\draw    (529.48,253.87) .. controls (529.48,245.36) and (512.22,240.82) .. (500.04,241.39) ;
\draw    (492.93,271.89) -- (493,317) ;

\draw (482.86,59.59) node [anchor=north west][inner sep=0.75pt]    {$=\ d_{x} \ d_{y}$};
\draw (572.87,61.51) node [anchor=north west][inner sep=0.75pt]    {$F_{RG}( z)$};
\draw (159.17,104.77) node [anchor=north west][inner sep=0.75pt]    {$z$};
\draw (98.16,33.24) node [anchor=north west][inner sep=0.75pt]    {$x$};
\draw (237,45.4) node [anchor=north west][inner sep=0.75pt]    {$F_{RG}$};
\draw (159.17,291.77) node [anchor=north west][inner sep=0.75pt]    {$z$};
\draw (24,240.4) node [anchor=north west][inner sep=0.75pt]    {$\sum _{w} N_{xy}^{w}$};
\draw (92,244.4) node [anchor=north west][inner sep=0.75pt]    {$w$};
\draw (317,240.4) node [anchor=north west][inner sep=0.75pt]    {$\sum _{w} N_{xy}^{w}$};
\draw (422.17,109.77) node [anchor=north west][inner sep=0.75pt]    {$F_{RG}( z)$};
\draw (325.16,38.24) node [anchor=north west][inner sep=0.75pt]    {$F_{RG}( x)$};
\draw (98,71.4) node [anchor=north west][inner sep=0.75pt]    {$y$};
\draw (324,75.4) node [anchor=north west][inner sep=0.75pt]    {$F_{RG}( y)$};
\draw (234,231.4) node [anchor=north west][inner sep=0.75pt]    {$F_{RG}$};
\draw (499.17,304.77) node [anchor=north west][inner sep=0.75pt]    {$F_{RG}( z)$};
\draw (389,244.4) node [anchor=north west][inner sep=0.75pt]    {$F_{RG}( w)$};
\draw (163.91,148.2) node [anchor=north west][inner sep=0.75pt]  [rotate=-91.5]  {$=$};

\end{tikzpicture}
    \caption{UV lines that are trivial in the IR are closed under fusion. Indeed, for the expressions in the two rows of the figure to agree, all $a\in F_{RG}(w)$ have to braid trivially with $F_{RG}(z)$ for all (topological and non-topological) line operators, $z$. As a result, all such $a$ are trivial in the IR, and $w\in \CC_{\rm triv}$ for all $w\in x\times y$.}
    \label{fig:IR trivial lines}
\end{figure}
$\CC_{\rm triv}$ is a fusion subcategory, ${\rm Rep}(G^{UV}/N)<{\rm Rep}(G^{UV})$, for some normal subgroup, $N\lhd G^{UV}$. Since the line operators in $\CC_{\rm triv}$ braid trivially with the genuine (topological or non-topological) line operators in the IR, the latter must have flux valued in the kernel of line operators in Rep$(G^{UV}/N)$. Demanding that the theory on a manifold with $T^2$ spatial slices is well-defined (we can also think of this requirement as a kind of generalization of the principle of remote detectability \cite{Kong:2014qka,Lan:2018bui,Lan:2018vjb,Johnson-Freyd:2020usu} for topological phases; see also \cite{Shi:2023kwr}) suggests\footnote{\label{fn:restriction}A formal argument for this result uses the fact that $F_{RG}: \CB_{UV} \to \CB_{IR}$ is a braided monoidal functor. In particular, it is a braided monoidal functor from Rep$(G^{UV})$ to $\CB_{IR}$. Any such functor is an embedding of Rep$(H)$ for some subgroup $H\le G$ in $\CB_{IR}$ (see, for example, \cite[Section 3.3.1]{Decoppet:2023rlx}). In our case, $N$ is the largest subgroup of $G^{UV}$ such that the restrictions of all representations in Rep$(G^{UV}/N)$ to $N$ result in (several copies of) the trivial representation of $N$. Indeed, since $F_{RG}$ trivializes the line operators in $\CC_{\rm triv}$, it is the restriction map from irreducible representation of $G^{UV}$ to $N$.}
\be
F_{RG}(\text{Rep}(G^{UV}))\cong\text{Rep}(N)~,
\ee
where $N=F_{RG}(G^{UV})$ is the normal subgroup of $G^{UV}$ described above, and 
\begin{equation}\label{BUVBIR}
F_{RG}(\CB_{UV})\cong\CB_{UV}/{\rm Rep}(G^{UV}/N)=\CB_{UV}/{\rm Rep}(G^{UV}/F_{RG}(G^{UV}))~.
\end{equation}
In particular, $F_{RG}(\CB_{UV})$ and $\CB_{UV}$ at most differ by anyon condensation.\footnote{\label{formaljust}We can further justify \eqref{BUVBIR} as follows. It is always possible to equivariantize with respect to a discrete $G$ symmetry by first equivariantizing with respect to a normal sugroup, $H\lhd G$, and then equivariantizing with respect to $G/H$ (e.g., see the discussion in \cite{Barkeshli:2014cna}). Proceeding in this way, first equivariantizing \eqref{DefRel} with respect to $F_{RG}(G^{UV})$ gives $F_{RG}(\CB_{UV})$. Further equivariantizing with respect to $G^{UV}/F_{RG}(G^{UV})$ gives $\CB_{UV}$. As a result, \eqref{BUVBIR} follows from condensation.} Equivalently, we can say that $\CI_{RG}$ acts on the BFB category, $\CB_{UV}$, as a (potentially trivial) surface implementing condensation from UV to IR combined with an invertible 0-form symmetry. 

So far, we have assumed that some simple UV lines are completely trivialized in the flow to the IR (i.e., the situation in which $F_{RG}(x)=d_x\cdot 1$). More generally, simple UV lines can potentially be partially trivialized in the following sense
\be\label{partialT}
F_{RG}(x)=N_{F_{RG}(x)}^1\cdot1+N_{F_{RG}(x)}^{a_0}\cdot a_0+\cdots~,\ \ \ N_{F_{RG}(x)}^1~,\ N_{F_{RG}(x)}^{a_0}>0~.
\ee
In writing \eqref{partialT}, we have assumed that $a_0$ is a non-trivial simple IR line, and the ellipses contain any additional non-trivial IR contributions to $F_{RG}(x)$. We say that $x$ is partially trivialized because $F_{RG}(x)$ contains a contribution from the trivial line. Such lines are part of a collection of lines we call $\CC_{\rm partial}$ (we can include $\CC_{\rm triv}$ as a potentially non-trivial subcategory). In general, $\CC_{\rm partial}$ is not closed under fusion. However, using the argument in footnote \ref{fn:restriction}, we obtain
\be
F_{RG}(\CZ_M^{\rm bos}(\CB_{UV}))=F_{RG}(\text{Rep}(G^{UV}))=\text{Rep}(H)~,
\ee
where $H=F_{RG}(G^{UV})$ is now a general subgroup of $G^{UV}$. This argument shows that $\CZ_M^{\rm bos}(F_{RG}(\CB_{UV}))=\text{Rep}(H)$. Moreover, $\CB_{UV}$ and $F_{RG}(\CB_{UV})$ are related by anyon condensation (the argument for this latter claim is similar to the one in footnote \ref{formaljust} but now $G^{UV}/H$ is a non-invertible symmetry; see also \cite[Theorems 2.1, 2.2]{Kirillov:2001ti}). To summarize, we have:

\bigskip
\noindent
{\bf Theorem 8:}  Consider an RG flow of the type discussed around \eqref{RGB} from a UV theory, $\CQ_{UV}$, with a UV BFB one-form symmetry category, $\CB_{UV}$, to an IR theory, $\CQ_{IR}$, with an IR BFB category, $\CB_{IR}$. Then, $F_{RG}(\CB_{UV})\le\CB_{IR}$ differs from $\CB_{UV}$ at most by anyon condensation.\footnote{Note that this statement, with its implicit ordering, is intuitively consistent with the $F$-theorem \cite{Casini:2012ei,Casini:2015woa}, since topological degrees of freedom contribute to this quantity.}

\bigskip
\noindent
In Section \ref{examples}, we will construct various examples that illustrate the above statement.\footnote{However, the examples of RG flows that we consider in this work do not include examples of the partial trivializations in \eqref{partialT}. It will be interesting to find examples of this potential phenomenon, but we leave this point for future work.}

Before continuing, let us make a comment on exactly marginal deformations in a continuous family of conformal field theories (e.g., the circle branch of the 2d compact boson or the gauge coupling fundamental domain in 4d $\CN=4$ super Yang-Mills). Such excursions are close cousins of RG flows and, instead of an RG interface, we can consider an interface with different values of the exactly marginal coupling on each side (such interfaces are sometimes known as \lq\lq Janus" interfaces \cite{DHoker:2006qeo}). In this case, we do not expect line operators to trivialize as we vary the exactly marginal parameter, and so it is natural to conjecture that if a CFT, $\CQ$, with a BFB category, $\CB<\CQ$, is part of some conformal manifold, $M_{CFT}$, then $\CB$ is an invariant of $M_{CFT}$.

In the next subsection we specialize to the case of gapped IR phases, where the long-distance physics is simpler, and we can therefore prove stronger statements.

\subsec{Gapped IR phases}\label{gapped}
Suppose the RG flow is such that the IR phase is gapped. Then, $\CB_{IR}$ must be a (super-) MTC, $\CM_{IR}$. In what follows, we will assume that the IR theory is an MTC. However, our results also extend simply to the case that $\CM_{IR}$ is a split super-MTC (i.e., it is of the form $\CM_{IR} \cong\widehat\CM_{IR}\boxtimes{\rm SVec}$, with $\widehat{\CM}_{IR}$ an MTC). The reason is that in this case we can focus on $\widehat{\CM}_{IR}$ and apply our logic below (which includes using results in \cite{muger2003structure} that give us a particularly strong handle on the IR phase).

There are various possibilities for the IR phase based on the fate of $\CZ^{\rm bos}_M(\CB_{UV})$ under the RG flow. Let us consider two extreme cases:
\begin{enumerate}
\item{Suppose the RG flow is such that all the lines in $\CZ^{\rm bos}_M(\CB_{UV})$ get mapped to the trivial line in the IR. Then, the RG flow condenses $\CZ_M^{\rm bos}(\CB_{UV})$ (we will see examples of this phenomenon in Section \ref{examples}). As a result,
\begin{equation}\label{GappedIRTriv}
\CB_{IR}=\CM_{IR}\cong\CM\boxtimes\CM'~,
\end{equation}
where $\CM'$ may (or, depending on the details of the RG flow, may not) be a non-trivial MTC. Here we see that (a factor of) $\CB_{IR}$ is the anomaly TQFT of $\CB_{UV}$.}

\item{On the other hand, suppose all line operators in $\CB_{UV}$ are mapped to simple IR lines. In particular, this statement implies that all the transparent lines in $\CZ_M^{\rm bos}(\CB_{UV})$ are non-trivial in the IR. Therefore, the IR MTC, $\CM_{IR}$, is a modular extension of $\CB_{UV}$.

Consider the image of $\CZ_M^{\rm bos}(\CB_{UV})$, $F_{RG}(\CZ_M^{\rm bos}(\CB_{UV}))<\CM_{IR}$. Suppose the set of line operators in $\CM_{IR}$ that braid trivially with all lines in $F_{RG}(\CZ_M^{\rm bos}(\CB_{UV}))$ is precisely $F_{RG}(\CB_{UV})$. In this case, the dimension of $\CM_{IR}$ and $\CB_{UV}$ are related as
\be
\text{dim}(\CM_{IR})= \text{dim}(\CB_{UV})\cdot \text{dim}(\CZ_M^{\rm bos}(\CB_{UV}))~.
\ee
To understand this relation, condense the bosons $F_{RG}(\CZ_M^{\rm bos}(\CB_{UV}))<\CM_{IR}$. Only the line operators in $\CM_{IR}$ which braid trivially with $F_{RG}(\CZ_M^{\rm bos}(\CB_{UV}))$ survive this condensation. After condensation, we therefore obtain the MTC, $\CM$. Therefore, $\CM_{IR}$ can be obtained from gauging a 0-form symmetry, $G^{UV}$, of $\CM$, and we have \cite{Barkeshli:2014cna}
\be
\text{dim}(\CM_{IR})=\text{dim}(\CM)\cdot |G^{UV}|^2~.
\ee
On the other hand, if we gauge the $G^{UV}$ symmetry of $\CM$ without including twisted sectors, we get the category $\CB_{UV}$, and so
\be
\text{dim}(\CB_{UV})=\text{dim}(\CM)\cdot |G^{UV}|~.
\ee
Combined with the equation above, we get
\be
\text{dim}(\CM_{IR})=\text{dim}(\CB_{UV})\cdot |G^{UV}|= \text{dim}(\CB_{UV})\cdot \text{dim}(\CZ_M^{\rm bos}(\CB_{UV})) ~.
\ee
In the second equality we used the fact that $F(\CZ_M^{\rm bos}(\CB_{UV}))\cong \text{Rep}(G^{UV})$ and that $\text{dim}(\CZ_M^{\rm bos}(\CB_{UV}))=\text{dim}(\text{Rep}(G^{UV}))=|G^{UV}|$. $\CM_{IR}$ is called a minimal modular extension of $\CB_{UV}$ \cite{muger2003structure} (see also \cite{lan2017modular}). All lines in $\CM_{IR}$ which are not in $F(\CB_{UV})$ have a non-trivial mixed anomaly with lines in $F(\CZ_M^{\rm bos}(\CB_{UV}))$.

However, not all braided fusion categories, $\CB_{UV}$, admit a minimal modular extension. Following \cite{bruillard2017fermionic,galindo2017categorical}, the obstruction to the existence of a minimal modular extension can be understood as follows. Condense $\CZ_M^{\rm bos}(\CB_{UV})=\CZ_M(\CB_{UV})$ to get the MTC, $\CM$, with a 0-form symmetry, $G^{UV}$. The action of $G^{UV}$ on the category $\CM$ is guaranteed to have trivial Postnikov class (recall that we are working with the case of a theory without transparent fermions). Now, $\CM$ can be extended to a $G^{UV}$-crossed braided category, $\CM_{G^{UV}}$, if and only if the anomaly of $G^{UV}$, $\omega\in H^4(G^{UV},U(1))$, is trivial. If the anomaly is trivial, then $\CM_{G^{UV}}$ is a consistent $G^{UV}$-crossed braided category with a $G^{UV}$ action. This $G^{UV}$ action can be gauged to get an MTC, and the resulting MTC is the required minimal modular extension. However, if $G^{UV}$ is anomalous, then $\CB_{UV}$ does not admit a minimal modular extension. Therefore, if $\CB_{UV}$ is such that the $G^{UV}$-symmetry of $\CM$ is anomalous, we see that the IR MTC, $\CM_{IR}$, necessarily hosts line operators having trivial mixed-anomaly with $F_{RG}(\CZ_M^{\rm bos}(\CB_{UV}))$ beyond those in $F_{RG}(\CB_{UV})$. As a result \cite{muger2003structure,galindo2017categorical},
\bea
\text{dim}(\CM_{IR})&=& {\rm dim}(C_{\CM_{IR}}(F_{RG}(Z_M^{\rm bos}(\CB_{UV}))))\cdot \text{dim}(F_{RG}(\CZ_M^{\rm bos}(\CB_{UV})))\cr&>&\text{dim}(\CB_{UV})\cdot \text{dim}(\CZ_M^{\rm bos}(\CB_{UV}))~,
\eea
where $C_{\CM_{IR}}(F_{RG}(Z_M^{\rm bos}(\CB_{UV})))$ is the centralizer of $F_{RG}(\CZ_M^{\rm bos}(\CB_{UV}))<\CM_{IR}$ (i.e., the subcategory of lines in $\CM_{IR}$ that braid trivially with $F_{RG}(\CZ_M^{\rm bos}(\CB_{UV}))$).\footnote{When $\CQ_{IR}$ is gapped, it is a $\CB_{UV}$-symmetric 2+1d TQFT.}

In fact, we can elaborate on the role of the additional lines, $\ell_i\in C_{\CM_{IR}}(F_{RG}(\CB_{UV}))$, that are not in $F_{RG}(\CB_{UV})$. It is easy to see that their images under anyon condensation are precisely the lines needed to cancel the $G^{UV}$ anomaly. Indeed, condensing $\CZ^{\rm bos}_{M}(F_{RG}(\CB_{UV}))$ yields
\begin{equation}
\CM_{IR}/\CZ_{M}^{\rm bos}(F_{RG}(\CB_{UV}))=\CM\boxtimes\CM'~,
\end{equation}
where $\CM'$ must be an MTC (coming from the images of the $\ell_i$ under condensation) with a $G^{UV}_{\psi}$ anomaly, $\omega'\in H^4(G^{UV},U(1))$, that cancels the anomaly, $\omega\in H^4(G^{UV},U(1))$, arising from $\CM$
\begin{equation}
\omega+\omega'=[0]\in H^4(G^{UV},U(1))~,
\end{equation}
so that we can gauge a diagonal $G^{UV}$ acting on $\CM$ and $\CM'$ to produce $\CM_{IR}$.}
\end{enumerate}

As we saw in Section \ref{RG}, the most general scenario is that the RG flow maps $\CZ_M(\CB_{UV})\cong{\rm Rep}(G^{UV})$ to $F_{RG}(\CZ_M^{\rm bos}(\CB_{UV}))\cong{\rm Rep}(F_{RG}(G^{UV}))$, where $F_{RG}(G^{UV})<G^{UV}$ is a subgroup (the first case above corresponds to $F_{RG}(G^{UV})=\mathbb{Z}_1$, and the second case corresponds to $F_{RG}(G^{UV})\cong G^{UV}$). 

We can repeat the analysis of the second case above in this more general setting. To that end, suppose that the set of line operators in $\CM_{IR}$ that braid trivially with all lines in $F_{RG}(\CZ_M^{\rm bos}(\CB_{UV}))$ is precisely $F_{RG}(\CB_{UV})$. In this case, the dimension of $\CM_{IR}$ and $\CB_{UV}$ are again related as
\be
\text{dim}(\CM_{IR})= \text{dim}(F_{RG}(\CB_{UV}))\cdot \text{dim}(F_{RG}(\CZ_M^{\rm bos}(\CB_{UV})))~.
\ee
We again have that $\CM_{IR}$ is a minimal modular extension of $F_{RG}(\CB_{UV})$ (although with respect to a subgroup, $F_{RG}(G^{UV})<G^{UV}$; if $F_{RG}(G^{UV})\cong\mathbb{Z}_1$, then $\CM_{IR}\cong\CM\cong F_{RG}(\CB_{UV})$).

However, as in the second case above, $F_{RG}(\CB_{UV})$ may not admit a minimal modular extension, and so
\bea
\text{dim}(\CM_{IR})&=& {\rm dim}(C_{\CM_{IR}}(F_{RG}(Z_M(\CB_{UV}))))\cdot \text{dim}(F_{RG}(\CZ_M^{\rm bos}(\CB_{UV})))\cr&>&\text{dim}(F_{RG}(\CB_{UV}))\cdot \text{dim}(F_{RG}(\CZ_M(\CB_{UV})))~.
\eea
Once more, $\CM_{IR}$ has line operators with trivial mixed-anomaly with $F_{RG}(\CZ_M^{\rm bos}(\CB_{UV}))$ beyond those in $F_{RG}(\CB_{UV})$.

The role of the additional lines, $\ell_i\in C_{\CM_{IR}}(F_{RG}(\CB_{UV}))$, that are not in $F_{RG}(\CB_{UV})$ generalizes their role in the previous case. Indeed, condensing $\CZ_{M}^{\rm bos}(F_{RG}(\CB_{UV}))$ again yields
\begin{equation}
\CM_{IR}/\CZ_{M}^{\rm bos}(F_{RG}(\CB_{UV}))=\CM\boxtimes\CM'~,
\end{equation}
where $\CM'$ must be an MTC (coming from the image of the $\ell_i$) with an $F_{RG}(G^{UV})$ anomaly, $\omega'\in H^4(F_{RG}(G^{UV}),U(1))$, cancelling the anomaly, $\omega\in H^4(F_{RG}(G^{UV}),U(1))$, arising from $\CM$
\begin{equation}
\omega+\omega'=[0]\in H^4(F_{RG}(G^{UV}),U(1))~,
\end{equation}
so that we can gauge a diagonal $F_{RG}(G^{UV})$ acting on $\CM$ and $\CM'$ to produce $\CM_{IR}$.

We can summarize the above discussion via the following theorem:

\bigskip
\noindent
{\bf Theorem 9:} Consider a QFT, $\CQ_{UV}$, with a one-form symmetry category, $\CB_{UV}$. Let us imagine an RG flow of the type described above \eqref{RGB} emanating from $\CQ_{UV}$ and ending in a gapped IR phase described by an MTC (or split super-MTC), $\CM_{IR}$. Then, we have 
\begin{equation}\label{MIRGapped}
\CM_{IR}/\CZ_{M}^{\rm bos}(F_{RG}(\CB_{UV}))=\CM\boxtimes\CM'~,
\end{equation}
where $\CM:=\CB_{UV}/\CZ_{M}^{\rm bos}(\CB_{UV})$. As a result, $\CM_{IR}$ is obtained by gauging a diagonal $F_{RG}(G^{UV})$ zero-form symmetry such that the corresponding 't Hooft anomalies of $\CM$ and $\CM'$ cancel.

\bigskip
Consider the special case when $\CB_{UV}$ is a BFB category. In this case, $F_{RG}(\CB_{UV})$ contains line operators with real spins. From Theorems 2 and 3, we know that a (super-) MTC with non-invertible line operators must contain line operators with complex spin. Therefore, when $F_{RG}(\CB_{UV})$ contains at least one non-invertible line operator, we find that the MTC $\CM_{IR}$, which is a modular extension of $F_{RG}(\CB_{UV})$, contains emergent line operators with complex spin.

\subsec{Lines of $\CM$ as 't Hooft Spectators}\label{spectators}
In this section, we would like to further comment on the role that $\CM:=\CB_{UV}/\CZ^{\rm bos}_M(\CB_{UV})$ plays in the above RG flow discussion. One aspect of $\CM$ that we have explained in great detail in Section \ref{RG} is that it encodes the one-form anomalies of $\CB_{UV}$ and $F_{RG}(\CB_{UV})$.

A more subtle role that $\CM$ plays is hinted at in Section \ref{gapped}: the lines of $\CM$ are zero-form anomaly \lq\lq spectators" reminiscent of the weakly coupled spectator fields appearing in the original argument for 't Hooft anomaly matching \cite{tHooft:1980xss}.

To understand this analogy, first recall that 't Hooft's spectators, $\CS$, are free fields charged under a global symmetry Lie group, $G$, that a QFT, $\CT_{UV}$, whose dynamics is being studied, is also charged under. Now, suppose that $\CT_{UV}$ has an anomaly $\CA^{(0)}(\CT_{UV})\ne0$ and that the spectators have cancelling anomaly, $\CA^{(0)}(\CS)=-\CA^{(0)}(\CT_{UV})$. Then, the diagonal $G$ acting on the combined $\CS$ and $\CT_{UV}$ degrees of freedom can be gauged. Moreover, we can consider gauging this symmetry in a parametrically weak way (i.e., taking the coupling at a given energy to be arbitrarily small). After initiating the RG flow, the resulting IR theory, $\CT_{IR}$, coming from $\CT_{UV}$ is generally very different (i.e., with emergent degrees of freedom), but $\CS$ will consist of the same weakly coupled UV fields in the IR because the gauging is parametrically weak (we can then consider \lq\lq ungauging" the symmetry to produce decoupled $\CS$ and $\CT_{IR}$ sectors without affecting the dynamics). Finally, since $\CA^{(0)}(\CS)$ does not change, we learn that $\CA^{(0)}(\CT_{UV})=\CA^{(0)}(\CT_{IR})$.

In our case, we can produce $\CB_{UV}<\CQ_{UV}$ by gauging a diagonal $G^{UV}$  symmetry of $\CM$ and $\CT_{UV}$. As in 't Hooft's discussion, we will generally have non-vanishing but cancelling anomalies for $G^{UV}$ in $\CM$ and $\CT_{UV}$. While we cannot weakly gauge a discrete symmetry as 't Hooft did for continuous symmetries, the fact that $\CM$ has non-vanishing one-form anomaly means that it is always present along the RG flow. More precisely, the logic around \eqref{MUVMIR} shows it is an invariant of an appropriate condensation / zero-form \lq\lq ungauging." Moreover, while the UV and IR zero-form symmetry groups $G^{UV}$ and $F_{RG}(G^{UV})\le G^{UV}$ can be different, the IR anomaly $F_{RG}(\omega)\in H^4(F_{RG}(G^{UV}),U(1))$ for $\CM$ is non-zero only if the UV anomaly $\omega\in H^4(G^{UV},U(1))$ is non-vanishing.

\subsec{Examples}\label{examples}
Let us now construct some simple examples that illustrate the discussion in Sections \ref{RG}, \ref{gapped}, and \ref{spectators}. Our examples are all UV-complete Poincar\'e-invariant theories coupled to topological degrees of freedom. We then study RG flows characterized by turning on vacuum expectation values and / or local deformations.

We choose examples of $\CQ_{UV}$ belonging to at least one of the following two classes of non-topological QFTs that contain topological lines (note that there is not always an invariant distinction between these classes; for example, dualities can relate them):
\begin{enumerate}
\item Theories built from gauging a diagonal discrete symmetry, $G_D$,  of a TQFT of the form \eqref{GenMTC} or \eqref{GenSMTC} and a CFT (the TQFT may also be trivial, in which case we are coupling a $G_D$ discrete gauge theory with Dijkgraaf-Witten twist, $\omega_3\in H^3(G_D,U(1))$, to the CFT). Depending on which representations of $G_D$ are present in the CFT, we will have different spectra of topological and non-topological lines. For example, if we have a local CFT operator transforming in representation $R\in{\rm Rep}(G_D)$, then the Wilson line, $W_R$, of charge $R$ can end on this operator. As a result, topological line operators that braid non-trivially with $W_R$ become non-topological \cite{Rudelius:2020orz} (since the bulk theory is a CFT, the topological lines that braid non-trivially with $W_R$ become one-dimensional defect CFTs). If $G_D$ is non-Abelian (or if $\omega_3$ is suitably chosen), we can have non-Abelian topological lines.
\item Theories corresponding to $G$ Yang-Mills (YM) theories with CS terms at level $\vec k$ (here $\vec k$ is a vector of levels that depends on $G$) coupled to charged massless matter (we will often refer to these theories as \lq\lq $G_{\vec k}$ QCD theories"). Such QFTs can be understood as relevant deformations of free gauge and matter fields. For generic matter representations, these QFTs have no topological lines. However, if the matter representations are neutral under the center of $G$, $Z(G)$, then there is an Abelian one-form symmetry group implemented by lines with fusion rules isomorphic to $Z(G)$ \cite{Gaiotto:2014kfa}. For even more special choices of gauge group, we sometimes have non-Abelian topological lines (e.g., this can happen if we gauge an appropriate outer automorphism of $G$ and if the Abelian topological lines form non-trivial orbits\footnote{In certain cases, this logic can be rephrased in terms of the methods described in \cite{Kaidi:2021xfk}.}). Note that we can also, as in the first class of theories, consider gauging a diagonal $G_D$ symmetry of $G_{\vec k}$ QCD and one of the TQFTs of the form \eqref{GenMTC} or \eqref{GenSMTC}.
\end{enumerate}

Starting from $\CQ_{UV}$ in (at least) one of the above classes of QFTs, we can then imagine turning on (at least) two different types of locality-preserving deformations (or a combination thereof; note that, under duality, these kinds of deformations are often exchanged):

\begin{enumerate}
\item Relevant deformations like mass terms (or more general ones). At energy scales small compared to the mass, we typically find various emergent symmetries. If the massive charged matter is fermionic, then the Chern-Simons levels and Dijkgraaf-Witten twists shift at one-loop with signs determined by the signs of the fermionic mass terms and magnitude determined by the amount of matter (measured by the Dynkin index, $T(R)$, corresponding to the representation, $R\in{\rm Rep}(G)$ that the matter is charged under).
\item We can sometimes turn on a vacuum expectation value for certain bosonic operators (often, dynamics of the QFT can result in certain non-zero vacuum expectation values being activated). Such deformations always exist in free bosonic CFTs and also in many supersymmetric theories because their potentials have (quantum protected) flat directions parameterized by gauge-invariant local operators. If $\CO$ is a defect endpoint operator for some topological line, $\ell$, we can often find an $n$ such that the normal-ordered product $:\CO^n:$ is a genuine (gauge-invariant) local operator.\footnote{Gauge invariance requires $\ell^{\times n}$ to contain the identity line. In other words, the $n$-fold product of $\ell$ satisfies
\begin{equation}
\ell\times\cdots\times\ell\ni 1~.
\end{equation}
}
Then, going to a vacuum in which $\langle:\CO^n:\rangle\ne0$ results in a condensation of $\ell$. This is one way in which the RG flow can trivialize topological lines (see also the discussion above \eqref{RGB}).
\end{enumerate}

Let us consider some of the simplest examples of the first type discussed above. To that end, we take a real scalar 
\begin{equation}
S(\phi)={1\over2}\int d^3x\ \partial^{\mu}\phi\partial_{\mu}\phi~,
\end{equation}
and gauge the $\mathbb{Z}_2\cong\langle g\rangle$ symmetry that acts on $\phi$ as $g(\phi)=-\phi$. We can think of this gauging as coupling a $\mathbb{Z}_2$ SPT characterized by $\omega_3\in H^3(\mathbb{Z}_2, U(1))\cong\mathbb{Z}_2$ to $\phi$
\begin{equation}\label{diagramQUV1}
\CQ_{UV}\ :=\ {\rm SPT}(\mathbb{Z}_2)_{\omega_3}\ \mathdash\ (\mathbb{Z}_2)\ \mathdash\ \CT_{UV}~, \ \ \ \CT_{UV}:=S(\phi)~.
\end{equation}
In \eqref{diagramQUV1}, \lq\lq $(\DZ_2)$" denotes the gauge group, and the links indicate that the degrees of freedom connected to $(\DZ_2)$ have a $\mathbb{Z}_2$ global symmetry that has been gauged. The resulting theory has $\CB_{UV}\cong{\rm Rep}(\mathbb{Z}_2)$ generated by a Wilson line, $W_{(1,0)}$, that can end on $\phi$. Therefore, the remaining lines, $W_{(0,1)}$ and $W_{(1,1)}$, become non-topological DCFTs, because they braid non-trivially with $W_{(1,0)}$. 

Let us now turn on a mass, $\delta\CL=-{1\over2}m^2\phi^2$, and flow to the deep IR. We find (up to various countertetms) a (twisted) $\mathbb{Z}_2$ discrete gauge theory, $D(\mathbb{Z}_2)_{\omega_3}$
\begin{equation}
\CQ_{IR}:={1\over4\pi}\int d^3x \vec a^T Kd\vec a~,\ \ \ K=\begin{pmatrix}
0 & 2 \\
2 & 2\omega_3 \\
\end{pmatrix}~,\ \ \ \omega_3=0, 1~.
\end{equation}
In this case, $F_{RG}(\CB_{UV})\cong\CB_{UV}\cong{\rm Rep}(\mathbb{Z}_2)$, and $\CB_{IR}\cong\CQ_{IR}\cong\CM_{IR}$ (anomalies match rather trivially between the UV one-form symmetry and the non-emergent part of the IR one-form symmetry). Therefore, it is clear that \eqref{MIRGapped} holds with $\CM_{IR}/F_{RG}(\CB_{UV})$ a trivial TQFT (i.e., both $\CM$ and $\CM'$ are trivial).\footnote{More precisely, we have $\CM_{IR}/F_{RG}(\CB_{UV})\cong{\rm SPT}(\mathbb{Z}_2)_{\omega_3}$.}

Since $\CQ_{UV}$ has a ray of vacua, $\CV\cong\mathbb{R}_{\ge0}$, parameterized by $\langle:\phi^2:\rangle$, we can consider going to points on this space with $\langle:\phi^2:\rangle\ne0$ (i.e., the second class of deformations described above). In this case, we condense $\CB_{UV}$ (i.e., $F_{RG}(\CB_{UV})\cong{\rm SPT}(\mathbb{Z}_2)_{\omega_3}$) and find that $\CB_{IR}\cong{\rm SPT}(\mathbb{Z}_2)_{\omega_3}$ is trivial as a theory of lines. This discussion is again consistent with \eqref{MIRGapped}.

Let us now consider examples where $\CB_{UV}$ has some non-trivial braiding. To that end, let us couple an untwisted $\mathbb{Z}_2$ discrete gauge theory, $D(\DZ_2)$, to a free real scalar
\begin{equation}\label{BraidingEx}
\CQ_{UV}\ :=\ D(\mathbb{Z}_2)\ \mathdash\ (\mathbb{Z}_2)_{\omega_3}\ \mathdash\ \CT_{UV}~, \ \ \ \CT_{UV}:=S(\phi)~,
\end{equation}
where the action of $\mathbb{Z}_2\cong\langle g\rangle$ is EM duality on $D(\mathbb{Z}_2)$ and sign-flip on $\phi$
\begin{equation}\label{Z2act}
g(\ell_{(e,m)})=\ell_{(m,e)}~,\ \ \ g(\phi)=-\phi~.
\end{equation}
The $\omega_3$ subscript in \eqref{BraidingEx} describes the $\mathbb{Z}_2$ SPT we turn on in the process of gauging the $\mathbb{Z}_2$ symmetry in \eqref{Z2act}.

In $\CQ_{UV}$, we have the following topological lines
\begin{equation}
\CB_{UV}\cong\left\{\ell_{(0,0),\pm}~, \ \ell_{(1,1),\pm}~,\ \ell_{[(1,0)]}\right\}~,\ \ \ \CZ_M(\CB_{UV})\cong\left\{\ell_{(0,0),\pm}\right\}\cong{\rm Rep}(\mathbb{Z}_2)~,
\end{equation}
where $\ell_{(0,0),-}$ is transparent to the other topological lines (i.e., it generates the M\"uger center of $\CB_{UV}$) and can end on $\phi$ (as described in Section \ref{BFB}, this fact guarantees that the twisted sector lines become non-topological), $\ell_{(1,1),\pm}$ are fermionic Abelian lines that come from the dyon under $\mathbb{Z}_2$ gauging, and $\ell_{[(1,0)]}$ is the quantum dimension-two line that comes from the $\mathbb{Z}_2$ orbit, $\ell_{(1,0)}\oplus\ell_{(0,1)}$. Clearly $\CB_{UV}$ is a $\mathbb{Z}_2\times\mathbb{Z}_2$ Tambara-Yamagami (TY) category with modular data
\begin{eqnarray}\label{STsubcat}
S=\begin{pmatrix}
1 & 1 & 1& 1 & 2\\
1 & 1 & 1& 1 & 2\\
1 & 1 & 1& 1 & -2\\
1 & 1 & 1& 1 & -2\\
2 & 2 & -2& -2 &0\\
\end{pmatrix}~,\ \ \ 
\theta(\ell_{(0,0),\pm})=\theta(\ell_{[(1,0)]})=1~,\ \theta(\ell_{(1,1),\pm})=-1~.
\end{eqnarray}

Now, let us consider turning on a mass, $\delta\CL=-{1\over2}m^2\phi^2$, and flowing to the IR. We find
\begin{equation}\label{BraidExIR}
\CB_{IR}\cong\CQ_{IR}\cong\CM_{IR}\cong{\rm Ising}^{(\nu)}\boxtimes\overline{\rm Ising^{(\nu)}}~,
\end{equation}
where $\nu=1,3$ label different theories related to the $\mathbb{Z}_2$ SPT we stack with in \eqref{BraidingEx} (we can choose $\nu=2\omega_3+1$). In particular, $\nu=1$ corresponds to what is typically referred to as the Ising TQFT (i.e., we have the topological spin $\theta(\sigma)=\exp(\pi i/8)$ for the non-invertible line), and $\nu=3$ corresponds to ${\rm Ising}^{(3)}\cong SU(2)_2$. More generally, $\nu=1,3,5,7$, but we have the identifications $\nu=1\sim\nu=7$ and $\nu=3\sim\nu=5$ in the IR product theory \eqref{BraidExIR}.

Clearly, we also have that $F_{RG}(\CB_{UV})\cong\CB_{UV}$, and so this example is consistent with the general non-Abelian anomaly matching discussion in the main text. In fact, we can re-write the UV theory as
\begin{equation}\label{BraidExRw}
\CQ_{UV}:={\rm Ising^{(\nu)}}\boxtimes\overline{\rm Ising^{(\nu)}}\ \mathdash\ S(\phi)~,
\end{equation}
where the coupling to ${\rm Ising}\boxtimes\overline{\rm Ising}$ is via the $(\epsilon,\overline{\epsilon})$ line.\footnote{In this analysis, $\epsilon$ and $\overline{\epsilon}$ are the fermionic lines in ${\rm Ising}$ and $\overline{\rm Ising}$ respectively.}

We can also consider turning on a VEV by moving out onto the $\CV\cong\mathbb{R}_{\ge0}$ moduli space of the free scalar by turning on $\langle:\phi^2:\rangle\ne0$. This maneuver gives us $D(\mathbb{Z}_2)\boxtimes{\rm SPT}(\DZ_2)_{\omega_3}\boxtimes S(\phi)$, and we find
\begin{equation}
\CB_{IR}=F_{RG}(\CB_{UV})=D(\mathbb{Z}_2)\cong \CB_{UV}/\CZ_{M}(\CB_{UV})~.
\end{equation}
This example is again consistent with our general discussion in the previous sections.

Next let us consider ${\rm Pin}^+(N)$ Yang-Mills (YM) theory with bare CS level $k_0$ coupled to $N_f$ adjoint Majorana fermions (e.g., see \cite{Cordova:2017vab,Kaidi:2021xfk} for recent discussions of this theory). 
 Let us define the shifted CS coupling $k:=k_0-(N-2)N_f/2$. Turning on $\delta\CL=-M\Psi_a\bar\Psi_a$ and taking $|M|\gg g^2$ large (where $g$ is the YM coupling), we have
\begin{equation}\label{RGgen}
\CB_{IR}\cong\CQ_{IR}\cong\CM_{IR}\cong\begin{cases} 
      {\rm {Pin}}^+(N)_{k_0} \boxtimes{\rm SVec}& M>0~, \\
       {\rm Pin}^+(N)_{k_0-(N-2)N_f} \boxtimes{\rm SVec}& M<0~, 
   \end{cases}
\end{equation}
where the IR TQFT arises through the one-loop shift in the CS coupling.

For simplicity, let us focus in more detail on the case of $N=4$ and $k=k_0-N_f$. 
We can understand ${\rm Pin}^+(4)_k$ as a gauging of the spin-exchange symmetry of ${\rm Spin}(4)_k\cong SU(2)_k\times SU(2)_k$ (i.e., the symmetry that sends lines $(j_1,j_2)\leftrightarrow(j_2,j_1)$ where $0\le j_i\le k/2$), and \eqref{RGgen} becomes
\begin{equation}\label{RGN4}
\CB_{IR}\cong\CQ_{IR}\cong\CM_{IR}\cong\begin{cases} 
      {\rm {Pin}}^+(4)_{k_0} \boxtimes{\rm SVec}& M>0~, \\
       {\rm Pin}^+(4)_{k_0-2N_f} \boxtimes{\rm SVec}& M<0~. 
   \end{cases}
\end{equation}

If we start with a ${\rm Spin}(4)_{k_0}$ theory coupled to $N_f$ adjoint Majorana fermions, the lines corresponding to the $\mathbb{Z}_2\times\mathbb{Z}_2$ center (i.e., $W_{(0,0)}$, $W_{(k_0/2,0)}$, $W_{(0,k_0/2)}$, and $W_{(k_0/2,k_0/2)}$) are topological even in the presence of the adjoint fermions. Then, so to are the lines obtained by gauging the spin exchange symmetry to produce the ${\rm Pin}^+(4)_{k_0}$ gauge theory coupled to $N_f$ adjoint fermions.\footnote{A different argument was used in \cite{Kaidi:2021xfk} for the cases of interest  there.} Indeed, we have a $\mathbb{Z}_2\times\mathbb{Z}_2$ Tambara-Yamagami (TY) category
\begin{equation}
W_{[(k_0/2,0)]}\times W_{[(k_0/2,0)]}=W_{(0,0),+}+W_{(0,0),-}+W_{(k_0/2,k_0/2),+}+W_{(k_0/2,k_0/2),-}~,
\end{equation} 
where $W_{(0,0),-}$ is the quantum Abelian line that arises from gauging the exchange symmetry of the ${\rm Spin}(4)_{k_0}$ theory, $W_{(k_0/2,k_0/2),\pm}$ are the images of the invariant $W_{(k_0/2,k_0/2)}$ ${\rm Spin}(4)$ line under gauging, and $W_{[(k_0/2,0)]}$ is the non-Abelian line that arises from the $W_{(k_0/2,0)}\oplus W_{(0,k_0/2)}$ orbit under the exchange symmetry (see also the related discussion in \cite{Borisov:1997nc}). Note that
\begin{equation}
\CB_{UV}\cong\left\{W_{(0,0),\pm}~,\ W_{(k_0/2,k_0/2),\pm}~,\ W_{[(k_0/2,0)]}\right\}\boxtimes{\rm SVec} ~,
\end{equation}
and we find the following modular data for the TY lines in \eqref{RGN4} (this subcategory is a BFB category for $k_0=2n$)
\begin{eqnarray}\label{STsubcat}
S&=&\begin{cases}\begin{pmatrix}
1 & 1 & 1& 1 & 2\\
1 & 1 & 1& 1 & 2\\
1 & 1 & 1& 1 & -2\\
1 & 1 & 1& 1 & -2\\
2 & 2 & -2& -2 &0\\
\end{pmatrix}& k_0\ {\rm odd}\\
\\
\begin{pmatrix}
1 & 1 & 1& 1 & 2\\
1 & 1 & 1& 1 & 2\\
1 & 1 & 1& 1 & 2\\
1 & 1 & 1& 1 & 2\\
2 & 2 & 2& 2 &4\\
\end{pmatrix}& k_0\ {\rm even}\\
\end{cases}~,\cr \cr
\theta\left(W_{[(k_0/2,0)]}\right)&=&\exp(\pi i k_0/2)~,\ \ \ \theta\left(W_{(0,0),\pm}\right)=1~,\ \ \ \theta\left(W_{(k_0/2,k_0/2),\pm}\right)=\exp(\pi i k_0)~.\ \ \ \ \ \ \ 
\end{eqnarray}
Crucially, we see that the one-form symmetry anomalies match since the two IR phases are related by a shift in the IR CS level by an even number (i.e., $\delta k_{IR}=-2N_f$). Note that if $N_f$ is odd, then there is a mismatch between $\theta(\ell)$ in the two IR phases with the different signs for $M$. However, this mismatch is cancelled by including the ${\rm SVec}$ factor (i.e., we map $\ell\leftrightarrow\ell\times\psi$ via a domain wall between the two IR phases, where $\psi\in{\rm SVec}$ is the transparent fermion).

As our final example, consider the following case that illustrates the role of $\CM\cong\CM_{UV}$ as a generalization of the 't Hooft spectators described in Section \ref{spectators}
\begin{eqnarray}
\CB_{UV}&:=&D(\mathbb{Z}_2)\boxtimes{\rm Rep}(\mathbb{Z}_2\times\mathbb{Z}_2)~,\  \CZ_M(\CB_{UV})={\rm Rep}(\mathbb{Z}_2\times\mathbb{Z}_2)~,\cr \CM&:=&\CB_{UV}/{\rm Rep}(\mathbb{Z}_2\times\mathbb{Z}_2)\cong D(\mathbb{Z}_2)~,\ \cr\CQ_{UV}&:=& D(\mathbb{Z}_2)\ \mathdash\ \left(\mathbb{Z}_2\times\mathbb{Z}_2\right)\ \mathdash\ \CT_{UV}~,\ \CT_{UV}= D(\mathbb{Z}_2)\boxtimes S(\phi)~,\ S(\phi):=\int d^3x|\partial_{\mu}\phi|^2~.\ \ \ \ \ \ \ \ 
\end{eqnarray}
Here our notation for $\CQ_{UV}$ means that we gauge the diagonal $\mathbb{Z}_2\times\mathbb{Z}_2\cong\langle g_1,g_2\rangle$ 0-form symmetry that acts on a complex free scalar as
\begin{equation}\label{compPhiAct}
\phi:={1\over\sqrt{2}}(\phi_1+i\phi_2)~,\ \ \ g_1(\phi)={1\over\sqrt{2}}(-\phi_1+i\phi_2)~,\ \ \ g_2(\phi)={1\over\sqrt{2}}(\phi_1-i\phi_2)~.
\end{equation}
This symmetry does not permute any of the lines in the $D(\mathbb{Z}_2)$ factors (note that one $D(\mathbb{Z}_2)$ factor is in $\CT_{UV}$, so we can think of $\CB_{UV}$ as a closed subcategory of the full UV one-form symmetry, $\CC_{UV}:=D(\mathbb{Z}_2)^{\boxtimes2}\boxtimes{\rm Rep}(\mathbb{Z}_2\times\mathbb{Z}_2)$) but has an $H^4(\mathbb{Z}_2\times\mathbb{Z}_2, U(1))\cong\mathbb{Z}_2\times\mathbb{Z}_2$ anomaly in each factor. Indeed, following the discussion in \cite{Kapustin:2014zva}, we can choose an anomaly $\omega^{(i)}\in H^4(\mathbb{Z}_2\times\mathbb{Z}_2,U(1))$ in the $i$th $D(\mathbb{Z}_2)$ sector specified by the 4d action
\begin{equation}
S_{4d} = -\frac{2}{(2\pi)^2}\int A^{(i)}_1\,A^{(i)}_2\,(dA^{(i)}_1+dA^{(i)}_2)~,
\end{equation}
where the $A_{1,2}^{(i)}$ are background gauge fields for the $G^{(i)}\cong\mathbb{Z}_2\times\mathbb{Z}_2$ symmetry acting on the $i$th $D(\mathbb{Z}_2)$ sector. By considering a diagonal symmetry (i.e., identifying $A_a^{(1)}=A_a^{(2)}=A_a$) we obtain a vanishing anomaly (the contribution from the scalar sector is vanishing), and we can gauge the diagonal $\mathbb{Z}_2\times\mathbb{Z}_2$ symmetry described in \eqref{compPhiAct}.

Now, consider deforming the complex scalar action by $\delta\CL=-m^2|\phi|^2$ and flowing to the IR. Via the RG flow, this maneuver renders the one-hundred-and-ninety-two UV DCFTs that we get from the twisted sector gauging of the two $D(\mathbb{Z}_2)$ theories topological in the IR, and we have an IR MTC with
\begin{equation}
|\CM_{IR}|=256~,\ \CM_{IR}\supset D(\mathbb{Z}_2)^{\boxtimes 2}\boxtimes{\rm Rep}(\mathbb{Z}_2\times\mathbb{Z}_2)~, \ C_{\CM_{IR}}({\rm Rep}(\mathbb{Z}_2\times\mathbb{Z}_2))\cong D(\mathbb{Z}_2)^{\boxtimes2}~.
\end{equation}
Here $\CM_{IR}$ is a non-minimal modular extension of $\CB_{UV}$.\footnote{A minimal modular extension would be of the form $SU(2)_1\boxtimes D_{\omega}(\mathbb{Z}_2\times\mathbb{Z}_2)$, where the second factor is a (twisted) Dijkgraaf-Witten theory.} Clearly, the additional lines that braid trivially with ${\rm Rep}(\mathbb{Z}_2\times\mathbb{Z}_2)$ come from the cancelling of the $H^4(\mathbb{Z}_2\times\mathbb{Z}_2,U(1))$ anomaly via $D(\mathbb{Z}_2)\subset\CT_{UV}$.

Now let us consider the example from the previous bullet but, instead of adding a mass term, let us move onto the moduli space, $\CV:=\mathbb{R}_+^2$, parameterized by $\langle :\phi_a^2:\rangle$. We find that $\CM_{UV}=D(\mathbb{Z}_2)\to D(\mathbb{Z}_2)^{\boxtimes2}=\CM_{IR}$. In particular, $\CM_{IR}$ still has accidental symmetries due to the anomaly. Note that the IR consists of a gapped phase stacked with a gapless one.

\newsec{Discussion}
In this paper, we have classified BFB symmetry categories and shown that they are closely related to groups. In particular, we proved that any BFB category, $\CB$, is weakly group theoretical and, if non-invertible, it is in fact non-intrinsically non-invertible. We have also showed that QFTs with BFB symmetries posses invariants associated with (locality and BFB-preserving) continuous deformations. In particular, Theorem 6 (i), (ii), Theorem 7, and Theorem 9 follow from natural assumptions on the RG flow preserving the fusion and braiding between line operators. Moreover, using the technical assumption that $F_{RG}$ is a braided monoidal functor, we derived an explicit expression for the anomaly matching condition for one-form symmetries (Theorem 6 (iii) and its Corollary). Under this assumption, we also derived Theorem 8 which showed that even in the general case when some UV line operators are only partially trivialized, the UV line operators are related to their IR images by (at most) a restricted form of anyon condensation. 

Our work leaves many open questions. For example:

\begin{itemize}
\item As alluded to in the previous paragraph, we associated a (super-) MTC, $\CM$, with each locality-preserving deformation of a QFT with BFB symmetry. Topological modular forms (TMFs) are also deformation invariants of large classes of QFTs (e.g., see \cite{stolz2004elliptic,stolz2011supersymmetric,Tachikawa:2021mvw}). Both our MTC and TMFs make use of modularity. Is there a deeper relation?
\item Can we use the $\CM$ invariants of the previous bullets to prove or conjecture new RG monotonicity theorems in $2+1$d?
\item  From  \cite{Buican:2021uyp,Buican:2023ehi}, we know that we can associate a quantum code with each Abelian MTC. Therefore, we realize a goal described in \cite{Buican:2021uyp,Buican:2023ehi} of associating quantum codes with deformation classes of QFTs. It would be interesting to further explore the implications of the resulting code map.
\item  We obtained our BFB categories by coupling to a non-topological QFT. It would be interesting to relate these ideas to the lattice constructions in \cite{Sohal:2024qvq,Ellison:2024svg} and mixed state topological order.
\item Our categories play important roles in dualities involving CS theories for groups, $G$, with $\mathfrak{so}(n)$ Lie algebras. Can our results be used to extend these dualities? Can we leverage known classification results involving metaplectic categories to further constrain these dualities?\footnote{It would also be interesting to understand if there is any relation between the simplicity of classifying metaplectic modular categories and BFB categories. One common feature is the close relation with CS theories having ${\rm Spin}(N)$ gauge group (although in the metaplectic case, $N$ is odd).}
\item In a similar spirit, BFB categories play an important role in the dualities and dynamics of 3d $\CN=1$ theories in \cite{Gaiotto:2018yjh}. Can our results shed further light on these QFTs?
\end{itemize}

We hope to return to some of these questions soon.

\ack{We thank Anindya Banerjee, Clement Delcamp, Zhihao Duan, Andrea Ferrari, Davide Gaiotto, Andrey Gromov, Hongliang Jiang, Kantaro Ohmori, Adrian Padellaro, Pavel Putrov, Sanjaye Ramgoolam, Brandon Rayhaun, Ingo Runkel, Nati Seiberg, Sahand Seifnashri, Shu-Heng Shao, Yifan Wang, and Yunqin Zheng for discussions, for comments, and (a proper subset of these colleagues) for collaborations. M.~B. and R.~R. thank ULB Brussels and the International Solvay Institutes for hospitality during the workshop  \lq\lq Symmetries, Anomalies and Dynamics of Quantum Field theory" during which part of this work was completed. M.~B. also thanks Osaka Metropolitan University (during the workshop \lq\lq QFT and Related Mathematical Aspects 2024," where part of this work was completed), New York University, and the Institute for Advanced Study for hospitality. R.~R. thanks Perimeter Institute for Theoretical Physics for hospitality during the workshop \lq\lq Higher Categorical Tools for Quantum Phases of Matter" and University of Sarajevo for hospitality during the workshop \lq\lq Paths to QFT 2024," where part of this work was completed. R.~R. thanks IMSc, Chennai and IHES for hospitality during a visit. M.~B.'s work was partly supported by The Royal Society under the grant “Relations, Transformations, and Emergence in Quantum Field Theory.” M.~B. and M.~B. were partly supported by STFC under the grant “Amplitudes, Strings and Duality.” No new data were generated or analyzed in this study.}

\newpage 

\begin{appendices}

\section{Non-unitary BFB (super-) MTCs}
\label{ap:non-unitary BFB}

The expression for the FS indicator given in \eqref{FS2final} is also valid in a non-unitary super-MTC. Therefore, we can repeat the arguments leading to Theorems 2 and 3 to show that in a non-unitary BFB (super-) MTC\footnote{In the non-unitary super-MTC case, we should be careful to rule out the existence of non-self dual lines. Such lines would have $d_{\ell_i}=\nu_2(\ell_i)=0$, but this situation is incompatible with semi-simplicity \cite[Proposition 4.8.4]{etingof2016tensor},\cite{geer2013topological}.}
\be
\label{eq:qdim FS}
d_{\ell_i}=\nu_2(\ell_i)=\pm 1~ \forall ~ \ell_i \in \CB~.
\ee
In order to show that all line operators are invertible, we have to show that FPdim$(\ell_i)=1$ for all $\ell_i\in \CB$.\footnote{In a non-unitary fusion category, non-invertible line operators can have quantum dimension $1$. For example, this is the case in the non-unitary Fibonacci $\boxtimes$ Lee-Yang MTC.} We will now show that this is indeed the case.

To that end, consider the modular tensor category, $\CZ(\CB)$, obtained from taking the Drinfeld centre of the super-MTC, $\CB$. We have the following relation 
\be
\label{eq:center dimension relations}
\text{FPdim}(\CZ(\CB))=\text{FPdim}(\CB)^2 ~, ~ D_{\CZ(\CB)}=D^2_{\CB}~,
\ee
where FPdim$(\CB):=\sum_{\ell_i\in \CB} \text{FPdim}(\ell_i)^2$, and $D_{\CB}:=\sqrt{\sum_{\ell_i\in \CB} d_{\ell_i}^2}$. Recall that both quantum dimensions and Frobenius-Perron dimensions of line operators in  $\CZ(\CB)$ are one-dimensional characters of the $\CZ(\CB)$ fusion ring. Since $\CZ(\CB)$ is modular, there exists some line operator, $\ell_j \in \CZ(\CB)$, such that 
\be
\text{FPdim}(\ell_i)=\frac{S_{\ell_i\ell_j}}{S_{1\ell_j}}~.
\ee
Then, we get 
\be
\text{FPdim}(\CZ(\CB))=\sum_{\ell_i} \text{FPdim}(\ell_i)^2 = \sum_{\ell_i} \frac{S^2_{\ell_i\ell_j}}{S^2_{1\ell_j}} = \frac{D^2_{\CZ(\CB)}}{S_{1\ell_j}^2}= \frac{D^2_{\CZ(\CB)}}{d_{\ell_j}^2}~. 
\ee
Using \eqref{eq:center dimension relations}, we get
\be
\label{eq: FPdim and D constraint}
\text{FPdim}(\CB)^2 =\frac{D_{\CB}^4}{d_{\ell_j}^2}~.
\ee
By assumption, the (super-) MTC, $\CB$, has line operators with real spins. Using \eqref{eq:qdim FS}  and \cite[Theorem 5.5]{ng2007frobenius}, we find that the order of the $T$-matrix of $\CZ(\CB)$ is $2$. Since the quantum dimensions of $\CZ(\CB)$ must live in the same number field as the topological spins \cite[Proposition 4]{Bantay:2001ni}, this implies that the quantum dimensions of line operators in  $\CZ(\CB)$ are integers. Therefore, the $d_{\ell_j}^2$ in \eqref{eq: FPdim and D constraint} is a positive integer. We get
\be
\text{FPdim}(\CB)^2\leq D_{\CB}^4 \implies \text{FPdim}(\CB)\leq D^2_{\CB}~,
\ee
where we have used the fact that FPdim$(\CB)$ is a positive number by definition, and $D_{\CB}$ is positive in our case since $d_{\ell_i}=\pm 1 ~ \forall ~\ell_i\in \CB$. Now, we use Proposition 8.21 in \cite{etingof2005fusion}, which states that $D^2_{\CB}\leq \text{FPdim}(\CB)$ for any fusion category $\CB$. Applying this to \eqref{eq: FPdim and D constraint}, we find that 
\be
\text{FPdim}(\CB)=D^2_{\CB}~.
\ee
Therefore, we find that any (super-) MTC with real spins is pseudo-unitary. By \cite[Proposition 8.23]{etingof2005fusion}, in any pseudo-unitary category there is a unique spherical structure such that quantum dimensions are positive and agree with the Frobenius-Perron dimensions. In our case, the allowed quantum dimensions are valued in $\pm 1$. As a result, there exists a spherical structure such that $d_{\ell_i}=1=\text{FPdim}(\ell_i) ~ \forall ~ \ell_i \in \CB$, and all line operators in $\CB$ are invertible.

\section{Details of the RG interface and braided monoidal functor}

\label{Ap:FRG}

In this appendix, we will explicitly describe the RG functor 
\be
F_{RG}: \CB_{UV} \to \CB_{IR}~,
\ee
and the assumptions that make it a braided monoidal functor. We will also give a proof of the anomaly matching condition.

\begin{figure}[h!]
\centering
\tikzset{every picture/.style={line width=0.75pt}} 

\begin{tikzpicture}[x=0.75pt,y=0.75pt,yscale=-1,xscale=1]

\draw  [color={rgb, 255:red, 37; green, 118; blue, 212 }  ,draw opacity=1 ][fill={rgb, 255:red, 30; green, 94; blue, 151 }  ,fill opacity=0.51 ] (280.99,178.47) -- (281.01,63.85) -- (358.99,41.35) -- (358.97,155.97) -- cycle ;
\draw [color={rgb, 255:red, 139; green, 6; blue, 24 }  ,draw opacity=1 ]   (209,110) -- (319.99,109.91) ;
\draw  [color={rgb, 255:red, 139; green, 6; blue, 24 }  ,draw opacity=1 ][fill={rgb, 255:red, 139; green, 6; blue, 24 }  ,fill opacity=1 ] (318.28,109.58) .. controls (318.28,108.66) and (319.04,107.91) .. (319.99,107.91) .. controls (320.93,107.91) and (321.7,108.66) .. (321.7,109.58) .. controls (321.7,110.5) and (320.93,111.25) .. (319.99,111.25) .. controls (319.04,111.25) and (318.28,110.5) .. (318.28,109.58) -- cycle ;
\draw    (321.7,109.58) -- (430,110) ;

\draw (206,55.4) node [anchor=north west][inner sep=0.75pt]    {$\CQ_{UV}$};
\draw (407,55.4) node [anchor=north west][inner sep=0.75pt]    {$\CQ_{IR}$};
\draw (332,51.4) node [anchor=north west][inner sep=0.75pt]    {$\CI_{RG}$};
\draw (211,113.4) node [anchor=north west][inner sep=0.75pt]    {$x$};
\draw (419,113.4) node [anchor=north west][inner sep=0.75pt]    {$a$};

\end{tikzpicture}
\caption{Junction of topological line operators on $\CI$.}
\label{fig:junction on RG interface}
\end{figure}
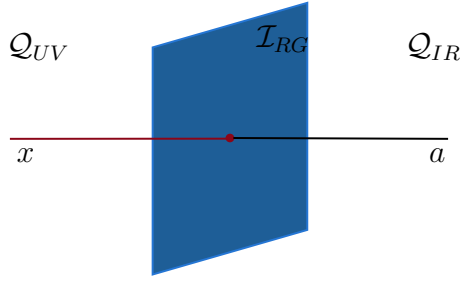

To that end, let $x,y,z$ be the simple line operators in $\CB_{UV}$, and let $a,b,c$ be the simple line operators in $\CB_{IR}$. As discussed in Section \ref{RG}, we have 
\be
F_{RG}(x)= \sum_a N_{F_{RG}(x)}^a ~ a~,
\ee
for some non-negative integers, $N_{F_{RG}(x)}^a$. It is useful to think of the RG flow as an interface, $\CI_{RG}$, between $\CQ_{UV}$ and $\CQ_{IR}$.  In particular, $\CI_{RG}$ tells us how the full spectrum of local and extended operators in $\CQ_{UV}$ get mapped to $\CQ_{IR}$. A subset of the data specifying $\CI_{RG}$ is the map, $F_{RG}$, that tells us how the topological lines operators of the UV QFT are mapped to the IR QFT. If $N_{F_{RG}(x)}^a\neq 0$, then the line operators $x$ and $a$ can form a non-trivial junction on $\CI_{RG}$ (see Fig. \ref{fig:junction on RG interface}). 
If $N_{xy}^z$ is non-zero, then the line operators $x,y,z$ can form a tri-valent junction, which can host point operators belonging to a fusion space, $V_{xy}^z$, of dimension $N_{xy}^z$. As we move this fusion space across $\CI_{RG}$, we get a fusion space, $V_{F_{RG}(x)F_{RG}(y)}^{F_{RG}(z)}$, in the IR QFT. Let us make the following assumptions about the RG interface: 
\begin{itemize}
\item Since the fusion of two topological line operators does not depend on the distance between them, the RG flow must preserve fusion rules of line operators in $\CB_{UV}$. Moreover, we require that it also preserve the topological junctions between line operators in $\CB_{UV}$. In particular, we require the isomorphisms (see Fig. \ref{fig:Phi map})
\be
\Phi_{x,y}: F_{RG}(x) \times F_{RG}(y) \xrightarrow{\sim} F_{RG}(x \times y)~.
\ee
These relations ensure that we have an invertible map relating the fusion spaces of $\CB_{UV}$ and their images under $F_{RG}$ in  $\CB_{IR}$.\footnote{Note that the RG interface $\CI_{RG}$ does not give an invertible map between operators in the UV and IR QFTs. However, by assumption, since the RG flow preserves the topological line operators and their junctions described by $\CB_{UV}$, it is natural to require the isomorphims, $\Phi$, between fusion spaces in $\CB_{UV}$ and their images under $F_{RG}$ in $\CB_{IR}$.} 
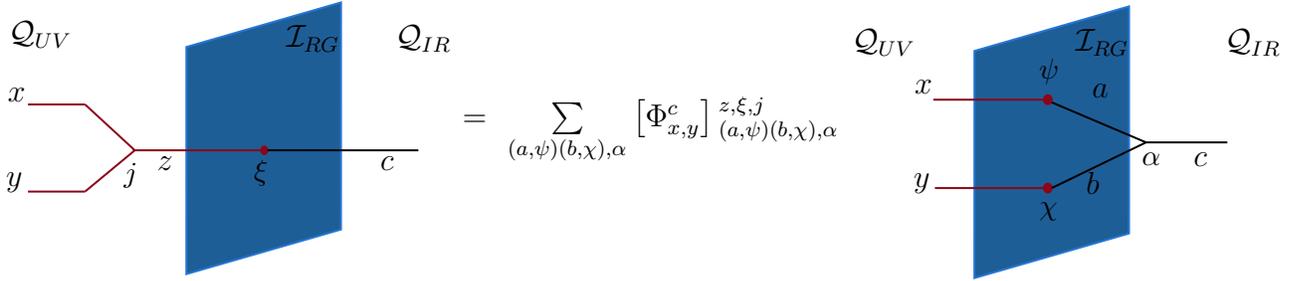
\begin{figure}[h!]
\centering

\tikzset{every picture/.style={line width=0.75pt}} 

\begin{tikzpicture}[x=0.75pt,y=0.75pt,yscale=-1,xscale=1]

\draw  [color={rgb, 255:red, 37; green, 118; blue, 212 }  ,draw opacity=1 ][fill={rgb, 255:red, 30; green, 94; blue, 151 }  ,fill opacity=0.51 ] (496.43,190.47) -- (496.45,75.85) -- (574.43,53.35) -- (574.41,167.97) -- cycle ;
\draw  [color={rgb, 255:red, 37; green, 118; blue, 212 }  ,draw opacity=1 ][fill={rgb, 255:red, 30; green, 94; blue, 151 }  ,fill opacity=0.51 ] (98.99,188.47) -- (99.01,73.85) -- (176.99,51.35) -- (176.97,165.97) -- cycle ;
\draw [color={rgb, 255:red, 139; green, 6; blue, 24 }  ,draw opacity=1 ]   (475.68,100.45) -- (535.42,100.45) ;
\draw [color={rgb, 255:red, 139; green, 6; blue, 24 }  ,draw opacity=1 ]   (476.5,145.05) -- (535.42,145.05) ;
\draw  [color={rgb, 255:red, 139; green, 6; blue, 24 }  ,draw opacity=1 ][fill={rgb, 255:red, 139; green, 6; blue, 24 }  ,fill opacity=1 ] (531.5,100.45) .. controls (531.5,99.22) and (532.38,98.22) .. (533.46,98.22) .. controls (534.54,98.22) and (535.42,99.22) .. (535.42,100.45) .. controls (535.42,101.68) and (534.54,102.68) .. (533.46,102.68) .. controls (532.38,102.68) and (531.5,101.68) .. (531.5,100.45) -- cycle ;
\draw  [color={rgb, 255:red, 139; green, 6; blue, 24 }  ,draw opacity=1 ][fill={rgb, 255:red, 139; green, 6; blue, 24 }  ,fill opacity=1 ] (531.5,145.05) .. controls (531.5,143.82) and (532.38,142.82) .. (533.46,142.82) .. controls (534.54,142.82) and (535.42,143.82) .. (535.42,145.05) .. controls (535.42,146.28) and (534.54,147.28) .. (533.46,147.28) .. controls (532.38,147.28) and (531.5,146.28) .. (531.5,145.05) -- cycle ;
\draw    (535.42,101.8) -- (583,122) ;
\draw    (535.42,145.05) -- (583,122) ;
\draw    (583,122) -- (624,122) ;
\draw [color={rgb, 255:red, 139; green, 6; blue, 24 }  ,draw opacity=1 ]   (19.07,102.86) -- (47.9,102.86) ;
\draw [color={rgb, 255:red, 139; green, 6; blue, 24 }  ,draw opacity=1 ]   (19.07,146.82) -- (47.9,146.82) ;
\draw  [color={rgb, 255:red, 139; green, 6; blue, 24 }  ,draw opacity=1 ][fill={rgb, 255:red, 139; green, 6; blue, 24 }  ,fill opacity=1 ] (136.73,125.99) .. controls (136.73,124.94) and (137.42,124.09) .. (138.27,124.09) .. controls (139.12,124.09) and (139.8,124.94) .. (139.8,125.99) .. controls (139.8,127.05) and (139.12,127.9) .. (138.27,127.9) .. controls (137.42,127.9) and (136.73,127.05) .. (136.73,125.99) -- cycle ;
\draw [color={rgb, 255:red, 139; green, 6; blue, 24 }  ,draw opacity=1 ]   (47.9,102.86) -- (73,125.99) ;
\draw [color={rgb, 255:red, 139; green, 6; blue, 24 }  ,draw opacity=1 ]   (47.9,146.82) -- (73,125.99) ;
\draw    (139.8,125.99) -- (216,126) ;
\draw [color={rgb, 255:red, 139; green, 6; blue, 24 }  ,draw opacity=1 ]   (72,125.99) -- (139.8,125.99) ;

\draw (8,59.4) node [anchor=north west][inner sep=0.75pt]    {$\CQ_{UV}$};
\draw (203,61.4) node [anchor=north west][inner sep=0.75pt]    {$\CQ_{IR}$};
\draw (464.9,88.91) node [anchor=north west][inner sep=0.75pt]    {$x$};
\draw (7.49,92.95) node [anchor=north west][inner sep=0.75pt]    {$x$};
\draw (464.29,136.36) node [anchor=north west][inner sep=0.75pt]    {$y$};
\draw (6.56,135.75) node [anchor=north west][inner sep=0.75pt]    {$y$};
\draw (82.81,127.7) node [anchor=north west][inner sep=0.75pt]    {$z$};
\draw (551.41,135.44) node [anchor=north west][inner sep=0.75pt]    {$b$};
\draw (554.36,91.04) node [anchor=north west][inner sep=0.75pt]    {$a$};
\draw (605.5,125.4) node [anchor=north west][inner sep=0.75pt]    {$c$};
\draw (195.33,127.76) node [anchor=north west][inner sep=0.75pt]    {$c$};
\draw (527.59,77.95) node [anchor=north west][inner sep=0.75pt]  [font=\small]  {$\psi $};
\draw (527.68,150.86) node [anchor=north west][inner sep=0.75pt]  [font=\small]  {$\chi $};
\draw (131.59,129.29) node [anchor=north west][inner sep=0.75pt]  [font=\small]  {$\xi $};
\draw (579.42,125.84) node [anchor=north west][inner sep=0.75pt]  [font=\small]  {$\alpha$};
\draw (65.97,131.07) node [anchor=north west][inner sep=0.75pt]  [font=\small]  {$j$};
\draw (236.84,97.4) node [anchor=north west][inner sep=0.75pt]    {$=\ \sum\limits_{( a,\psi )( b,\chi ) ,\alpha}\left[ \Phi _{x,y}^{c}\right]{_{( a,\psi )( b,\chi ) ,\alpha}^{z,\xi  ,j}}$};
\draw (433.44,63.4) node [anchor=north west][inner sep=0.75pt]    {$\CQ_{UV}$};
\draw (621.44,63.4) node [anchor=north west][inner sep=0.75pt]    {$\CQ_{IR}$};
\draw (148,61.4) node [anchor=north west][inner sep=0.75pt]    {$\CI_{RG}$};
\draw (546,63.4) node [anchor=north west][inner sep=0.75pt]    {$\CI_{RG}$};

\end{tikzpicture}
\caption{Map between fusion spaces under $\CI_{RG}$.}
\label{fig:Phi map}
\end{figure}
We must also demand compatibility with the $F$ matrices of both $\CB_{UV}$ and $\CB_{IR}$ (see Fig. \ref{fig:I associativity condition}).
\begin{figure}
    \centering

\tikzset{every picture/.style={line width=0.75pt}} 

\begin{tikzpicture}[x=0.75pt,y=0.75pt,yscale=-1,xscale=1]

\draw  [color={rgb, 255:red, 37; green, 118; blue, 212 }  ,draw opacity=1 ][fill={rgb, 255:red, 30; green, 94; blue, 151 }  ,fill opacity=0.51 ] (529.14,629.37) -- (529.16,511.35) -- (607.79,490.05) -- (607.77,608.06) -- cycle ;
\draw  [color={rgb, 255:red, 37; green, 118; blue, 212 }  ,draw opacity=1 ][fill={rgb, 255:red, 30; green, 94; blue, 151 }  ,fill opacity=0.51 ] (515.14,358.37) -- (515.16,240.35) -- (593.79,219.05) -- (593.77,337.06) -- cycle ;
\draw  [color={rgb, 255:red, 37; green, 118; blue, 212 }  ,draw opacity=1 ][fill={rgb, 255:red, 30; green, 94; blue, 151 }  ,fill opacity=0.51 ] (326.14,845.4) -- (326.16,727.38) -- (404.79,706.08) -- (404.77,824.09) -- cycle ;
\draw  [color={rgb, 255:red, 37; green, 118; blue, 212 }  ,draw opacity=1 ][fill={rgb, 255:red, 30; green, 94; blue, 151 }  ,fill opacity=0.51 ] (104.14,632.4) -- (104.16,514.38) -- (182.79,493.08) -- (182.77,611.09) -- cycle ;
\draw  [color={rgb, 255:red, 37; green, 118; blue, 212 }  ,draw opacity=1 ][fill={rgb, 255:red, 30; green, 94; blue, 151 }  ,fill opacity=0.51 ] (107.14,371.4) -- (107.16,253.38) -- (185.79,232.08) -- (185.77,350.09) -- cycle ;
\draw [color={rgb, 255:red, 139; green, 6; blue, 24 }  ,draw opacity=1 ]   (96.29,261.45) -- (166.83,261.45) ;
\draw [color={rgb, 255:red, 139; green, 6; blue, 24 }  ,draw opacity=1 ]   (96.01,297.97) -- (167.59,297.97) ;
\draw  [color={rgb, 255:red, 139; green, 6; blue, 24 }  ,draw opacity=1 ][fill={rgb, 255:red, 139; green, 6; blue, 24 }  ,fill opacity=1 ] (163.41,261.45) .. controls (163.41,260.53) and (164.17,259.78) .. (165.12,259.78) .. controls (166.06,259.78) and (166.83,260.53) .. (166.83,261.45) .. controls (166.83,262.38) and (166.06,263.12) .. (165.12,263.12) .. controls (164.17,263.12) and (163.41,262.38) .. (163.41,261.45) -- cycle ;
\draw  [color={rgb, 255:red, 139; green, 6; blue, 24 }  ,draw opacity=1 ][fill={rgb, 255:red, 139; green, 6; blue, 24 }  ,fill opacity=1 ] (164.16,297.97) .. controls (164.16,297.05) and (164.93,296.3) .. (165.88,296.3) .. controls (166.82,296.3) and (167.59,297.05) .. (167.59,297.97) .. controls (167.59,298.89) and (166.82,299.64) .. (165.88,299.64) .. controls (164.93,299.64) and (164.16,298.89) .. (164.16,297.97) -- cycle ;
\draw [color={rgb, 255:red, 139; green, 6; blue, 24 }  ,draw opacity=1 ]   (447.55,523.98) -- (502.1,561.9) ;
\draw  [color={rgb, 255:red, 139; green, 6; blue, 24 }  ,draw opacity=1 ][fill={rgb, 255:red, 139; green, 6; blue, 24 }  ,fill opacity=1 ] (558.55,562.24) .. controls (558.55,561.32) and (559.3,560.58) .. (560.23,560.58) .. controls (561.17,560.58) and (561.92,561.32) .. (561.92,562.24) .. controls (561.92,563.15) and (561.17,563.89) .. (560.23,563.89) .. controls (559.3,563.89) and (558.55,563.15) .. (558.55,562.24) -- cycle ;
\draw [color={rgb, 255:red, 139; green, 6; blue, 24 }  ,draw opacity=1 ]   (446.79,561.13) -- (474.75,580.81) ;
\draw [color={rgb, 255:red, 139; green, 6; blue, 24 }  ,draw opacity=1 ]   (447.87,598.84) -- (502.1,561.9) ;
\draw    (561.92,562.24) -- (635,561.5) ;
\draw [color={rgb, 255:red, 139; green, 6; blue, 24 }  ,draw opacity=1 ]   (500.51,562.24) -- (561.92,562.24) ;
\draw [color={rgb, 255:red, 139; green, 6; blue, 24 }  ,draw opacity=1 ]   (446.4,252.76) -- (500.94,290.69) ;
\draw  [color={rgb, 255:red, 139; green, 6; blue, 24 }  ,draw opacity=1 ][fill={rgb, 255:red, 139; green, 6; blue, 24 }  ,fill opacity=1 ] (557.39,291.02) .. controls (557.39,290.11) and (558.15,289.37) .. (559.08,289.37) .. controls (560.01,289.37) and (560.77,290.11) .. (560.77,291.02) .. controls (560.77,291.93) and (560.01,292.67) .. (559.08,292.67) .. controls (558.15,292.67) and (557.39,291.93) .. (557.39,291.02) -- cycle ;
\draw [color={rgb, 255:red, 139; green, 6; blue, 24 }  ,draw opacity=1 ]   (445.64,289.91) -- (473.67,271.72) ;
\draw [color={rgb, 255:red, 139; green, 6; blue, 24 }  ,draw opacity=1 ]   (446.72,327.62) -- (500.94,290.69) ;
\draw    (560.77,291.02) -- (623,289.5) ;
\draw [color={rgb, 255:red, 139; green, 6; blue, 24 }  ,draw opacity=1 ]   (499.36,291.02) -- (560.77,291.02) ;
\draw [color={rgb, 255:red, 139; green, 6; blue, 24 }  ,draw opacity=1 ]   (96.77,336.66) -- (168.34,336.66) ;
\draw  [color={rgb, 255:red, 139; green, 6; blue, 24 }  ,draw opacity=1 ][fill={rgb, 255:red, 139; green, 6; blue, 24 }  ,fill opacity=1 ] (164.92,336.66) .. controls (164.92,335.74) and (165.69,334.99) .. (166.63,334.99) .. controls (167.58,334.99) and (168.34,335.74) .. (168.34,336.66) .. controls (168.34,337.59) and (167.58,338.33) .. (166.63,338.33) .. controls (165.69,338.33) and (164.92,337.59) .. (164.92,336.66) -- cycle ;
\draw [color={rgb, 255:red, 0; green, 0; blue, 0 }  ,draw opacity=1 ]   (166.47,261.99) -- (221.02,299.91) ;
\draw [color={rgb, 255:red, 0; green, 0; blue, 0 }  ,draw opacity=1 ]   (165.71,299.13) -- (193.74,280.95) ;
\draw [color={rgb, 255:red, 0; green, 0; blue, 0 }  ,draw opacity=1 ]   (166.79,336.85) -- (221.02,299.91) ;
\draw [color={rgb, 255:red, 0; green, 0; blue, 0 }  ,draw opacity=1 ]   (221.02,299.91) -- (252,300) ;
\draw  [color={rgb, 255:red, 37; green, 118; blue, 212 }  ,draw opacity=1 ][fill={rgb, 255:red, 30; green, 94; blue, 151 }  ,fill opacity=0.51 ] (328.14,162.4) -- (328.16,44.38) -- (406.79,23.08) -- (406.77,141.09) -- cycle ;
\draw [color={rgb, 255:red, 139; green, 6; blue, 24 }  ,draw opacity=1 ]   (281.23,51.13) -- (308.51,70.09) ;
\draw  [color={rgb, 255:red, 139; green, 6; blue, 24 }  ,draw opacity=1 ][fill={rgb, 255:red, 139; green, 6; blue, 24 }  ,fill opacity=1 ] (369.34,70.42) .. controls (369.34,69.5) and (370.1,68.76) .. (371.03,68.76) .. controls (371.96,68.76) and (372.72,69.5) .. (372.72,70.42) .. controls (372.72,71.33) and (371.96,72.07) .. (371.03,72.07) .. controls (370.1,72.07) and (369.34,71.33) .. (369.34,70.42) -- cycle ;
\draw [color={rgb, 255:red, 139; green, 6; blue, 24 }  ,draw opacity=1 ]   (280.47,88.28) -- (308.51,70.09) ;
\draw [color={rgb, 255:red, 139; green, 6; blue, 24 }  ,draw opacity=1 ]   (283.83,118.25) -- (371.03,116.85) ;
\draw    (414,88) -- (444,88) ;
\draw [color={rgb, 255:red, 139; green, 6; blue, 24 }  ,draw opacity=1 ]   (308.51,70.09) -- (371.03,70.42) ;
\draw  [color={rgb, 255:red, 139; green, 6; blue, 24 }  ,draw opacity=1 ][fill={rgb, 255:red, 139; green, 6; blue, 24 }  ,fill opacity=1 ] (369.32,116.85) .. controls (369.32,115.93) and (370.08,115.18) .. (371.03,115.18) .. controls (371.97,115.18) and (372.74,115.93) .. (372.74,116.85) .. controls (372.74,117.77) and (371.97,118.52) .. (371.03,118.52) .. controls (370.08,118.52) and (369.32,117.77) .. (369.32,116.85) -- cycle ;
\draw    (372.72,70.42) -- (414,88) ;
\draw    (372.74,116.85) -- (414,88) ;
\draw [color={rgb, 255:red, 139; green, 6; blue, 24 }  ,draw opacity=1 ]   (93.29,524.99) -- (166.83,524.99) ;
\draw [color={rgb, 255:red, 139; green, 6; blue, 24 }  ,draw opacity=1 ]   (93.01,561.51) -- (167.59,561.51) ;
\draw  [color={rgb, 255:red, 139; green, 6; blue, 24 }  ,draw opacity=1 ][fill={rgb, 255:red, 139; green, 6; blue, 24 }  ,fill opacity=1 ] (163.41,524.99) .. controls (163.41,524.07) and (164.17,523.32) .. (165.12,523.32) .. controls (166.06,523.32) and (166.83,524.07) .. (166.83,524.99) .. controls (166.83,525.92) and (166.06,526.66) .. (165.12,526.66) .. controls (164.17,526.66) and (163.41,525.92) .. (163.41,524.99) -- cycle ;
\draw  [color={rgb, 255:red, 139; green, 6; blue, 24 }  ,draw opacity=1 ][fill={rgb, 255:red, 139; green, 6; blue, 24 }  ,fill opacity=1 ] (164.16,561.51) .. controls (164.16,560.58) and (164.93,559.84) .. (165.88,559.84) .. controls (166.82,559.84) and (167.59,560.58) .. (167.59,561.51) .. controls (167.59,562.43) and (166.82,563.18) .. (165.88,563.18) .. controls (164.93,563.18) and (164.16,562.43) .. (164.16,561.51) -- cycle ;
\draw [color={rgb, 255:red, 139; green, 6; blue, 24 }  ,draw opacity=1 ]   (93.77,600.2) -- (168.34,600.2) ;
\draw  [color={rgb, 255:red, 139; green, 6; blue, 24 }  ,draw opacity=1 ][fill={rgb, 255:red, 139; green, 6; blue, 24 }  ,fill opacity=1 ] (164.92,600.2) .. controls (164.92,599.28) and (165.69,598.53) .. (166.63,598.53) .. controls (167.58,598.53) and (168.34,599.28) .. (168.34,600.2) .. controls (168.34,601.13) and (167.58,601.87) .. (166.63,601.87) .. controls (165.69,601.87) and (164.92,601.13) .. (164.92,600.2) -- cycle ;
\draw [color={rgb, 255:red, 0; green, 0; blue, 0 }  ,draw opacity=1 ]   (166.47,525.52) -- (221.02,563.45) ;
\draw [color={rgb, 255:red, 0; green, 0; blue, 0 }  ,draw opacity=1 ]   (165.71,562.67) -- (193.91,581.92) ;
\draw [color={rgb, 255:red, 0; green, 0; blue, 0 }  ,draw opacity=1 ]   (166.79,600.39) -- (221.02,563.45) ;
\draw [color={rgb, 255:red, 0; green, 0; blue, 0 }  ,draw opacity=1 ]   (221.02,563.45) -- (259,563) ;
\draw  [color={rgb, 255:red, 139; green, 6; blue, 24 }  ,draw opacity=1 ][fill={rgb, 255:red, 139; green, 6; blue, 24 }  ,fill opacity=1 ] (377.91,749.62) .. controls (377.91,748.71) and (378.67,747.97) .. (379.6,747.97) .. controls (380.53,747.97) and (381.29,748.71) .. (381.29,749.62) .. controls (381.29,750.53) and (380.53,751.27) .. (379.6,751.27) .. controls (378.67,751.27) and (377.91,750.53) .. (377.91,749.62) -- cycle ;
\draw    (415.21,769.74) -- (446,770) ;
\draw [color={rgb, 255:red, 139; green, 6; blue, 24 }  ,draw opacity=1 ]   (297.78,749.62) -- (379.6,749.62) ;
\draw  [color={rgb, 255:red, 139; green, 6; blue, 24 }  ,draw opacity=1 ][fill={rgb, 255:red, 139; green, 6; blue, 24 }  ,fill opacity=1 ] (377.89,796.06) .. controls (377.89,795.13) and (378.66,794.39) .. (379.6,794.39) .. controls (380.55,794.39) and (381.31,795.13) .. (381.31,796.06) .. controls (381.31,796.98) and (380.55,797.73) .. (379.6,797.73) .. controls (378.66,797.73) and (377.89,796.98) .. (377.89,796.06) -- cycle ;
\draw    (381.29,749.62) -- (415.21,769.74) ;
\draw    (381.31,796.06) -- (415.21,769.74) ;
\draw [color={rgb, 255:red, 139; green, 6; blue, 24 }  ,draw opacity=1 ]   (293.99,776.71) -- (321.95,796.39) ;
\draw [color={rgb, 255:red, 139; green, 6; blue, 24 }  ,draw opacity=1 ]   (295.07,814.42) -- (321.95,796.39) ;
\draw [color={rgb, 255:red, 139; green, 6; blue, 24 }  ,draw opacity=1 ]   (322.02,796.06) -- (379.6,796.06) ;
\draw    (198,204.5) -- (258.55,146.88) ;
\draw [shift={(260,145.5)}, rotate = 136.42] [color={rgb, 255:red, 0; green, 0; blue, 0 }  ][line width=0.75]    (10.93,-3.29) .. controls (6.95,-1.4) and (3.31,-0.3) .. (0,0) .. controls (3.31,0.3) and (6.95,1.4) .. (10.93,3.29)   ;
\draw    (473,132.5) -- (523.79,198.91) ;
\draw [shift={(525,200.5)}, rotate = 232.59] [color={rgb, 255:red, 0; green, 0; blue, 0 }  ][line width=0.75]    (10.93,-3.29) .. controls (6.95,-1.4) and (3.31,-0.3) .. (0,0) .. controls (3.31,0.3) and (6.95,1.4) .. (10.93,3.29)   ;
\draw    (560,379.5) -- (559.02,473.5) ;
\draw [shift={(559,475.5)}, rotate = 270.6] [color={rgb, 255:red, 0; green, 0; blue, 0 }  ][line width=0.75]    (10.93,-3.29) .. controls (6.95,-1.4) and (3.31,-0.3) .. (0,0) .. controls (3.31,0.3) and (6.95,1.4) .. (10.93,3.29)   ;
\draw    (191,377.5) -- (190.02,471.5) ;
\draw [shift={(190,473.5)}, rotate = 270.6] [color={rgb, 255:red, 0; green, 0; blue, 0 }  ][line width=0.75]    (10.93,-3.29) .. controls (6.95,-1.4) and (3.31,-0.3) .. (0,0) .. controls (3.31,0.3) and (6.95,1.4) .. (10.93,3.29)   ;
\draw    (229,636.5) -- (279.79,702.91) ;
\draw [shift={(281,704.5)}, rotate = 232.59] [color={rgb, 255:red, 0; green, 0; blue, 0 }  ][line width=0.75]    (10.93,-3.29) .. controls (6.95,-1.4) and (3.31,-0.3) .. (0,0) .. controls (3.31,0.3) and (6.95,1.4) .. (10.93,3.29)   ;
\draw    (467,702.5) -- (527.55,644.88) ;
\draw [shift={(529,643.5)}, rotate = 136.42] [color={rgb, 255:red, 0; green, 0; blue, 0 }  ][line width=0.75]    (10.93,-3.29) .. controls (6.95,-1.4) and (3.31,-0.3) .. (0,0) .. controls (3.31,0.3) and (6.95,1.4) .. (10.93,3.29)   ;

\draw (81.98,250.9) node [anchor=north west][inner sep=0.75pt]    {$x$};
\draw (435.83,550.4) node [anchor=north west][inner sep=0.75pt]    {$y$};
\draw (80.46,287.05) node [anchor=north west][inner sep=0.75pt]    {$y$};
\draw (435.56,589.01) node [anchor=north west][inner sep=0.75pt]    {$z$};
\draw (480.61,575.7) node [anchor=north west][inner sep=0.75pt]  [font=\normalsize]  {$u$};
\draw (619.42,562.5) node [anchor=north west][inner sep=0.75pt]    {$d$};
\draw (158.23,263.61) node [anchor=north west][inner sep=0.75pt]  [font=\small]  {$\psi $};
\draw (158.07,299.44) node [anchor=north west][inner sep=0.75pt]  [font=\small]  {$\chi $};
\draw (554.49,567.65) node [anchor=north west][inner sep=0.75pt]  [font=\small]  {$\sigma $};
\draw (469.13,585) node [anchor=north west][inner sep=0.75pt]  [font=\small]  {$i$};
\draw (45.38,135.11) node [anchor=north west][inner sep=0.75pt]    {$\sum \limits_{( a,\psi )( b,\chi ) ,\alpha }\left[ \Phi _{x,y}^{e}\right]{_{( a,\psi )( b,\chi ) ,\alpha }^{v,\varphi ,k}}$};
\draw (437.34,514.03) node [anchor=north west][inner sep=0.75pt]    {$x$};
\draw (434.67,279.19) node [anchor=north west][inner sep=0.75pt]    {$y$};
\draw (434.41,317.79) node [anchor=north west][inner sep=0.75pt]    {$z$};
\draw (518,290.2) node [anchor=north west][inner sep=0.75pt]    {$w$};
\draw (600.26,293.29) node [anchor=north west][inner sep=0.75pt]    {$d$};
\draw (552.34,293.44) node [anchor=north west][inner sep=0.75pt]  [font=\small]  {$\sigma $};
\draw (436.19,242.81) node [anchor=north west][inner sep=0.75pt]    {$x$};
\draw (508.64,564.64) node [anchor=north west][inner sep=0.75pt]    {$w$};
\draw (482.64,261.92) node [anchor=north west][inner sep=0.75pt]    {$v$};
\draw (496.41,564.65) node [anchor=north west][inner sep=0.75pt]  [font=\small]  {$j$};
\draw (470.07,276.8) node [anchor=north west][inner sep=0.75pt]  [font=\small]  {$k$};
\draw (493.58,295.7) node [anchor=north west][inner sep=0.75pt]  [font=\small]  {$l$};
\draw (572.92,399.12) node [anchor=north west][inner sep=0.75pt]    {$\sum \limits_{k,v,l}\left[ F_{xyz}^{w}\right]_{k,v,l}^{i,u,j}$};
\draw (81.22,326.75) node [anchor=north west][inner sep=0.75pt]    {$z$};
\draw (159.31,341.13) node [anchor=north west][inner sep=0.75pt]  [font=\small]  {$\xi $};
\draw (175.44,296.73) node [anchor=north west][inner sep=0.75pt]    {$b$};
\draw (193.06,317.5) node [anchor=north west][inner sep=0.75pt]    {$c$};
\draw (178.96,257.36) node [anchor=north west][inner sep=0.75pt]    {$a$};
\draw (231.93,280.9) node [anchor=north west][inner sep=0.75pt]    {$d$};
\draw (207.02,273.26) node [anchor=north west][inner sep=0.75pt]    {$e$};
\draw (187.51,284.09) node [anchor=north west][inner sep=0.75pt]  [font=\small]  {$\alpha $};
\draw (213.78,303.76) node [anchor=north west][inner sep=0.75pt]  [font=\small]  {$\beta $};
\draw (378.36,30.9) node [anchor=north west][inner sep=0.75pt]  [font=\small]  {$\CI_{RG}$};
\draw (269.51,77.56) node [anchor=north west][inner sep=0.75pt]    {$y$};
\draw (271.52,108.42) node [anchor=north west][inner sep=0.75pt]    {$z$};
\draw (386.07,106.23) node [anchor=north west][inner sep=0.75pt]    {$c$};
\draw (364.03,73.78) node [anchor=north west][inner sep=0.75pt]  [font=\small]  {$\varphi $};
\draw (271.02,41.18) node [anchor=north west][inner sep=0.75pt]    {$x$};
\draw (315.56,52) node [anchor=north west][inner sep=0.75pt]    {$v$};
\draw (303.14,74.17) node [anchor=north west][inner sep=0.75pt]  [font=\small]  {$k$};
\draw (364.63,122.38) node [anchor=north west][inner sep=0.75pt]  [font=\small]  {$\xi $};
\draw (424.64,69.43) node [anchor=north west][inner sep=0.75pt]    {$d$};
\draw (528.31,135.3) node [anchor=north west][inner sep=0.75pt]    {$\sum\limits_{( e,\varphi )( c,\xi ) ,i}\left[ \Phi _{v,z}^{d}\right]{_{( e,\varphi )( c,\xi ) ,\beta }^{w,\sigma ,l}}$};
\draw (388.09,58.34) node [anchor=north west][inner sep=0.75pt]    {$e$};
\draw (408.92,94.62) node [anchor=north west][inner sep=0.75pt]  [font=\small]  {$\beta $};
\draw (157.47,527.05) node [anchor=north west][inner sep=0.75pt]  [font=\small]  {$\psi $};
\draw (159.07,565.98) node [anchor=north west][inner sep=0.75pt]  [font=\small]  {$\chi $};
\draw (159.31,605.67) node [anchor=north west][inner sep=0.75pt]  [font=\small]  {$\xi $};
\draw (183.23,556.82) node [anchor=north west][inner sep=0.75pt]    {$b$};
\draw (197.87,529.73) node [anchor=north west][inner sep=0.75pt]    {$a$};
\draw (235.93,542.44) node [anchor=north west][inner sep=0.75pt]    {$d$};
\draw (188.26,585.55) node [anchor=north west][inner sep=0.75pt]  [font=\small]  {$\gamma $};
\draw (218.3,564.75) node [anchor=north west][inner sep=0.75pt]  [font=\small]  {$\delta $};
\draw (78.7,515.22) node [anchor=north west][inner sep=0.75pt]    {$x$};
\draw (79.19,552.37) node [anchor=north west][inner sep=0.75pt]    {$y$};
\draw (79.95,591.06) node [anchor=north west][inner sep=0.75pt]    {$z$};
\draw (53.62,397.82) node [anchor=north west][inner sep=0.75pt]    {$\sum\limits_{\alpha ,e,\beta }\left[ F_{abc}^{d}\right]_{\alpha ,e,\beta }^{\gamma ,f,\delta }$};
\draw (202.96,573.07) node [anchor=north west][inner sep=0.75pt]    {$f$};
\draw (57.77,680.35) node [anchor=north west][inner sep=0.75pt]    {$\sum \limits_{( b,\chi )( c,\xi ) ,\gamma }\left[ \Phi _{y,z}^{f}\right]{_{( b,\chi )( c,\xi ) ,\gamma }^{u,\rho ,i}}$};
\draw (393.64,784.53) node [anchor=north west][inner sep=0.75pt]    {$f$};
\draw (372.6,751.98) node [anchor=north west][inner sep=0.75pt]  [font=\small]  {$\psi $};
\draw (427.22,752.64) node [anchor=north west][inner sep=0.75pt]    {$d$};
\draw (489.46,685.8) node [anchor=north west][inner sep=0.75pt]    {$\sum\limits_{( a,\psi )( f,\rho ) ,\delta }\left[ \Phi _{x,u}^{d}\right]{_{( a,\psi )( f,\rho ) ,\delta }^{w,\sigma ,j}}$};
\draw (402.67,741.54) node [anchor=north west][inner sep=0.75pt]    {$a$};
\draw (411.49,773.82) node [anchor=north west][inner sep=0.75pt]  [font=\small]  {$\delta $};
\draw (286.47,739.31) node [anchor=north west][inner sep=0.75pt]    {$x$};
\draw (284.54,766.76) node [anchor=north west][inner sep=0.75pt]    {$y$};
\draw (284.27,803.82) node [anchor=north west][inner sep=0.75pt]    {$z$};
\draw (317.09,800.36) node [anchor=north west][inner sep=0.75pt]  [font=\small]  {$i$};
\draw (335.62,778.09) node [anchor=north west][inner sep=0.75pt]    {$u$};
\draw (373.6,798.81) node [anchor=north west][inner sep=0.75pt]  [font=\small]  {$\rho $};
\draw (174.45,592.69) node [anchor=north west][inner sep=0.75pt]    {$c$};
\draw (158.36,238.9) node [anchor=north west][inner sep=0.75pt]  [font=\small]  {$\CI_{RG}$};
\draw (155.36,501.9) node [anchor=north west][inner sep=0.75pt]  [font=\small]  {$\CI_{RG}$};
\draw (375.36,715.9) node [anchor=north west][inner sep=0.75pt]  [font=\small]  {$\CI_{RG}$};
\draw (565.36,227.87) node [anchor=north west][inner sep=0.75pt]  [font=\small]  {$\CI_{RG}$};
\draw (580.36,499.87) node [anchor=north west][inner sep=0.75pt]  [font=\small]  {$\CI_{RG}$};
\draw (406,224.4) node [anchor=north west][inner sep=0.75pt]    {$\CQ_{UV}$};
\draw (437.44,21.4) node [anchor=north west][inner sep=0.75pt]    {$\CQ_{IR}$};
\draw (228,21.4) node [anchor=north west][inner sep=0.75pt]    {$\CQ_{UV}$};
\draw (418,486.4) node [anchor=north west][inner sep=0.75pt]    {$\CQ_{UV}$};
\draw (262,712.4) node [anchor=north west][inner sep=0.75pt]    {$\CQ_{UV}$};
\draw (44,492.4) node [anchor=north west][inner sep=0.75pt]    {$\CQ_{UV}$};
\draw (53,228.4) node [anchor=north west][inner sep=0.75pt]    {$\CQ_{UV}$};
\draw (628.44,224.4) node [anchor=north west][inner sep=0.75pt]    {$\CQ_{IR}$};
\draw (627.44,486.4) node [anchor=north west][inner sep=0.75pt]    {$\CQ_{IR}$};
\draw (433.44,709.4) node [anchor=north west][inner sep=0.75pt]    {$\CQ_{IR}$};
\draw (237.44,494.4) node [anchor=north west][inner sep=0.75pt]    {$\CQ_{IR}$};
\draw (242.44,229.4) node [anchor=north west][inner sep=0.75pt]    {$\CQ_{IR}$};

\end{tikzpicture}
    \caption{Compatibility of $\CI_{RG}$ with the fusion of line operators.}
    \label{fig:I associativity condition}
\end{figure}
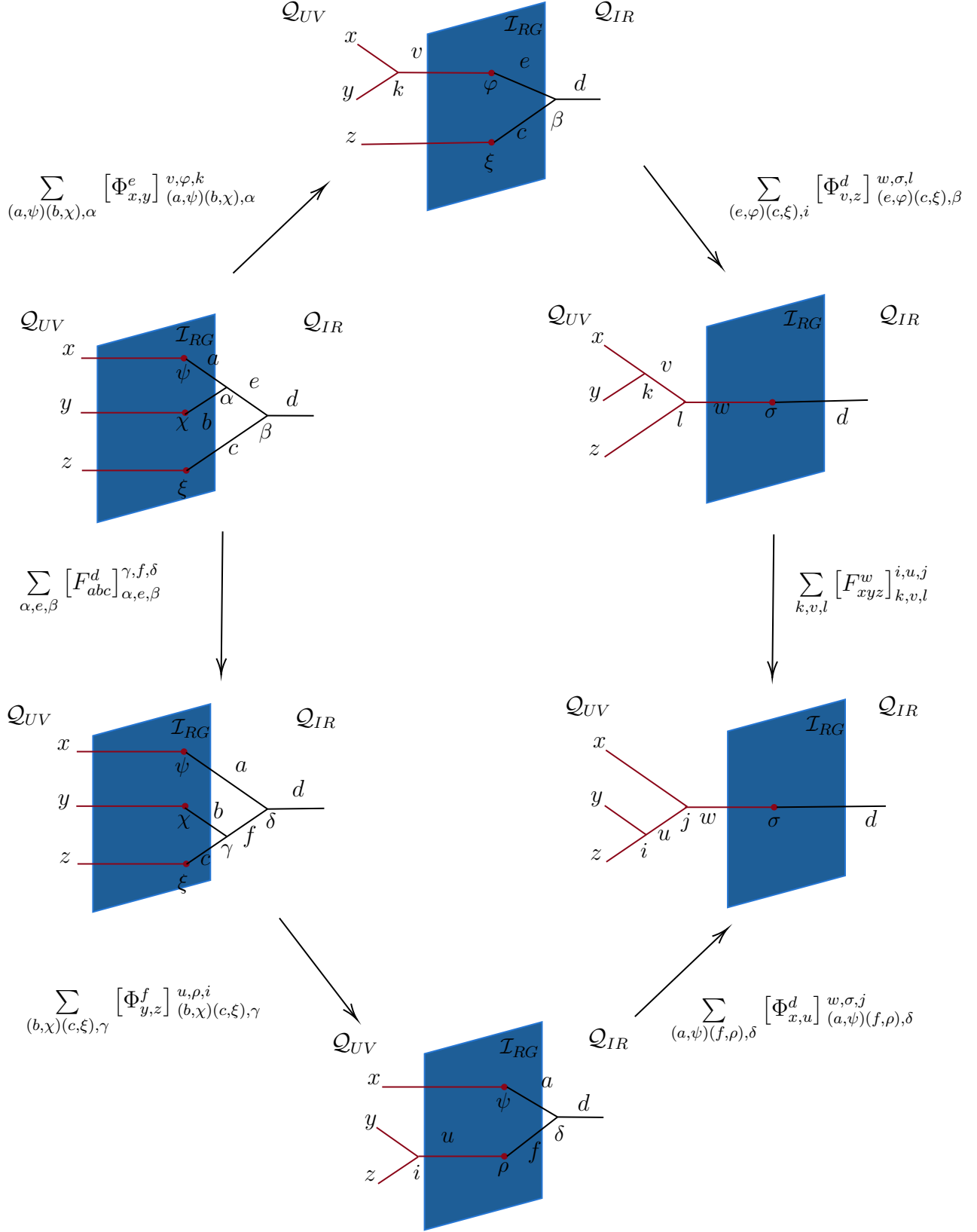

\item Since the braiding of two line operators in $\CB_{UV}$ does not depend on the distance between the line operators, the RG flow must preserve the braidings.\footnote{These conditions mean that, by definition, $F_{RG}$ is a braided monoidal functor \cite{baez2004some}.}
(see Fig. \ref{fig:braiding constraint})
\be
R_{x,y} \circ \Phi_{x,y} = \Phi_{y,x} \circ R_{F_{RG}(x),F_{RG}(y)}~.
\ee
By choosing a basis for the fusion spaces, this condition can be written explicitly as in Fig \ref{fig:braiding constraint}.
\begin{figure}
\centering

\tikzset{every picture/.style={line width=0.75pt}} 

\begin{tikzpicture}[x=0.75pt,y=0.75pt,yscale=-1,xscale=1]

\draw  [color={rgb, 255:red, 37; green, 118; blue, 212 }  ,draw opacity=1 ][fill={rgb, 255:red, 30; green, 94; blue, 151 }  ,fill opacity=0.51 ] (526.43,324.47) -- (526.45,209.85) -- (604.43,187.35) -- (604.41,301.97) -- cycle ;
\draw  [color={rgb, 255:red, 37; green, 118; blue, 212 }  ,draw opacity=1 ][fill={rgb, 255:red, 30; green, 94; blue, 151 }  ,fill opacity=0.51 ] (500.99,148.47) -- (501.01,33.85) -- (578.99,11.35) -- (578.97,125.97) -- cycle ;
\draw [color={rgb, 255:red, 139; green, 6; blue, 24 }  ,draw opacity=1 ]   (505.68,234.45) -- (565.42,234.45) ;
\draw [color={rgb, 255:red, 139; green, 6; blue, 24 }  ,draw opacity=1 ]   (506.5,279.05) -- (565.42,279.05) ;
\draw  [color={rgb, 255:red, 139; green, 6; blue, 24 }  ,draw opacity=1 ][fill={rgb, 255:red, 139; green, 6; blue, 24 }  ,fill opacity=1 ] (561.5,234.45) .. controls (561.5,233.22) and (562.38,232.22) .. (563.46,232.22) .. controls (564.54,232.22) and (565.42,233.22) .. (565.42,234.45) .. controls (565.42,235.68) and (564.54,236.68) .. (563.46,236.68) .. controls (562.38,236.68) and (561.5,235.68) .. (561.5,234.45) -- cycle ;
\draw  [color={rgb, 255:red, 139; green, 6; blue, 24 }  ,draw opacity=1 ][fill={rgb, 255:red, 139; green, 6; blue, 24 }  ,fill opacity=1 ] (561.5,279.05) .. controls (561.5,277.82) and (562.38,276.82) .. (563.46,276.82) .. controls (564.54,276.82) and (565.42,277.82) .. (565.42,279.05) .. controls (565.42,280.28) and (564.54,281.28) .. (563.46,281.28) .. controls (562.38,281.28) and (561.5,280.28) .. (561.5,279.05) -- cycle ;
\draw    (565.42,235.8) -- (613,256) ;
\draw    (565.42,279.05) -- (613,256) ;
\draw    (613,256) -- (654,256) ;
\draw [color={rgb, 255:red, 139; green, 6; blue, 24 }  ,draw opacity=1 ]   (421.07,62.86) -- (449.9,62.86) ;
\draw [color={rgb, 255:red, 139; green, 6; blue, 24 }  ,draw opacity=1 ]   (421.07,106.82) -- (449.9,106.82) ;
\draw  [color={rgb, 255:red, 139; green, 6; blue, 24 }  ,draw opacity=1 ][fill={rgb, 255:red, 139; green, 6; blue, 24 }  ,fill opacity=1 ] (538.73,85.99) .. controls (538.73,84.94) and (539.42,84.09) .. (540.27,84.09) .. controls (541.12,84.09) and (541.8,84.94) .. (541.8,85.99) .. controls (541.8,87.05) and (541.12,87.9) .. (540.27,87.9) .. controls (539.42,87.9) and (538.73,87.05) .. (538.73,85.99) -- cycle ;
\draw [color={rgb, 255:red, 139; green, 6; blue, 24 }  ,draw opacity=1 ]   (449.9,62.86) -- (475,85.99) ;
\draw [color={rgb, 255:red, 139; green, 6; blue, 24 }  ,draw opacity=1 ]   (449.9,106.82) -- (475,85.99) ;
\draw    (541.8,85.99) -- (618,86) ;
\draw [color={rgb, 255:red, 139; green, 6; blue, 24 }  ,draw opacity=1 ]   (474,85.99) -- (541.8,85.99) ;
\draw  [color={rgb, 255:red, 37; green, 118; blue, 212 }  ,draw opacity=1 ][fill={rgb, 255:red, 30; green, 94; blue, 151 }  ,fill opacity=0.51 ] (101.43,146.47) -- (101.45,31.85) -- (179.43,9.35) -- (179.41,123.97) -- cycle ;
\draw  [color={rgb, 255:red, 139; green, 6; blue, 24 }  ,draw opacity=1 ][fill={rgb, 255:red, 139; green, 6; blue, 24 }  ,fill opacity=1 ] (139.18,83.99) .. controls (139.18,82.94) and (139.86,82.09) .. (140.71,82.09) .. controls (141.56,82.09) and (142.24,82.94) .. (142.24,83.99) .. controls (142.24,85.05) and (141.56,85.9) .. (140.71,85.9) .. controls (139.86,85.9) and (139.18,85.05) .. (139.18,83.99) -- cycle ;
\draw    (142.24,83.99) -- (218.44,84) ;
\draw [color={rgb, 255:red, 139; green, 6; blue, 24 }  ,draw opacity=1 ]   (74.45,83.99) -- (142.24,83.99) ;
\draw [color={rgb, 255:red, 139; green, 6; blue, 24 }  ,draw opacity=1 ]   (25,66) .. controls (44,91) and (46,108) .. (74.45,83.99) ;
\draw [color={rgb, 255:red, 139; green, 6; blue, 24 }  ,draw opacity=1 ]   (43,85) .. controls (51,82) and (55.45,68.99) .. (74.45,83.99) ;
\draw [color={rgb, 255:red, 139; green, 6; blue, 24 }  ,draw opacity=1 ]   (23,98) .. controls (25,98) and (32,93) .. (37,90) ;
\draw  [color={rgb, 255:red, 37; green, 118; blue, 212 }  ,draw opacity=1 ][fill={rgb, 255:red, 30; green, 94; blue, 151 }  ,fill opacity=0.51 ] (528.93,756.47) -- (528.95,641.85) -- (606.93,619.35) -- (606.91,733.97) -- cycle ;
\draw [color={rgb, 255:red, 139; green, 6; blue, 24 }  ,draw opacity=1 ]   (508.18,666.45) -- (567.92,666.45) ;
\draw [color={rgb, 255:red, 139; green, 6; blue, 24 }  ,draw opacity=1 ]   (509,711.05) -- (567.92,711.05) ;
\draw  [color={rgb, 255:red, 139; green, 6; blue, 24 }  ,draw opacity=1 ][fill={rgb, 255:red, 139; green, 6; blue, 24 }  ,fill opacity=1 ] (564,666.45) .. controls (564,665.22) and (564.88,664.22) .. (565.96,664.22) .. controls (567.04,664.22) and (567.92,665.22) .. (567.92,666.45) .. controls (567.92,667.68) and (567.04,668.68) .. (565.96,668.68) .. controls (564.88,668.68) and (564,667.68) .. (564,666.45) -- cycle ;
\draw  [color={rgb, 255:red, 139; green, 6; blue, 24 }  ,draw opacity=1 ][fill={rgb, 255:red, 139; green, 6; blue, 24 }  ,fill opacity=1 ] (564,711.05) .. controls (564,709.82) and (564.88,708.82) .. (565.96,708.82) .. controls (567.04,708.82) and (567.92,709.82) .. (567.92,711.05) .. controls (567.92,712.28) and (567.04,713.28) .. (565.96,713.28) .. controls (564.88,713.28) and (564,712.28) .. (564,711.05) -- cycle ;
\draw    (567.92,667.8) -- (615.5,688) ;
\draw    (567.92,711.05) -- (615.5,688) ;
\draw    (615.5,688) -- (656.5,688) ;
\draw  [color={rgb, 255:red, 37; green, 118; blue, 212 }  ,draw opacity=1 ][fill={rgb, 255:red, 30; green, 94; blue, 151 }  ,fill opacity=0.51 ] (527.93,567.69) -- (527.95,453.07) -- (605.93,430.57) -- (605.91,545.2) -- cycle ;
\draw [color={rgb, 255:red, 139; green, 6; blue, 24 }  ,draw opacity=1 ]   (507.18,477.67) -- (566.92,477.67) ;
\draw [color={rgb, 255:red, 139; green, 6; blue, 24 }  ,draw opacity=1 ]   (508,522.28) -- (566.92,522.28) ;
\draw  [color={rgb, 255:red, 139; green, 6; blue, 24 }  ,draw opacity=1 ][fill={rgb, 255:red, 139; green, 6; blue, 24 }  ,fill opacity=1 ] (563,477.67) .. controls (563,476.44) and (563.88,475.44) .. (564.96,475.44) .. controls (566.04,475.44) and (566.92,476.44) .. (566.92,477.67) .. controls (566.92,478.9) and (566.04,479.9) .. (564.96,479.9) .. controls (563.88,479.9) and (563,478.9) .. (563,477.67) -- cycle ;
\draw  [color={rgb, 255:red, 139; green, 6; blue, 24 }  ,draw opacity=1 ][fill={rgb, 255:red, 139; green, 6; blue, 24 }  ,fill opacity=1 ] (563,522.28) .. controls (563,521.05) and (563.88,520.05) .. (564.96,520.05) .. controls (566.04,520.05) and (566.92,521.05) .. (566.92,522.28) .. controls (566.92,523.51) and (566.04,524.51) .. (564.96,524.51) .. controls (563.88,524.51) and (563,523.51) .. (563,522.28) -- cycle ;
\draw    (612.45,500.99) -- (653.45,500.99) ;
\draw    (565.96,479.9) .. controls (584.96,504.9) and (585,525) .. (613.45,500.99) ;
\draw    (583,502) .. controls (589,497) and (594.45,485.99) .. (613.45,500.99) ;
\draw    (564.96,521.05) .. controls (566.96,521.05) and (574,511) .. (579,508) ;
\draw  [color={rgb, 255:red, 37; green, 118; blue, 212 }  ,draw opacity=1 ][fill={rgb, 255:red, 30; green, 94; blue, 151 }  ,fill opacity=0.51 ] (102.43,553.47) -- (102.45,438.85) -- (180.43,416.35) -- (180.41,530.97) -- cycle ;
\draw  [color={rgb, 255:red, 139; green, 6; blue, 24 }  ,draw opacity=1 ][fill={rgb, 255:red, 139; green, 6; blue, 24 }  ,fill opacity=1 ] (140.18,490.99) .. controls (140.18,489.94) and (140.86,489.09) .. (141.71,489.09) .. controls (142.56,489.09) and (143.24,489.94) .. (143.24,490.99) .. controls (143.24,492.05) and (142.56,492.9) .. (141.71,492.9) .. controls (140.86,492.9) and (140.18,492.05) .. (140.18,490.99) -- cycle ;
\draw    (143.24,490.99) -- (219.44,491) ;
\draw [color={rgb, 255:red, 139; green, 6; blue, 24 }  ,draw opacity=1 ]   (75.45,490.99) -- (143.24,490.99) ;
\draw [color={rgb, 255:red, 139; green, 6; blue, 24 }  ,draw opacity=1 ]   (26,473) .. controls (45,498) and (47,515) .. (75.45,490.99) ;
\draw [color={rgb, 255:red, 139; green, 6; blue, 24 }  ,draw opacity=1 ]   (44,492) .. controls (52,489) and (56.45,475.99) .. (75.45,490.99) ;
\draw [color={rgb, 255:red, 139; green, 6; blue, 24 }  ,draw opacity=1 ]   (24,505) .. controls (26,505) and (33,500) .. (38,497) ;

\draw (410,19.4) node [anchor=north west][inner sep=0.75pt]    {$\CQ_{UV}$};
\draw (605,21.4) node [anchor=north west][inner sep=0.75pt]    {$\CQ_{IR}$};
\draw (494.9,222.91) node [anchor=north west][inner sep=0.75pt]    {$x$};
\draw (409.49,52.95) node [anchor=north west][inner sep=0.75pt]    {$x$};
\draw (494.29,270.36) node [anchor=north west][inner sep=0.75pt]    {$y$};
\draw (408.56,95.75) node [anchor=north west][inner sep=0.75pt]    {$y$};
\draw (484.81,88.7) node [anchor=north west][inner sep=0.75pt]    {$z$};
\draw (581.41,269.44) node [anchor=north west][inner sep=0.75pt]    {$b$};
\draw (584.36,225.04) node [anchor=north west][inner sep=0.75pt]    {$a$};
\draw (635.5,259.4) node [anchor=north west][inner sep=0.75pt]    {$c$};
\draw (597.33,85.76) node [anchor=north west][inner sep=0.75pt]    {$c$};
\draw (557.59,211.95) node [anchor=north west][inner sep=0.75pt]  [font=\small]  {$\psi $};
\draw (557.68,284.86) node [anchor=north west][inner sep=0.75pt]  [font=\small]  {$\chi $};
\draw (533.59,89.29) node [anchor=north west][inner sep=0.75pt]  [font=\small]  {$\xi $};
\draw (609.42,259.84) node [anchor=north west][inner sep=0.75pt]  [font=\small]  {$\alpha $};
\draw (467.97,91.07) node [anchor=north west][inner sep=0.75pt]  [font=\small]  {$j$};
\draw (233.84,231.4) node [anchor=north west][inner sep=0.75pt]    {$=\sum\limits_{( a,\psi )( b,\chi ) ,\alpha ,j}\left[ R_{xy}^{z}\right]_{j}^{k}\left[ \Phi _{x,y}^{c}\right]{_{( a,\psi )( b,\chi ) ,\alpha }^{ z,\xi  ,j}}$};
\draw (463.44,197.4) node [anchor=north west][inner sep=0.75pt]    {$\CQ_{UV}$};
\draw (622.44,191.4) node [anchor=north west][inner sep=0.75pt]    {$\CQ_{IR}$};
\draw (10.44,17.4) node [anchor=north west][inner sep=0.75pt]    {$\CQ_{UV}$};
\draw (205.44,19.4) node [anchor=north west][inner sep=0.75pt]    {$\CQ_{IR}$};
\draw (9.93,50.95) node [anchor=north west][inner sep=0.75pt]    {$x$};
\draw (9,93.75) node [anchor=north west][inner sep=0.75pt]    {$y$};
\draw (85.26,86.7) node [anchor=north west][inner sep=0.75pt]    {$z$};
\draw (197.77,83.76) node [anchor=north west][inner sep=0.75pt]    {$c$};
\draw (134.03,87.29) node [anchor=north west][inner sep=0.75pt]  [font=\small]  {$\xi $};
\draw (68.41,89.07) node [anchor=north west][inner sep=0.75pt]  [font=\small]  {$k$};
\draw (234,58.4) node [anchor=north west][inner sep=0.75pt]    {$=\ \sum _{j}\left[ R_{xy}^{z}\right]_{j}^{k}$};
\draw (497.4,654.91) node [anchor=north west][inner sep=0.75pt]    {$x$};
\draw (496.79,702.36) node [anchor=north west][inner sep=0.75pt]    {$y$};
\draw (583.91,705.44) node [anchor=north west][inner sep=0.75pt]    {$b$};
\draw (586.86,657.04) node [anchor=north west][inner sep=0.75pt]    {$a$};
\draw (638,691.4) node [anchor=north west][inner sep=0.75pt]    {$c$};
\draw (560.09,643.95) node [anchor=north west][inner sep=0.75pt]  [font=\small]  {$\psi $};
\draw (560.18,716.86) node [anchor=north west][inner sep=0.75pt]  [font=\small]  {$\chi $};
\draw (611.92,691.84) node [anchor=north west][inner sep=0.75pt]  [font=\small]  {$\alpha $};
\draw (222.34,654.4) node [anchor=north west][inner sep=0.75pt]    {$=\sum\limits_{( a,\psi )( b,\chi ) ,\beta ,\alpha }\left[ \Phi _{y,x}^{c}\right]_{( b,\chi )( a,\psi ) ,\beta }^{ z,\xi  ,k}\left[ R_{ab}^{c}\right]_{\alpha }^{\beta }$};
\draw (465.94,629.4) node [anchor=north west][inner sep=0.75pt]    {$\CQ_{UV}$};
\draw (624.94,623.4) node [anchor=north west][inner sep=0.75pt]    {$\CQ_{IR}$};
\draw (496.4,466.13) node [anchor=north west][inner sep=0.75pt]    {$x$};
\draw (495.79,513.59) node [anchor=north west][inner sep=0.75pt]    {$y$};
\draw (586.91,511.67) node [anchor=north west][inner sep=0.75pt]    {$a$};
\draw (588.86,472.26) node [anchor=north west][inner sep=0.75pt]    {$b$};
\draw (635,500.62) node [anchor=north west][inner sep=0.75pt]    {$c$};
\draw (559.09,455.17) node [anchor=north west][inner sep=0.75pt]  [font=\small]  {$\psi $};
\draw (559.18,528.08) node [anchor=north west][inner sep=0.75pt]  [font=\small]  {$\chi $};
\draw (610.45,503.39) node [anchor=north west][inner sep=0.75pt]  [font=\small]  {$\beta $};
\draw (464.94,440.62) node [anchor=north west][inner sep=0.75pt]    {$\CQ_{UV}$};
\draw (623.94,434.62) node [anchor=north west][inner sep=0.75pt]    {$\CQ_{IR}$};
\draw (230.84,465.4) node [anchor=north west][inner sep=0.75pt]    {$=\sum\limits_{( a,\psi )( b,\chi ) ,\beta }\left[ \Phi _{y,x}^{c}\right]{_{( b,\chi )( a,\psi ) ,\beta }^{ z,\xi  ,k}}$};
\draw (11.44,424.4) node [anchor=north west][inner sep=0.75pt]    {$\CQ_{UV}$};
\draw (206.44,426.4) node [anchor=north west][inner sep=0.75pt]    {$\CQ_{IR}$};
\draw (10.93,457.95) node [anchor=north west][inner sep=0.75pt]    {$x$};
\draw (10,500.75) node [anchor=north west][inner sep=0.75pt]    {$y$};
\draw (86.26,493.7) node [anchor=north west][inner sep=0.75pt]    {$z$};
\draw (198.77,490.76) node [anchor=north west][inner sep=0.75pt]    {$c$};
\draw (135.03,494.29) node [anchor=north west][inner sep=0.75pt]  [font=\small]  {$\xi $};
\draw (69.41,496.07) node [anchor=north west][inner sep=0.75pt]  [font=\small]  {$k$};
\draw (311,362) node [anchor=north west][inner sep=0.75pt]   [align=left] {(i)};
\draw (307,787) node [anchor=north west][inner sep=0.75pt]   [align=left] {(ii)};
\draw (150,17.4) node [anchor=north west][inner sep=0.75pt]    {$\CI_{RG}$};
\draw (549,19.4) node [anchor=north west][inner sep=0.75pt]    {$\CI_{RG}$};
\draw (575,197.4) node [anchor=north west][inner sep=0.75pt]    {$\CI_{RG}$};
\draw (150,425.4) node [anchor=north west][inner sep=0.75pt]    {$\CI_{RG}$};
\draw (577,438.4) node [anchor=north west][inner sep=0.75pt]    {$\CI_{RG}$};
\draw (578,627.4) node [anchor=north west][inner sep=0.75pt]    {$\CI_{RG}$};

\end{tikzpicture}
\caption{Compatibility of $\CI_{RG}$ with the braiding of line operators.}
\label{fig:braiding constraint}
\end{figure}
Comparing the expressions in Fig \ref{fig:braiding constraint} (i) and (ii), we get the equation
\be
\label{eq:cosntraint from braiding}
\sum _{j}\left[ R_{xy}^{z}\right]_{j}^{k}\left[ \Phi _{x,y}^{c}\right]{_{( a,\psi )( b,\chi ) ,\alpha }^{ z,\xi  ,j}}=\sum _{\beta}\left[ \Phi _{y,x}^{c}\right]_{( b,\chi )( a,\psi ) ,\beta }^{ z,\xi ,k}\left[ R_{ab}^{c}\right]_{\alpha }^{\beta }~.
\ee
\end{itemize}

Now, we are ready to prove Theorem 6 from the main text which we rephrase in light of the above discussion:

\vspace{0.3cm}

\noindent {\bf Theorem 6: }Consider an RG interface, $\CI_{RG}$, such that the topological data is mapped via $F_{RG}$ satisfying the assumptions above. Then, we have 
\begin{enumerate}
\item[(i)] $N_{F_{RG}(x)F_{RG}(y)}^{F_{RG}(z)}=N_{xy}^z$.
\item[(ii)] $d_{F_{RG}(x)}=d_x$.
\item[(iii)] If $N_{F_{RG}(x)}^a\neq 0$, then $\theta_{a}=\theta_{x}$~.
\end{enumerate}

\noindent \textbf{Proof:} \begin{itemize}
\item[(i)] By assumption, $\Phi_{x,y}$ are isomorphisms between fusion spaces $V_{xy}^z$ and $V_{F_{RG}(x)F_{RG}(y)}^{F_{RG}(z)}$. Therefore, we have $N_{F_{RG}(x)F_{RG}(y)}^{F_{RG}(z)}=N_{xy}^z~.$
\item[(ii)]Let $K_{UR}$ be the ring formed by the  fusion of line operators in $\CB_{UV}$. The ring $K_{IR}$ is defined similarly. From $(i)$, we know that 
\be
F_{RG}: K_{UV} \to K_{IR}~,
\ee
is a ring homomorphism. A ring homomorphism preserves Frobenius-Perron dimension of line operators \cite[Proposition 3.3.13]{etingof2016tensor}. 
\be
\text{FPdim}(F_{RG}(x))=\text{FPdim}(x)~.
\ee
We will assume that both $\CQ_{UV}$ and $\CQ_{IR}$ are unitary quantum field theories. Therefore, the braided fusion categories $\CB_{UV}$ and $\CB_{IR}$ are unitary. In a unitary braided fusion category, the Frobenius-Perron dimensions and quantum dimensions agree \cite[Proposition 9.5.1]{etingof2016tensor}. Therefore,
\be
d_{F_{RG}(x)}=d_{d_x} ~ \forall x \in \CB_{UV}~.
\ee
Note that this result is independent of the existence of $\Phi_{x,y}$ and only depends on $F_{RG}$ being a ring homomorphism between $K_{UV}$ and $K_{IR}$. 
\item[(iii)] Using the fact that $\Phi_{x,y}$ is an isomorphism, we can rewrite \eqref{eq:cosntraint from braiding} as
\be
\left[ R_{xy}^{z}\right]_{j}^{k}=\sum _{\alpha, \beta, (a,\psi), (b,\chi)} \left[ \Phi _{y,x}^{c}\right]_{( b,\chi )( a,\psi ) ,\beta }^{ z,\xi ,k}\left[ R_{ab}^{c}\right]_{\alpha }^{\beta } \left[ (\Phi_{x,y}^c)^{-1} \right]_{z,\xi,j}^{(a,\psi),(b,\chi),\alpha}~.
\ee
Then,
\bea
\sum_j \left[ R_{xx}^{z}\right]_{j}^{j}= \sum _{j,\alpha, \beta, (a,\psi), (b,\chi)} \left[ \Phi _{x,x}^{c}\right]_{( b,\chi )( a,\psi ) ,\beta }^{ z,\xi ,j}\left[ R_{ab}^{c}\right]_{\alpha }^{\beta } \left[ (\Phi_{x,x}^c)^{-1} \right]_{z,\xi,j}^{(a,\psi),(b,\chi),\alpha}~.
\eea
Note that the LHS is independent of both $c$ and $\xi$. Therefore, we have 
\bea
&& \sum_{c, \xi} d_c \sum _{j,\alpha, \beta, (a,\psi), (b,\chi)} \left[ \Phi _{x,x}^{c}\right]_{( b,\chi )( a,\psi ) ,\beta }^{ z,\xi ,j}\left[ R_{ab}^{c}\right]_{\alpha }^{\beta } \left[ (\Phi_{x,x}^c)^{-1} \right]_{z,\xi,j}^{(a,\psi),(b,\chi),\alpha} \nonumber \\
&=& \sum_{c, \xi} d_c \left[ R_{xx}^{z}\right]_{j}^{j} = d_z \left[ R_{xx}^{z}\right]_{j}^{j}~,
\eea
where, in the last equality, we used $\sum_{c,\xi} d_c = \sum_{c} N_{F_{RG}(z)}^{c} c = d_z$. Rearranging the sums in the equation above, we can write
\bea
\sum_{z,j}  \left[ R_{xx}^{z}\right]_{j}^{j}&=& \sum_c  d_c  \sum _{\alpha, \beta, (a,\psi), (b,\chi)} \sum_{z,\xi,j} \left[ \Phi _{x,x}^{c}\right]_{( b,\chi )( a,\psi ) ,\beta }^{ z,\xi ,j}\left[ R_{ab}^{c}\right]_{\alpha }^{\beta } \left[ (\Phi_{x,x}^c)^{-1} \right]_{z,\xi,j}^{(a,\psi),(b,\chi),\alpha} \nonumber \\
&=& \sum_c  d_c  \sum_{\alpha,\beta,(a,\psi),(b,\chi)} \left[ R_{ab}^{c}\right]_{\alpha }^{\beta } \delta_{(a,\psi),(b,\chi)} \delta_{\alpha,\beta} \nonumber \\
&=& \sum_c  d_c  \sum_{\alpha,(a,\psi)} \left[ R_{aa}^{c}\right]_{\alpha }^{\alpha}   = N_{F_{RG}(x)}^a  \sum_{c, \alpha} d_c  \left[ R_{aa}^{c}\right]_{\alpha }^{\alpha} ~,
\eea
where, in the second equality, we used the invertibility of $\Phi_{x,x}^c$. Therefore, using $\theta_x =\frac{1}{d_x} \sum_{z,j} \left[ R_{xx}^x \right ]_j^j$ and $\theta_a =\frac{1}{d_a} \sum_{c,\alpha} \left[ R_{aa}^c \right ]_\alpha^\alpha$, we get
\be
d_x \theta_x = \sum_{a} N_{F_{RG}(x)}^a d_a \theta_a~. 
\ee
Regarranging the terms, we get
\be
\sum_a \left ( N_{F_{RG}(x)}^a \frac{d_a}{d_x} \right) \frac{\theta_a}{\theta_x} =1~.
\ee
The sum is over phases with real non-negative coefficients. Moreover, $\sum_a N_{F_{RG}(x)}^a \frac{d_a}{d_x} = 1$. Therefore, we have 
\be
\theta_a = \theta_x ~ \forall ~ a \text{ such that } N_{F_{RG}(x)}^a\neq 0~.
\ee
\end{itemize}
\hfill $\square$
\vspace{0.2cm}

\noindent {\bf Corollary (one-form symmetry 't Hooft anomaly matching):} The braiding of line operators in $\CB_{UV}$ and $\CB_{IR}$ are related as follows
\be
\label{eq:anomaly matching}
S_{xy} = \sum_{a,b} N_{F_{RG}(x)}^a N_{F_{RG}(y)}^b~ S_{ab}~.
\ee

\noindent \textbf{Proof:} We have
\be
S_{x,y}= \sum_{z} N_{xy}^z \frac{\theta_z}{\theta_x \theta_y} d_z~.
\ee
Using Theorem 6, we find that
\bea
\theta_x \theta_y S_{x,y}&=& \sum_{z} N_{xy}^z \theta_z d_z = \sum_{z} N_{xy}^z  \sum_{c} N_{F_{RG}(z)}^c d_c \theta_c \\
&=& \sum_{a,b,c} N_{F_{RG}(x)}^a N_{F_{RG}(y)}^b  N_{ab}^c d_c \theta_c = \sum_{a,b} N_{F_{RG}(x)}^a N_{F_{RG}(y)}^b\theta_a \theta_b S_{a,b}~,
\eea
where, in the third equality, we used the isomorphism between $F_{RG}(x) \times F_{RG}(y)$ and $F_{RG}(x\times y)$. Using $\theta_a = \theta_x$ for all a such that $N_{F_{RG}(x)}^a\neq 0$ and $\theta_b = \theta_y$ for all a such that $N_{F_{RG}(y)}^b\neq 0$, we obtain
\be
S_{x,y} = \sum_{a,b} N_{F_{RG}(x)}^a N_{F_{RG}(y)}^b S_{a,b} ~.
\ee
\hfill $\square$

\end{appendices}

\newpage
\bibliography{chetdocbib}

\end{document}